\documentclass[11pt]{article}
\usepackage{jheppub}

\usepackage{graphicx}
\usepackage{dcolumn}
\usepackage{bm}
\usepackage{amssymb}
\usepackage{amsmath}
\usepackage{amsthm}
\usepackage{mathtools}
\usepackage[usenames,dvipsnames]{xcolor}
\usepackage{tikz}
\usepackage{tikz-cd}
\usetikzlibrary{arrows,shapes,positioning}
\usetikzlibrary{shapes.geometric, arrows}
\usepackage{appendix}
\usepackage[utf8]{inputenc}
\usepackage{dsfont}
\usepackage{cancel}
\usepackage{comment}
\usepackage{graphicx}
\usepackage{graphbox}
\usepackage{mathrsfs}
\usepackage{fontawesome5}
\usepackage[normalem]{ulem}
\usepackage{lscape}
\usepackage{stmaryrd}
\usepackage{esvect}
\usepackage{adjustbox}
\usetikzlibrary{external}
\usepackage[multiple]{footmisc}

\usepackage{enumitem}
\setlist{  
  listparindent=\parindent,
  parsep=0pt,
}

\usepackage{hyperref}
\hypersetup{
	colorlinks = True,
   	citecolor = BrickRed,
   	linkcolor = MidnightBlue,
        urlcolor = MidnightBlue
}

\usepackage[most]{tcolorbox}
\tcbset{
    enhanced jigsaw, 
    colback=gray!5!white,
    colframe=gray!20!white,
    coltitle=black,
    fonttitle=\bfseries
}

\newcommand{\Vbr}{E_{\mathfrak{g}}^{\begin{tikzpicture}[scale=0.7]
        \coordinate (A) at (0,0);
        \coordinate (B) at (2/5,0);
        \draw[thick,dotted] (A) --  (B);
        \fill[black] (A) circle (1.5pt);
        \fill[black] (B) circle (1.5pt);
\end{tikzpicture}}}
\newcommand{\Vso}{E_{\mathfrak{g}}^{\begin{tikzpicture}[scale=0.7]
        \coordinate (A) at (0,0);
        \coordinate (B) at (2/5,0);
        \draw[thick,double] (A) --  (B);
        \fill[black] (A) circle (1.5pt);
        \fill[black] (B) circle (1.5pt);
\end{tikzpicture}}}

\makeatletter
\newcommand{\github}{%
   {\href{https://github.com/ASphericalCow/KinematicFlowFromCuts.git}{\faGithub}}%
}
\makeatother

\newcommand{\nn}{\nonumber}
\renewcommand{\d}{\text{d}}

\newcommand{\res}{\text{Res}}

\newcommand{\vphi}{\varphi}

\newcommand{\la}{\langle}
\newcommand{\ra}{\rangle}
\newcommand{\mbf}[1]{\mathbf{#1}}
 
\newcommand{\mat}[1]{\underline{\boldsymbol{#1}}}
\newcommand{\del}[1]{\nabla}

\newcommand{\cdel}[1]{{\check{\nabla}}}

\newcommand{\kin}{\mathrm{kin}}

\newcommand{\B}{\mathcal{B}}

\newcommand{\dlog}{\mathrm{dlog}}

\newcommand{\C}{\mathcal{C}}
\newcommand\fnsep{\textsuperscript{,}}
\newcommand{\Vsf}{\mathsf{V}}

\newcommand{\T}{\mathcal{T}}
\renewcommand{\max}{\mathrm{max}}
\newcommand{\g}{\mathfrak{g}}
\newcommand{\gp}{{\mathfrak{g}^\prime}}
\newcommand{\sgn}{\mathrm{sgn}}

\newcommand{\lfrak}{\mathfrak{l}}
\newcommand{\lfrakp}{\mathfrak{l}^\prime}

\newcommand{\al}{\tikz \filldraw[scale=0.1, rotate=90] (0,0) -- (1,0) -- (0.5,0.866) -- cycle;}
\newcommand{\alb}{\tikz \filldraw[scale=0.15, rotate=90] (0,0) -- (1,0) -- (0.5,0.866) -- cycle;}
\newcommand{\albb}{\tikz \filldraw[scale=0.13, rotate=90] (0,0) -- (1,0) -- (0.5,0.866) -- cycle;}
\newcommand{\ar}{\tikz[] \filldraw[scale=0.1, rotate=-90] (0,0) -- (1,0) -- (0.5,0.866) -- cycle;}
\newcommand{\arbb}{\tikz[] \filldraw[scale=0.13, rotate=-90] (0,0) -- (1,0) -- (0.5,0.866) -- cycle;}
\newcommand{\arb}{\tikz[] \filldraw[scale=0.15, rotate=-90] (0,0) -- (1,0) -- (0.5,0.866) -- cycle;}

 \newcommand{\new}[1]{{\color{Green}#1}}

\newcommand{\be}{\begin{equation}\begin{aligned}}
\newcommand{\ee}{\end{aligned}\end{equation}}

\definecolor{darkCyan}{RGB}{0, 139, 139}
\definecolor{darkMagenta}{RGB}{139, 0, 139}

\title{Kinematic flow from the flow of cuts}

\author[]{Ross Glew$^1$ and}\emailAdd{r.glew@herts.ac.uk}
\author[]{Andrzej Pokraka$^2$}\emailAdd{andrzej\_pokraka@brown.edu.}

\affiliation[1]{Department of Physics, Astronomy and Mathematics, \\ University of Hertfordshire, \\  Hatfield, Hertfordshire, AL10 9AB, United Kingdom}
\affiliation[2]{Department of Physics, Brown University, \\ 182 Hope Street, Providence, RI 02912, U.S.A.}

\abstract{%
The wavefunction coefficients of conformally coupled scalars in power-law FRW cosmologies satisfy differential equations governed by a set of simple combinatorial rules known as the {\it kinematic flow}. 
In this paper we derive the kinematic flow, expressed using a set of differential forms referred to as the \emph{cut basis}, from a geometric perspective, relying solely on the cosmological hyperplane arrangement and without invoking bulk physics.
Each element of the cut basis corresponds to the \emph{positive geometry} associated to an independent cut of the physical FRW-form and can be labeled by decorating (minors of) the truncated Feynman graph with an acyclic orientation. 
We provide a straightforward prescription to associate a \emph{logarithmic} differential form to each element of the cut basis by considering its corresponding decorated graph. 
Moreover, we show that the residues of the physical FRW-form are canonical forms of certain \emph{graphical zonotopes} labeled by the same set of decorated graphs.  
These zonotopes control the cut combinatorics---\emph{flow of cuts}---of the physical FRW-form and the cut basis (by construction). 
Using the theory of \emph{relative twisted cohomology} and \emph{intersection theory}, we derive a closed form formula for the differential equations of the cut basis. 
We also introduce combinatorial rules that compute the kinematic differential of any basis element without explicit calculation.
The combinatorics of our differential equations is a natural consequence of the flow of cuts and is equivalent (up to rescaling) to the kinematic flow for the recently studied time integral basis. 
In particular, our differential equations decouple into exponentially many sectors, one for each way of cutting a subset of edges of the graph.
}

\begin{document}
\maketitle
\newpage

\section{Introduction}
\label{sec:intro}
In recent years, significant advances have been made in the study of scattering amplitudes driven by their connection to geometry. From the geometric point of view, certain scattering amplitudes are represented as canonical differential forms associated with positive geometries defined directly in the kinematic space of the underlying theory \cite{Arkani-Hamed:2017tmz}. Traditional properties of scattering amplitudes, such as factorization, then emerge naturally from the boundary structure of these geometries. The first example of this phenomenon was the amplituhedron whose associated canonical form calculates the tree-level scattering amplitudes and all loop integrand of planar $\mathcal{N}=4$ SYM \cite{Arkani-Hamed:2013jha}. Since then similar discoveries have been made for scattering amplitudes in other theories most notably $\text{tr}(\phi^3)$ and ABJM  \cite{Arkani-Hamed:2017mur,He:2023rou}. An exciting question moving forward is whether the techniques developed for the study of scattering amplitudes can be applied more generally to other physical quantities. 

One such family of observables receiving increased attention by the amplitudes community is that of {\it cosmological correlation functions}. Just as the cross section in particle physics is derived from the more fundamental scattering amplitudes, these cosmological observables are also derived from a more fundamental object, poetically named the {\it wavefunction of the universe}. An initial connection between the wavefunction and geometry was made by the discovery of the cosmological polytope, the canonical form of which computes a contribution to the flat space wavefunction coefficients arising from a single Feynman graph \cite{Arkani-Hamed:2017fdk}. 

Importantly, the flat space wavefunction coefficients serve as the fundamental ingredient in the computation of wavefunction coefficients in cosmologies beyond flat spacetime. The most commonly studied toy model, also the focus of this work, consists of conformally coupled scalars with polynomial interactions in a Friedmann-Robertson-Walker (FRW) cosmology. In this broader context, the wavefunction coefficients are no longer simple rational functions, but rather complicated analytic functions represented as certain twisted integrals; the behavior of which is governed by the theory of twisted cohomology. While the integrand of the wavefunction coefficient defines only one element of the twisted cohomology, it plays a special role in physics and our story. 
Therefore, we will refer to this distinguished element of the twisted cohomology as the \emph{physical FRW-form}. 

Moreover, since the theory of twisted cohomology is a finite dimensional vector space, as is familiar from the study of loop amplitudes \cite{Henn:2013pwa},
it follows that the basis of twisted integrals satisfy coupled linear differential equations written schematically as 
\begin{align}
\d_\kin \vv{\mathcal{I}}_G  = {\bf A}_G \cdot \vv{\mathcal{I}}_G. 
\end{align}
Here, $\vv{\mathcal{I}}_G$ is an appropriate choice of basis for the twisted cohomology into which the contribution to the wavefunction coefficient associated to a given graph can be expanded, $\d_\kin$ is the exterior derivative with respect to the kinematic variables and ${\bf A}_G$ is the connection matrix encoding the differential equations. 
A particular basis was first explored in \cite{Arkani-Hamed:2023bsv,Arkani-Hamed:2023kig}, where it was observed that the resulting differential equations obey a set of simple combinatorial rules known as the {\it kinematic flow}, expressed in terms of graph tubings. The kinematic flow has since been studied in detail in \cite{Arkani-Hamed:2023bsv, Baumann:2024mvm, Hang:2024xas, De:2024zic, He:2024olr, Baumann:2025qjx,Capuano:2025ehm}. 

In this paper, we refine our understanding of the {\it cut basis} proposed in \cite{De:2024zic} and provide a purely geometric derivation of kinematic flow.%
\footnote{
    Mathematically, cutting a propagator (denominator of a rational function) should be understood as taking the residue on the vanishing loci of that denominator. The jargon ``cut'' and ``propagator'' come from the specific physical significance of these denominators and residues.  
} 
The cut basis is intimately tied to the structure of the untwisted hyperplane arrangement of the propagators and the associated positive geometry on each maximal cut%
\footnote{%
    The maximal cut of an integral, is one where every propagator (special denominator) is cut. 
} 
of the physical FRW integral.  
It is constructed by leveraging a combinatorial understanding of the non-trivial sequential cuts/residues of the physical FRW integrand (flow of cuts) to isolate the independent cuts of the physical FRW form. 
This construction guarantees that the cut basis is a basis for the \emph{minimal} subspace of integrals that couple to the physical FRW integral in the differential equations.
Specifically, the physical sequential residue operators---those that do not annihilate the physical integrand---can be used to project onto the physical subspace providing an invariant definition for the subspace of interest \cite{De:2024zic}.%
\footnote{%
    In mathematics, such a subspace is called a monodromy invariant subspace.
    This is because the discontinuities (related to monodromies) of the physical FRW integral have integral representations where one replaces the physical integration contour with a contour that has a sequential residue as a component. 
}\fnsep\footnote{%
    In physics, using sequential residues to project integrals onto one another is called generalized unitarity.
    However, there are many situations where the sequential residue does not fully localize the integral. 
    Intersection theory provides an algorithmic way to construct residues operators that localize any remaining integration and project one integral onto another. 
}
Like the kinematic flow, differential equations for the cut basis can be written down without explicit calculation by considering the combinatorics of certain decorated graphs; we derive this combinatorial formula using simple residue calculus originating from intersection theory. 

It is important to note that the time integral basis used in the most recent formulation of the kinematics flow is equivalent to the cut basis up to a trivial rescaling after performing integration-by-parts \cite{He:2024olr, Baumann:2025qjx}. However, it suffers from two disadvantages, neither of which are present in our construction. First, as its name suggests, the construction of the time integral basis relies on bulk physics. This contradicts the original intent of the kinematic flow and positive geometry program: to formulate physics in such a way that bulk physics is emergent rather than an input. 
Second, many of the integrals in \cite{Baumann:2025qjx} have integrands that appear with higher-order poles. 
While these are integration-by-parts equivalent to integrands with simple poles, this extra integration-by-parts step obscures the connection to the underlying positive geometry. 

\section*{Summary of results}
In the remainder of the introduction we summarize the main results of the paper, including our choice of basis and its combinatorial interpretation, as well as the structure of the resulting differential equations and their connection to the geometry of zonotopes.

\begin{figure}
\centering
\includegraphics[width=\textwidth]{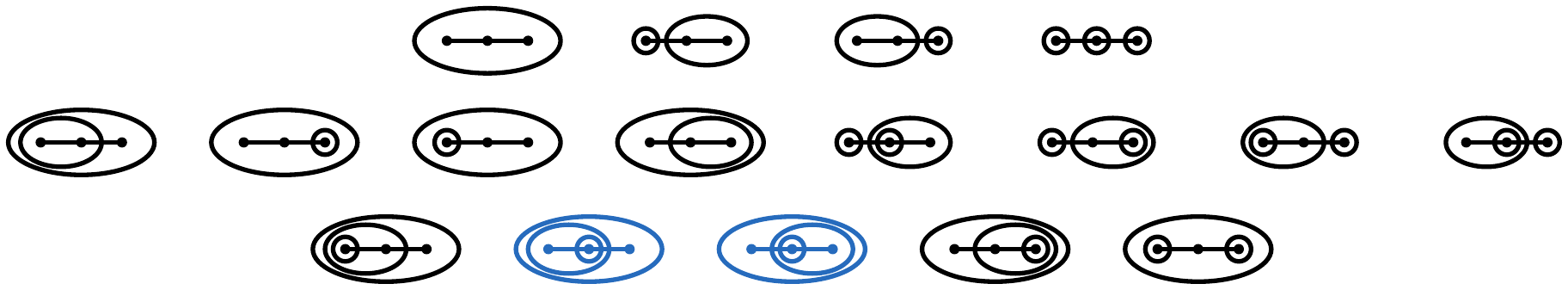}
    \caption{The cut tubings for the path graph on three vertices. The two cut tubings highlighted in blue are degenerate.}
    \label{fig:all_cuts}
\end{figure}

\paragraph{Cut basis from positive geometries.} 
Of central importance to our analysis is the cosmological hyperplane arrangement, $\B$, a collection of hyperplanes labeled by connected subgraphs (also known as tubes) of the truncated Feynman graph $G$. In addition to these, the coordinate hyperplanes, $\T$, play a distinguished role as twisted planes.  
We will often refer to the cosmological hyperplane arrangement, $\B$, as the untwisted hyperplane arrangement and the coordinate hyperplanes, $\T$, as the twisted hyperplane arrangement.
For example, in the case of the three-site chain, $G=\includegraphics[align=c]{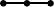}$, the arrangement consists of nine planes: the three (twisted) coordinate hyperplanes and six (untwisted) planes corresponding to the tubes. 
Each untwisted hyperplane is the vanishing loci of a polynomial labeled by $B_\bullet$: 
\begin{align}\begin{aligned}
    \Big\{ B_{ \begin{tikzpicture}[scale=0.8]
        \coordinate (A) at (0,0);
        \coordinate (B) at (1/2,0);
        \coordinate (C) at (1,0);
        \coordinate (D) at (3/2,0);
        \draw[thick] (A) -- (B) -- (C);
        \draw[black,thick] (A) circle (5pt);
        \fill[black] (A) circle (2pt);
        \fill[black] (B) circle (2pt);
        \fill[black] (C) circle (2pt);
    \end{tikzpicture} } 
    &,
    &
    B_{ \begin{tikzpicture}[scale=0.8]
        \coordinate (A) at (0,0);
        \coordinate (B) at (1/2,0);
        \coordinate (C) at (1,0);
        \coordinate (D) at (3/2,0);
        \draw[thick] (A) -- (B) -- (C);
        \draw[black,thick] (B) circle (5pt);
        \fill[black] (A) circle (2pt);
        \fill[black] (B) circle (2pt);
        \fill[black] (C) circle (2pt);
    \end{tikzpicture} } 
    &,  
    & 
    B_{ \begin{tikzpicture}[scale=0.8]
        \coordinate (A) at (0,0);
        \coordinate (B) at (1/2,0);
        \coordinate (C) at (1,0);
        \coordinate (D) at (3/2,0);
        \draw[thick] (A) -- (B) -- (C);
        \draw[black,thick] (C) circle (5pt);
        \fill[black] (A) circle (2pt);
        \fill[black] (B) circle (2pt);
        \fill[black] (C) circle (2pt);
    \end{tikzpicture} } 
    &,
    &
    B_{ \begin{tikzpicture}[scale=0.8]
        \coordinate (A) at (0,0);
        \coordinate (B) at (1/2,0);
        \coordinate (C) at (1,0);
        \coordinate (D) at (3/2,0);
        \draw[thick, black] (0.25,0) ellipse (0.42cm and 0.20cm);
        \draw[thick] (A) -- (B) -- (C);
        \fill[black] (A) circle (2pt);
        \fill[black] (B) circle (2pt);
        \fill[black] (C) circle (2pt);
    \end{tikzpicture} } 
    &, 
    &
    B_{ \begin{tikzpicture}[scale=0.8]
        \coordinate (A) at (0,0);
        \coordinate (B) at (1/2,0);
        \coordinate (C) at (1,0);
        \coordinate (D) at (3/2,0);
        \draw[thick, black] (0.75,0) ellipse (0.42cm and 0.2cm);
        \draw[thick] (A) -- (B) -- (C);
        \fill[black] (A) circle (2pt);
        \fill[black] (B) circle (2pt);
        \fill[black] (C) circle (2pt);
    \end{tikzpicture} } 
    &, 
    &
    B_{ \begin{tikzpicture}[scale=0.75]
        \coordinate (A) at (0,0);
        \coordinate (B) at (1/2,0);
        \coordinate (C) at (1,0);
        \coordinate (D) at (3/2,0);
        \draw[thick, black] (0.5,0) ellipse (0.75cm and 0.25cm);
        \draw[thick] (A) -- (B) -- (C);
        \fill[black] (A) circle (2pt);
        \fill[black] (B) circle (2pt);
        \fill[black] (C) circle (2pt);
    \end{tikzpicture} } 
    & \Big\}.
\end{aligned}\end{align} 
The basis of integrals used in this paper is motivated by the combinatorics of the cuts of the physical FRW-form presented in \cite{De:2024zic}. 
There, it was shown that the set of (physical) cuts
can be indexed by a combinatorial construction on the graph referred to as {\it cut-tubings}. 
A cut tubing,  $\tau$, is a set of tubes which satisfy compatibility conditions (see \cite{De:2024zic} for details). 
Each cut tubing, $\tau$, indexes the intersection $\B_\tau := \cap_{t\in\tau} \Vsf(B_t) \subset \B$%
\footnote{%
    We use $\Vsf$ for the vanishing symbol to distinguish it from $V_G$, which is used for the set of vertices of a graph $G$.
    Given a polynomial $P$, $\Vsf(P)$ is the variety corresponding to the vanishing loci of $P$. 
}
such that, given a global ordering for the $B_\bullet$, the sequential residue is compatible with this ordering and \emph{does not} annihilate the physical FRW integrand. The cut tubings for the three-site chain are shown in figure~\ref{fig:all_cuts}.

As argued in \cite{De:2024zic}, restricting the twisted arrangement $\mathcal{T}$ (coordinate hyperplanes) to the intersection of untwisted planes labeled by a cut tubing (i.e., $\B_\tau$), defines a single bounded chamber. 
This bounded chamber is an example of a {\it positive geometry $\Gamma_\tau$} \cite{Arkani-Hamed:2017tmz}, and, as such, comes equipped with a natural logarithmic differential form $\Omega_{\Gamma_\tau}$ called the canonical form. 
Each element of the cut basis is a differential form generated from two pieces of information: 
the positive geometry of a cut and the cut istelf. 
Schematically, every element of the cut basis can be bought to the form $\dlog_\tau  \propto \bigwedge_{t\in\tau} \dlog B_t \wedge  \tilde\Omega_{\Gamma_\tau}$ such that $\dlog_\tau$ becomes the canonical form of ${\Gamma_\tau}$ after cutting all $B_{t\in\tau}$ (i.e., taking the maximal cut of $\dlog_\tau$)
\be
    \Omega_{\Gamma_\tau}= \res_{B_{t_{|\tau|}} = 0} {\circ} {\cdots} {\circ} \res_{B_{t_1} = 0} [ 
		\dlog B_{t_1} {\wedge} {\cdots} {\wedge} \dlog B_{t_{|\tau|}} {\wedge} \tilde\Omega_{\Gamma\tau}
    ]
    \new{%
    = \tilde{\Omega}_{\Gamma_\tau} \vert_{B_{t\in\tau}=0}
    }
    \,.
\ee 
The positive geometry ${\Gamma_\tau}$ is extremely simple: it is the cartesian product of simplicies. 
This simplifies the construction of $\tilde\Omega_{\Gamma_\tau}$ and the residue calculus from intersection theory. 
Moreover, each element of the cut basis can be conveniently packaged into a simple one-line combinatorial formula presented in the main text.

\begin{figure}
\centering
\includegraphics[width=0.9\textwidth]{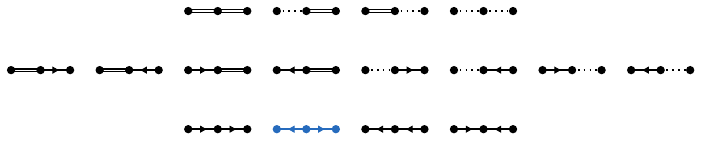}
    \caption{The acyclic minors for the path graph on three vertices. The acyclic minor highlighted in blue is associated to two degenerate cut tubings. }
    \label{fig:all_mins}
\end{figure}

\paragraph{Resolving degeneracies.}
One disadvantage of working with the set of cut tubings is that they are generally {\it degenerate}, in the sense that multiple cut tubings can label the same geometric space $\B_\tau = \B_{\tau'}$. 
This stems from linear dependencies between the hyperplanes and leads to linear relations among the forms $\dlog_\tau$. 
For example, the cut tubings highlighted in blue on figure~\ref{fig:all_cuts} are degenerate and label the same cut
\be
\B_{\raisebox{0cm}{\begin{tikzpicture}[scale=0.8]
        \coordinate (A) at (0,0);
        \coordinate (B) at (1/2,0);
        \coordinate (C) at (1,0);
        \coordinate (D) at (3/2,0);
        \draw[thick] (A) -- (B) --(C);
        \draw[thick, black] (0.27,0) ellipse (0.44cm and 0.25cm);
        \draw[thick, black] (0.5,0) ellipse (0.78cm and 0.34cm);
        \draw[black,thick] (B) circle (4pt);
        \fill[black] (B) circle (2pt);
        \fill[black] (A) circle (2pt);
        \fill[black] (C) circle (2pt);
    \end{tikzpicture}}} 
    &=  \Vsf( B_{\raisebox{-0.1cm}{\begin{tikzpicture}[scale=0.8]
        \coordinate (A) at (0,0);
        \coordinate (B) at (1/2,0);
        \coordinate (C) at (1,0);
        \draw[thick] (A) -- (B) --(C);
        \draw[thick, black] (0.25,0) ellipse (0.44cm and 0.22cm);
        \fill[black] (B) circle (2pt);
        \fill[black] (A) circle (2pt);
        \fill[black] (C) circle (2pt);
    \end{tikzpicture}}})  \cap 
    \Vsf( B_{\raisebox{-0.02cm}{\begin{tikzpicture}[scale=0.8]
        \coordinate (A) at (0,0);
        \coordinate (B) at (1/2,0);
        \coordinate (C) at (1,0);
        \coordinate (D) at (3/2,0);
        \draw[thick] (A) -- (B) --(C);
        \draw[black,thick] (B) circle (4pt);
        \fill[black] (B) circle (2pt);
        \fill[black] (A) circle (2pt);
        \fill[black] (C) circle (2pt);
    \end{tikzpicture}}}) 
    \cap \Vsf( B_{\raisebox{-0.1cm}{\begin{tikzpicture}[scale=0.8]
        \coordinate (A) at (0,0);
        \coordinate (B) at (1/2,0);
        \coordinate (C) at (1,0);
        \coordinate (D) at (3/2,0);
        \draw[thick] (A) -- (B) --(C);
        \draw[thick, black] (0.5,0) ellipse (0.75cm and 0.22cm);
        \fill[black] (B) circle (2pt);
        \fill[black] (A) circle (2pt);
        \fill[black] (C) circle (2pt);
    \end{tikzpicture}}})
   \\&=  \Vsf( B_{\raisebox{-0.1cm}{\begin{tikzpicture}[scale=0.8]
        \coordinate (A) at (0,0);
        \coordinate (B) at (1/2,0);
        \coordinate (C) at (1,0);
        \coordinate (D) at (3/2,0);
        \draw[thick] (A) -- (B) --(C);
        \draw[thick, black] (0.5,0) ellipse (0.75cm and 0.22cm);
        \fill[black] (B) circle (2pt);
        \fill[black] (A) circle (2pt);
        \fill[black] (C) circle (2pt);
    \end{tikzpicture}}})  \cap \Vsf( B_{\raisebox{-0.1cm}{\begin{tikzpicture}[scale=0.8]
        \coordinate (A) at (0,0);
        \coordinate (B) at (1/2,0);
        \coordinate (C) at (1,0);
        \draw[thick] (A) -- (B) --(C);
        \draw[thick, black] (0.75,0) ellipse (0.44cm and 0.22cm);
        \fill[black] (B) circle (2pt);
        \fill[black] (A) circle (2pt);
        \fill[black] (C) circle (2pt);
    \end{tikzpicture}}})  \cap \Vsf(B_{\raisebox{-0.02cm}{\begin{tikzpicture}[scale=0.8]
        \coordinate (A) at (0,0);
        \coordinate (B) at (1/2,0);
        \coordinate (C) at (1,0);
        \coordinate (D) at (3/2,0);
        \draw[thick] (A) -- (B) --(C);
        \draw[black,thick] (B) circle (4pt);
        \fill[black] (B) circle (2pt);
        \fill[black] (A) circle (2pt);
        \fill[black] (C) circle (2pt);
    \end{tikzpicture}}})
    = \B_{\raisebox{0cm}{\begin{tikzpicture}[scale=0.8]
        \coordinate (A) at (0,0);
        \coordinate (B) at (1/2,0);
        \coordinate (C) at (1,0);
        \coordinate (D) at (3/2,0);
        \draw[thick] (A) -- (B) --(C);
        \draw[thick, black] (0.73,0) ellipse (0.44cm and 0.25cm);
        \draw[thick, black] (0.5,0) ellipse (0.78cm and 0.34cm);
        \draw[black,thick] (B) circle (4pt);
        \fill[black] (B) circle (2pt);
        \fill[black] (A) circle (2pt);
        \fill[black] (C) circle (2pt);
    \end{tikzpicture}}}
    \,.
\ee
A solution this degeneracy problem is to work instead with the set of {\it acyclic minors} of the graph \cite{Glew:2025arc, Glew:2025ugf}. An acyclic minor $\g$ is a graph formed by first deleting and/or contracting a subset of edges of $G$, then assigning an acyclic orientation to the remaining edges. Remarkably, the set of acyclic minors counts each degenerate family of cut tubings exactly once! Building on this observation, it is natural to assign a differential form directly to the acyclic minor given by
\begin{align}
	\phi_\g = \sum_{\tau \in \mathcal{C}_\g} \dlog_\tau,
\end{align}
where $\C_\g$ denotes the set of degenerate cuts associated to the acyclic minor $\g$. We refer to the set of forms generated by the acyclic minors $\{ \phi_{\g} \}$ as the cut basis (these match the definitions in \cite{De:2024zic} up to signs).

Returning to the example of the three-site chain, the complete set of acyclic minors are shown in figure~\ref{fig:all_mins}. The oriented graph highlighted in \textcolor{NavyBlue}{blue} is associated to the two highlighted cut tubings in figure~\ref{fig:all_cuts}. Therefore, the form associated to this degenerate cut is
\begin{align}
\phi_{\begin{tikzpicture}[scale=0.8]
        \coordinate (A) at (0,0);
        \coordinate (B) at (1/2,0);
        \coordinate (C) at (1,0);
        \draw[thick] (B) --  node {\ar} (C);
        \draw[thick] (A) --  node {\al} (B);
        \fill[black] (A) circle (2pt);
        \fill[black] (B) circle (2pt);
        \fill[black] (C) circle (2pt);
    \end{tikzpicture}}  =  \dlog_{\raisebox{0cm}{\begin{tikzpicture}[scale=0.8]
        \coordinate (A) at (0,0);
        \coordinate (B) at (1/2,0);
        \coordinate (C) at (1,0);
        \coordinate (D) at (3/2,0);
        \draw[thick] (A) -- (B) --(C);
        \draw[thick, black] (0.27,0) ellipse (0.44cm and 0.25cm);
        \draw[thick, black] (0.5,0) ellipse (0.78cm and 0.34cm);
        \draw[black,thick] (B) circle (4pt);
        \fill[black] (B) circle (2pt);
        \fill[black] (A) circle (2pt);
        \fill[black] (C) circle (2pt);
    \end{tikzpicture}}}+\dlog_{\raisebox{0cm}{\begin{tikzpicture}[scale=0.8]
        \coordinate (A) at (0,0);
        \coordinate (B) at (1/2,0);
        \coordinate (C) at (1,0);
        \coordinate (D) at (3/2,0);
        \draw[thick] (A) -- (B) --(C);
        \draw[thick, black] (0.73,0) ellipse (0.44cm and 0.25cm);
        \draw[thick, black] (0.5,0) ellipse (0.78cm and 0.34cm);
        \draw[black,thick] (B) circle (4pt);
        \fill[black] (B) circle (2pt);
        \fill[black] (A) circle (2pt);
        \fill[black] (C) circle (2pt);
    \end{tikzpicture}}} &= \dlog  \raisebox{-0.1cm}{\begin{tikzpicture}[scale=0.8]
        \coordinate (A) at (0,0);
        \coordinate (B) at (1/2,0);
        \coordinate (C) at (1,0);
        \draw[thick] (A) -- (B) --(C);
        \draw[thick, black] (0.25,0) ellipse (0.44cm and 0.22cm);
        \fill[black] (B) circle (2pt);
        \fill[black] (A) circle (2pt);
        \fill[black] (C) circle (2pt);
    \end{tikzpicture}}  \wedge \dlog \  \raisebox{-0.02cm}{\begin{tikzpicture}[scale=0.8]
        \coordinate (A) at (0,0);
        \coordinate (B) at (1/2,0);
        \coordinate (C) at (1,0);
        \coordinate (D) at (3/2,0);
        \draw[thick] (A) -- (B) --(C);
        \draw[black,thick] (B) circle (4pt);
        \fill[black] (B) circle (2pt);
        \fill[black] (A) circle (2pt);
        \fill[black] (C) circle (2pt);
    \end{tikzpicture}} \wedge \dlog \  \raisebox{-0.1cm}{\begin{tikzpicture}[scale=0.8]
        \coordinate (A) at (0,0);
        \coordinate (B) at (1/2,0);
        \coordinate (C) at (1,0);
        \coordinate (D) at (3/2,0);
        \draw[thick] (A) -- (B) --(C);
        \draw[thick, black] (0.5,0) ellipse (0.75cm and 0.22cm);
        \fill[black] (B) circle (2pt);
        \fill[black] (A) circle (2pt);
        \fill[black] (C) circle (2pt);
    \end{tikzpicture}} \notag \\
    &+\dlog  \ \raisebox{-0.1cm}{\begin{tikzpicture}[scale=0.8]
        \coordinate (A) at (0,0);
        \coordinate (B) at (1/2,0);
        \coordinate (C) at (1,0);
        \coordinate (D) at (3/2,0);
        \draw[thick] (A) -- (B) --(C);
        \draw[thick, black] (0.5,0) ellipse (0.75cm and 0.22cm);
        \fill[black] (B) circle (2pt);
        \fill[black] (A) circle (2pt);
        \fill[black] (C) circle (2pt);
    \end{tikzpicture}}  \wedge \dlog \ \raisebox{-0.1cm}{\begin{tikzpicture}[scale=0.8]
        \coordinate (A) at (0,0);
        \coordinate (B) at (1/2,0);
        \coordinate (C) at (1,0);
        \draw[thick] (A) -- (B) --(C);
        \draw[thick, black] (0.75,0) ellipse (0.44cm and 0.22cm);
        \fill[black] (B) circle (2pt);
        \fill[black] (A) circle (2pt);
        \fill[black] (C) circle (2pt);
    \end{tikzpicture}}  \wedge \dlog \ \raisebox{-0.02cm}{\begin{tikzpicture}[scale=0.8]
        \coordinate (A) at (0,0);
        \coordinate (B) at (1/2,0);
        \coordinate (C) at (1,0);
        \coordinate (D) at (3/2,0);
        \draw[thick] (A) -- (B) --(C);
        \draw[black,thick] (B) circle (4pt);
        \fill[black] (B) circle (2pt);
        \fill[black] (A) circle (2pt);
        \fill[black] (C) circle (2pt);
    \end{tikzpicture}}
    \,.
\end{align}
The remaining (black) diagrams in figures \ref{fig:all_cuts} and \ref{fig:all_mins} are in one-to-one correspondence and have been ordered accordingly.

\begin{figure}
    \centering
    \includegraphics[width=.45\textwidth,align=c]{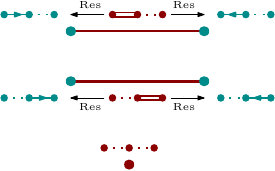}
    \qquad
    \includegraphics[width=.45\textwidth,align=c]{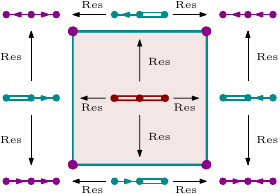}
    \caption{%
    Flow of cuts for the 3-site chain graph. 
    The top-dimensional facets of the zonotope are {\color{BrickRed}red}, the codimension-1 facets are {\color{darkCyan}cyan} and the codimension-2 facets are {\color{darkMagenta}magenta}. 
    }
    \label{fig:3chainZonosIntro}
\end{figure}

\paragraph{Emergence of zonotopes.} 
The physical FRW-form is represented in terms of the cut basis as
\begin{align}\begin{aligned}
    \Psi_{\includegraphics[scale=0.75]{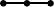}} {=}&
    ( 
        \phi_{\includegraphics[scale=0.75]{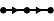}}
        {+}\phi_{\includegraphics[scale=0.75]{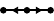}}
        {+}\phi_{\includegraphics[scale=0.75]{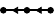}} 
        {+}\phi_{\includegraphics[scale=0.75]{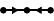}}
    )
    {-}(
        \phi_{\includegraphics[scale=0.75]{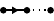}}
        {+}\phi_{\includegraphics[scale=0.75]{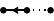}}
    )
    {-}(
        \phi_{\includegraphics[scale=0.75]{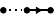}} 
        {+}\phi_{\includegraphics[scale=0.75]{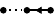}} 
    )
    {+} ( \phi_{\includegraphics[scale=0.75]{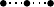}} )
    \,.  
\end{aligned}\end{align}
Each term in this expression can be translated to a distinct point ${\bf x}^*_\g = \B_{\tau\in\C_\g}$ in the cosmological hyperplane arrangement. Remarkably, taking the convex hull of the points grouped as above yields a family of graphical zonotopes
\be
\mathcal{Z}_{\includegraphics[scale=.8,align=c]{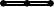}} 
	&= \text{conv} \{ {\bf x}^*_{\includegraphics[scale=0.75]{figures/amp_t2}}
        ,{\bf x}^*_{\includegraphics[scale=0.75]{figures/amp_t3}}
        ,{\bf x}^*_{\includegraphics[scale=0.75]{figures/amp_t4}} 
        ,{\bf x}^*_{\includegraphics[scale=0.75]{figures/amp_t5}}
         \}, 
        \\
	\mathcal{Z}_{\includegraphics[scale=.8,align=c]{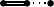}} 
	&= \text{conv} \{ {\bf x}^*_{\includegraphics[scale=0.75]{figures/amp_t8}}
        ,{\bf x}^*_{\includegraphics[scale=0.75]{figures/amp_t9}}
        \}, 
        \\
	\mathcal{Z}_{\includegraphics[scale=.8,align=c]{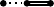}} 
	&= \text{conv} \{ {\bf x}^*_{\includegraphics[scale=0.75]{figures/amp_t6}}
        ,{\bf x}^*_{\includegraphics[scale=0.75]{figures/amp_t7}}
         \}, 
         \\
\mathcal{Z}_{\includegraphics[scale=.8,align=c]{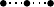}} &= \text{conv} \{ {\bf x}^*_{\includegraphics[scale=0.75]{figures/amp_t10}} \}.
\ee
As we will see, the combinatorics of compatible cuts—referred to as the {\it flow of cuts}—is encoded in these zonotopes. This can be seen by noting that the residues of the physical FRW-form, evaluated on any collection of tubes that partition the vertices of the graph, coincide with the canonical form of a zonotope or one of its boundaries
\begin{align}\begin{aligned}
    \res_{\includegraphics[scale=.8,align=c]{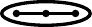}}[\Psi_{\includegraphics[scale=.8]{figures/amp_t1}}]
    &= \Omega[\mathcal{Z}_{
            \includegraphics[scale=.8,align=c]{figures/delta_cc.pdf}
        }]
    \,, &
    \res_{\includegraphics[scale=.8,align=c]{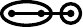}}[\Psi_{\includegraphics[scale=.8]{figures/amp_t1}}]
    &= \Omega[\mathcal{Z}_{
            \includegraphics[scale=.8,align=c]{figures/delta_cb.pdf}
        }]
    \,,
    \\
    \res_{\includegraphics[scale=.8,align=c]{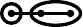}}[\Psi_{\includegraphics[scale=.8]{figures/amp_t1}}]
    &= \Omega[\mathcal{Z}_{
            \includegraphics[scale=.8,align=c]{figures/delta_bc.pdf}
        }]
    \,, &
    \res_{\includegraphics[scale=.8,align=c]{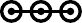}}[\Psi_{\includegraphics[scale=.8]{figures/amp_t1}}]
    &= \Omega[
        \mathcal{Z}_{
            \includegraphics[scale=.8,align=c]{figures/delta_bb.pdf}
        }
    ]
    \,.
\end{aligned}\end{align}
Further sequential residues of the physical FRW-form are governed by the combinatorics of these zonotopes, giving rise to the flow of cuts illustrated in figure \ref{fig:3chainZonosIntro}. 

\paragraph{Kinematic flow.}
It turns out that the kinematic flow inherits the reverse flow structure defined by the flow of cuts. As a result, the differential equations decouple into exponentially many sectors, one for each subset of edges in the graph. Within a given block, a function $\phi_\g$ couples to another $\phi_{\g'}$ if and only if the zonotope $\mathcal{Z}_{\g'}$ forms a codimension-one boundary of $\mathcal{Z}_\g$. For instance, the differential equations for the path graph decompose into four blocks. Consider the zonotope represented by the line segment at the top left of figure~\ref{fig:3chainZonosIntro}, the corresponding differential equations are given explicitly by 
\begin{align}
-\d_\kin \phi_{\begin{tikzpicture}[scale=0.8]
        \coordinate (A) at (0,0);
        \coordinate (B) at (1/2,0);
        \coordinate (C) at (1,0);
        \coordinate (D) at (3/2,0);
        \draw[thick,dotted] (B) -- (C);
        \draw[thick] (A) -- node {\ar} (B);
        \fill[black] (A) circle (2pt);
        \fill[black] (B) circle (2pt);
        \fill[black] (C) circle (2pt);
    \end{tikzpicture}}  
    &\simeq
    \left( \alpha_1 \raisebox{0.05cm}{\begin{tikzpicture}[scale=0.8]
        \coordinate (A) at (0,0);
        \coordinate (B) at (1/2,0);
        \coordinate (C) at (1,0);
        \coordinate (D) at (3/2,0);
        \draw[thick] (A) -- (B);
        \draw[thick,dotted] (B) -- (C);
        \fill[Red] (A) circle (2.4pt);
        \fill[black] (B) circle (2pt);
        \fill[black] (C) circle (2pt);
    \end{tikzpicture}}+\alpha_2\raisebox{-0.1cm}{\begin{tikzpicture}[scale=0.8]
        \coordinate (A) at (0,0);
        \coordinate (B) at (1/2,0);
        \coordinate (C) at (1,0);
        \coordinate (D) at (3/2,0);
        \draw[ultra thick,Red] (A) -- node {\arbb} (B);
        \draw[thick,dotted] (B) -- (C);
        \fill[black] (A) circle (2pt);
        \fill[Red] (B) circle (2.4pt);
        \fill[black] (C) circle (2pt);
        \end{tikzpicture}}+\alpha_3\raisebox{-0.1cm}{\begin{tikzpicture}[scale=0.8]
        \coordinate (A) at (0,0);
        \coordinate (B) at (1/2,0);
        \coordinate (C) at (1,0);
        \coordinate (D) at (3/2,0);
        \draw[thick] (A) -- node {\ar} (B);
        \draw[thick,dotted] (B) -- (C);
        \fill[black] (A) circle (2pt);
        \fill[black] (B) circle (2pt);
        \fill[Red] (C) circle (2.4pt);
    \end{tikzpicture}} \right) \phi_{\begin{tikzpicture}[scale=0.8]
        \coordinate (A) at (0,0);
        \coordinate (B) at (1/2,0);
        \coordinate (C) at (1,0);
        \coordinate (D) at (3/2,0);
        \draw[thick,dotted] (B) -- (C);
        \draw[thick] (A) -- node {\ar} (B);
        \fill[black] (A) circle (2pt);
        \fill[black] (B) circle (2pt);
        \fill[black] (C) circle (2pt);
    \end{tikzpicture}} + \frac{\alpha_1\alpha_2}{\alpha_1+\alpha_2}\left(\raisebox{-0.1cm}{\begin{tikzpicture}[scale=0.8]
        \coordinate (A) at (0,0);
        \coordinate (B) at (1/2,0);
        \coordinate (C) at (1,0);
        \coordinate (D) at (3/2,0);
        \draw[ultra thick,Red] (A) -- node {\arbb} (B);
        \draw[thick,dotted] (B) -- (C);
        \fill[black] (A) circle (2pt);
        \fill[Red] (B) circle (2.4pt);
        \fill[black] (C) circle (2pt);
    \end{tikzpicture}}- \raisebox{-0.1cm}{\begin{tikzpicture}[scale=0.8]
        \coordinate (A) at (0,0);
        \coordinate (B) at (1/2,0);
        \coordinate (C) at (1,0);
        \coordinate (D) at (3/2,0);
        \draw[thick] (A) -- node {\ar} (B);
        \draw[thick,dotted] (B) -- (C);
        \fill[Red] (A) circle (2.4pt);
        \fill[black] (B) circle (2pt);
        \fill[black] (C) circle (2pt);
    \end{tikzpicture}} \right) \phi_{\begin{tikzpicture}[scale=0.8]
        \coordinate (A) at (0,0);
        \coordinate (B) at (1/2,0);
        \coordinate (C) at (1,0);
        \coordinate (D) at (3/2,0);
        \draw[thick,dotted] (B) -- (C);
        \draw[thick,double] (A) -- (B);
        \fill[black] (A) circle (2pt);
        \fill[black] (B) circle (2pt);
        \fill[black] (C) circle (2pt);
    \end{tikzpicture}}, \notag \\
-\d_\kin \phi_{\begin{tikzpicture}[scale=0.8]
        \coordinate (A) at (0,0);
        \coordinate (B) at (1/2,0);
        \coordinate (C) at (1,0);
        \coordinate (D) at (3/2,0);
        \draw[thick,dotted] (B) -- (C);
        \draw[thick] (A) -- node {\al} (B);
        \fill[black] (A) circle (2pt);
        \fill[black] (B) circle (2pt);
        \fill[black] (C) circle (2pt);
    \end{tikzpicture}}  
    &\simeq
    \left( \alpha_1 \raisebox{-0.1cm}{\begin{tikzpicture}[scale=0.8]
        \coordinate (A) at (0,0);
        \coordinate (B) at (1/2,0);
        \coordinate (C) at (1,0);
        \coordinate (D) at (3/2,0);
        \draw[ultra thick,Red] (A) -- node {\albb} (B);
        \draw[thick,dotted] (B) -- (C);
        \fill[Red] (A) circle (2.4pt);
        \fill[black] (B) circle (2pt);
        \fill[black] (C) circle (2pt);
    \end{tikzpicture}}+ \alpha_2 \raisebox{-0.1cm}{\begin{tikzpicture}[scale=0.8]
        \coordinate (A) at (0,0);
        \coordinate (B) at (1/2,0);
        \coordinate (C) at (1,0);
        \coordinate (D) at (3/2,0);
        \draw[thick] (A) -- node {\al} (B);
        \draw[thick,dotted] (B) -- (C);
        \fill[black] (A) circle (2pt);
        \fill[Red] (B) circle (2.4pt);
        \fill[black] (C) circle (2pt);
    \end{tikzpicture}}+ \alpha_3 \raisebox{-0.1cm}{\begin{tikzpicture}[scale=0.8]
        \coordinate (A) at (0,0);
        \coordinate (B) at (1/2,0);
        \coordinate (C) at (1,0);
        \coordinate (D) at (3/2,0);
        \draw[thick] (A) -- node {\al} (B);
        \draw[thick,dotted] (B) -- (C);
        \fill[black] (A) circle (2pt);
        \fill[black] (B) circle (2pt);
        \fill[Red] (C) circle (2.4pt);
    \end{tikzpicture}} \right) \phi_{\begin{tikzpicture}[scale=0.8]
        \coordinate (A) at (0,0);
        \coordinate (B) at (1/2,0);
        \coordinate (C) at (1,0);
        \coordinate (D) at (3/2,0);
        \draw[thick,dotted] (B) -- (C);
        \draw[thick] (A) -- node {\al} (B);
        \fill[black] (A) circle (2pt);
        \fill[black] (B) circle (2pt);
        \fill[black] (C) circle (2pt);
    \end{tikzpicture}}+\frac{\alpha_1\alpha_2}{\alpha_1+\alpha_2}\left(\raisebox{-0.1cm}{\begin{tikzpicture}[scale=0.8]
        \coordinate (A) at (0,0);
        \coordinate (B) at (1/2,0);
        \coordinate (C) at (1,0);
        \coordinate (D) at (3/2,0);
        \draw[ultra thick,Red] (A) -- node {\albb} (B);
        \draw[thick,dotted] (B) -- (C);
        \fill[Red] (A) circle (2.4pt);
        \fill[black] (B) circle (2pt);
        \fill[black] (C) circle (2pt);
    \end{tikzpicture}}- \raisebox{-0.1cm}{\begin{tikzpicture}[scale=0.8]
        \coordinate (A) at (0,0);
        \coordinate (B) at (1/2,0);
        \coordinate (C) at (1,0);
        \coordinate (D) at (3/2,0);
        \draw[thick] (A) -- node {\al} (B);
        \draw[thick,dotted] (B) -- (C);
        \fill[black] (A) circle (2pt);
        \fill[Red] (B) circle (2.4pt);
        \fill[black] (C) circle (2pt);
    \end{tikzpicture}} \right) \phi_{\begin{tikzpicture}[scale=0.8]
        \coordinate (A) at (0,0);
        \coordinate (B) at (1/2,0);
        \coordinate (C) at (1,0);
        \coordinate (D) at (3/2,0);
        \draw[thick,dotted] (B) -- (C);
        \draw[thick,double] (A) -- (B);
        \fill[black] (A) circle (2pt);
        \fill[black] (B) circle (2pt);
        \fill[black] (C) circle (2pt);
    \end{tikzpicture}}, \notag \\
-\d_\kin \phi_{\begin{tikzpicture}[scale=0.8]
        \coordinate (A) at (0,0);
        \coordinate (B) at (1/2,0);
        \coordinate (C) at (1,0);
        \coordinate (D) at (3/2,0);
        \draw[thick,dotted] (B) -- (C);
        \draw[thick,double] (A) -- (B);
        \fill[black] (A) circle (2pt);
        \fill[black] (B) circle (2pt);
        \fill[black] (C) circle (2pt);
    \end{tikzpicture}} 
    &\simeq
    \left((\alpha_1+\alpha_2) \ \raisebox{0.05cm}{\begin{tikzpicture}[scale=0.8]
        \coordinate (A) at (0,0);
        \coordinate (B) at (1/2,0);
        \coordinate (C) at (1,0);
        \coordinate (D) at (3/2,0);
        \draw[very thick,Red,double] (A) -- (B) ;
        \draw[thick,dotted] (B) -- (C);
        \fill[Red] (A) circle (2.4pt);
        \fill[Red] (B) circle (2.4pt);
        \fill[black] (C) circle (2pt);
    \end{tikzpicture}}+\alpha_3 \raisebox{0.05cm}{\begin{tikzpicture}[scale=0.8]
        \coordinate (A) at (0,0);
        \coordinate (B) at (1/2,0);
        \coordinate (C) at (1,0);
        \coordinate (D) at (3/2,0);
        \draw[thick,double] (A) -- (B);
        \draw[thick,dotted] (B) -- (C);
        \fill[black] (A) circle (2pt);
        \fill[black] (B) circle (2pt);
        \fill[Red] (C) circle (2.4pt);
    \end{tikzpicture}} \right) \phi_{\begin{tikzpicture}[scale=0.8]
        \coordinate (A) at (0,0);
        \coordinate (B) at (1/2,0);
        \coordinate (C) at (1,0);
        \coordinate (D) at (3/2,0);
        \draw[thick,dotted] (B) -- (C);
        \draw[thick,double] (A) -- (B);
        \fill[black] (A) circle (2pt);
        \fill[black] (B) circle (2pt);
        \fill[black] (C) circle (2pt);
    \end{tikzpicture}},
\end{align}
where the graphs highlighted in red correspond to $\dlog$'s whose arguments are functions of the graph variables referred to as \textcolor{Red}{letters}. These same phenomena were recently observed in \cite{Baumann:2025qjx}, albeit from a more physics-inspired approach, through the study of the time integral representation of the wavefunction. 
In contrast, our approach reveals the geometric origin of zonotopes in the kinematic flow, without invoking any knowledge of bulk physics.

\section*{Outline}
In Section \ref{sec:wf} we introduce the central object of interest: the wavefunction of the universe. 
This will include the combinatorial interpretation of the wavefunction in terms of tubings on graphs and a minimal discussion on twisted cohomology/intersection theory relevant for computing the differential equations.
Next, in section \ref{sec:cutBasis}, we describe how to construct the cut basis and deal with the degeneracy of the cosmological hyperplane arrangement $\B$. 
We also introduce important combinatorial objects, called letters, that characterize the geometry and combinatorics of the cuts as well as the singularities that enter the differential equations. 
After providing a simple combinatorial formula for the elements of the cut basis, we connect these forms to a positive geometry on the cut. 
In section \ref{sec:zono} we show how the residues of the physical FRW-form are canonical forms of a certain family of zonotopes. 
From this, we conclude that the combinatorics of the sequential residues of the physical FRW-form is governed by the geometry of certain zonotopes which we call the ``flow of cuts''.
Finally, in section \ref{sec:flowOfCuts}, we derive the kinematic flow rules from the residue calculus of intersection theory. 
In this derivation, the flow of cuts provides important selection rules that control the combinatorics/organization of the differential equations.
In fact, the kinematic flow is simply the cut flow in reverse. 
The remainder of section \ref{sec:flowOfCuts} is dedicated to examples.

\section{Cosmological wavefunction}
\label{sec:wf}
In this work, we consider a toy model for quantum cosmology: conformally-coupled scalars in a power-law FRW cosmology with non-conformal polynomial interactions. 
This model is characterized by the following action in a $(d+1)$-dimensional space-time
\begin{align}
S= \int  \d^d x\; \d \eta\; \sqrt{-g} \left[ -\frac{1}{2} g^{\mu\nu} \partial_\mu \phi \partial_\nu \phi -\frac{d-1}{8d} R \phi^2 - \sum_{p\geq 3} \frac{\lambda_p}{p!} \phi^p\right].
\end{align}
Here, the invariant line element of the FRW spacetime in comoving coordinates with conformal time $\eta \in (-\infty,0]$ is $\d s^2 = a^2(\eta) \left( -\d \eta^2 + \d x_i \d x^i\right)$ where the index $i \in \{ 1,\ldots,d\}$ runs over spatial dimensions.
Furthermore, we assume that the scale factor is a power law $a(\eta)=(\eta/\eta_0)^{-(1+\epsilon)}$, with constant parameters $\eta_0$ and $\epsilon$. 
Depending on the choice of the cosmological parameter $\epsilon$ many cosmologies of interest can be recovered including: inflationary ($\epsilon \approx 0$), de Sitter ($\epsilon = 0$)  and flat-space ($\epsilon = -1$) for example.

In a cosmological setting, a key set of physical observables are the {\it cosmological correlation functions} at fixed conformal time $\eta=0$. Much like expectation values in standard quantum mechanics, cosmological correlators are computed via the Born rule
\begin{align}
\langle \Phi({\bf x_1}) {\cdots} \Phi({\bf x_n})  \rangle = \int \mathcal{D}\Phi\ \Phi({\bf x_1}) \cdots \Phi({\bf x_n}) |\Psi[\Phi]|^2,
\end{align}
where $\Phi({\bf x})=\phi({\bf x},0)$ is the boundary configuration of the field $\phi({\bf x},\eta)$ and the squared modulus of the {\it wavefunction of the universe} $|\Psi[\Phi]|^2$ is a probability distribution over the space of field configurations. The wavefunction $\Psi[\Phi]$ is defined formally as the path integral over all field configurations $\phi({\bf x},\eta)$ interpolating between the Bunch-Davies vacuum in the far past, $\eta=-\infty(1-i\epsilon)$, and a specified field configuration $\phi({\bf x},0)=\Phi({\bf x})$ at the boundary. 
Importantly, the wavefunction admits an expansion in Fourier space
\begin{align}
    \Psi[\Phi] {=} 
    \!\exp\!\! 
    \left[
    \sum_{n} 
    \frac{i}{n!} 
    {\int} 
    \frac{d^d {\bf k}_1}{(2\pi)^d} 
    {\cdots} 
    \frac{d^d {\bf k}_n}{(2\pi)^d} 
    \Phi({\bf k}_1)
    {\cdots} 
    \Phi({\bf k}_n)   
    \psi^{(n)}(\{{\bf k}_a\}_{a{=}1}^n)
    (2\pi)^d
    \delta^{(d)}\!( {\bf k}_1{+}{\ldots}{+}{\bf k}_n)
\right]\!\!, 
\end{align}
where the integration kernels $\psi^{(n)}$ are the so called {\it wavefunction coefficients} that depend on the $n$ spatial momenta $\{{\bf k}_i\}$ associated to field insertions on the late-time boundary ($\eta=0$). The magnitudes of the spatial momenta $|{\bf k}_i|$ are commonly referred to as {\it energies}.

\begin{figure}[]
\centering
\includegraphics[scale=1.2]{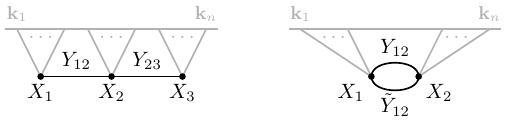}
\caption{Examples of Feynman diagrams contributing to the wavefunction coefficient at tree-level and one-loop.}
\label{fig:feynman}
\end{figure}

These wavefunction coefficients can be computed perturbatively as a sum over Feynman graphs; tree-level and one-loop examples are displayed in figure~\ref{fig:feynman}. 
For a given Feynman graph, the contribution to the wavefunction coefficient depends on the total external energies entering each vertex, represented by $X_v$, and the internal energies flowing through each propagator, denoted by $Y_e$. 
As such it is useful to work at the level of {\it truncated} graphs, obtained by removing all external (gray on figure~\ref{fig:feynman}) legs from the original Feynman graph. 
However, as reviewed in the next section, the full Feynman rules are not needed to compute the contribution, $\psi_G$, to the wavefunction coefficient from a given truncated graph $G$.%
\footnote{%
    Readers interested in the additional details may consult \cite{Baumann:2025qjx}.
}
Importantly, the FRW wavefunction coefficients $\psi_G$ can be constructed by shifting their flat-space counterparts $\hat\Omega_G$ and integrating against a specific kernel. 
Therefore, we begin by reviewing some key features of the flat-space wavefunction coefficients.

\subsection{The flat-space wavefunction}
For a given graph $G$ the cosmological polytope $\mathcal{P}_G$ is a positive geometry whose canonical differential form $\Omega_G$ encodes the corresponding contribution to the flat-space wavefunction coefficient \cite{Arkani-Hamed:2017fdk}. In this section, we review how to compute the canonical form of the cosmological polytope by considering the nested structure of subgraphs of $G$.

Given a graph $G=(V_G,E_G)$ with vertex set $V_G$ and edge set $E_G$, we refer to the set of all {\it connected subgraphs} as {\it tubes}. In explicit examples we will represent a tube graphically on $G$ by encircling its corresponding subgraph. A tube $t=(V_t,E_t)$ is said to cross an edge $v  \overset{e}{\text{---}}  v' \in E_G$ if $e \not \in E_t$ and $V_t \cap \{ v,v'\} \neq \emptyset$. We say two tubes are {\it compatible} if one is a subgraph of the other or they do not intersect on any vertices. We refer to a $k$-element set of pairwise compatible tubes as a $k$-tubing. A tubing is {\it maximal} if no more compatible tubes can be added.%
\footnote{Specifically, each maximal tubing contains exactly $|V_G|+|E_G|$ many tubes.} 
We denote the set of all tubes of a graph by $T_G$
and the set of all maximal tubings by $\mathcal{T}^{\text{max}}_G$.

The canonical form of the cosmological polytope $\Omega_G$, equivalently the flat-space wavefunction coefficient $\hat{\Omega}_G$, has a simple expression in terms of maximal tubings 
\begin{align}
    \Omega_G&=\hat{\Omega}_G({\bf X},{\bf Y}) \frac{\d^{|V_G|} {\bf X} \wedge \d^{|E_G|} {\bf Y}}{\mathrm{GL}(1)}, 
    && 
    \hat{\Omega}_G({\bf X},{\bf Y})=  \prod_{e\in E_G} 2 Y_e \sum_{\tau \in \mathcal{T}_G^{\text{max}}} \frac{1}{S_\tau}.
\label{eq:lin_form}
\end{align}
Here, $S_\tau= \prod_{t\in \tau} S_{t}$ and the $S_t$ are linear functions (``propagators'') associated to each tube 
\begin{align}
S_t  = \sum_{v \in V_t} X_v + \sum_{e \text{ cross }t}  Y_{e},
\label{eq:lin_func}
\end{align}
where the first sum is over all vertices contained in the tube and the second sum is over all edges crossed by the tube counted with multiplicity. 
Kinematic variables ${\bf X} = \{ X_v : v \in V_G\}$ and ${\bf Y} = \{ Y_e : e\in E_G\}$ are assigned to each vertex and edge of the graph; $X_v$ is the sum over all external energies, $|\mbf{k}_\bullet|$, connected to the vertex $v$ while $Y_e$ is the energy of the exchanged particle of edge $e$. 

To understand these definitions, consider the flat-space wavefunction coefficient for the three-site chain, with variables assigned as in figure~\ref{fig:feynman}. 
Using \eqref{eq:lin_form} and noting that there are two maximal tubings, one finds 
\begin{align}\begin{aligned}
\hat{\Omega}_{\begin{tikzpicture}[scale=0.8]
        \coordinate (A) at (0,0);
        \coordinate (B) at (1/2,0);
        \coordinate (C) at (1,0);
        \coordinate (D) at (3/2,0);
        \draw[thick] (A) -- (B) -- (C);
        \fill[black] (A) circle (1.8pt);
        \fill[black] (B) circle (1.8pt);
        \fill[black] (C) circle (1.8pt);
    \end{tikzpicture}} 
    &= \frac{4 Y_{12} Y_{23}}{S_{ \begin{tikzpicture}[scale=0.7]
            \coordinate (A) at (0,0);
            \coordinate (B) at (1/2,0);
            \coordinate (C) at (1,0);
            \coordinate (D) at (3/2,0);
            \draw[thick, black] (0.5,0) ellipse (0.9cm and 0.35cm);
            \draw[thick, black] (0.25,0) ellipse (0.5cm and 0.25cm);
            \draw[black,thick] (A) circle (4pt);
            \draw[black,thick] (B) circle (4pt);
            \draw[black,thick] (C) circle (4pt);
            \draw[thick] (A) -- (B) -- (C);
            \fill[black] (A) circle (2pt);
            \fill[black] (B) circle (2pt);
            \fill[black] (C) circle (2pt);
        \end{tikzpicture} }}+\frac{4 Y_{12} Y_{23}}{S_{ \begin{tikzpicture}[scale=0.7]
        \coordinate (A) at (0,0);
            \coordinate (B) at (1/2,0);
            \coordinate (C) at (1,0);
            \coordinate (D) at (3/2,0);
            \draw[thick, black] (0.5,0) ellipse (0.9cm and 0.35cm);
            \draw[thick, black] (0.75,0) ellipse (0.5cm and 0.25cm);
            \draw[black,thick] (A) circle (4pt);
            \draw[black,thick] (B) circle (4pt);
            \draw[black,thick] (C) circle (4pt);
            \draw[thick] (A) -- (B) -- (C);
            \fill[black] (A) circle (2pt);
            \fill[black] (B) circle (2pt);
            \fill[black] (C) circle (2pt);
    \end{tikzpicture} }}
    = \frac{
        4 Y_{12} Y_{23}
        (S_{ \begin{tikzpicture}[scale=0.7]
            \coordinate (A) at (0,0);
            \coordinate (B) at (1/2,0);
            \coordinate (C) at (1,0);
            \coordinate (D) at (3/2,0);
            \draw[thick, black] (0.25,0) ellipse (0.41cm and 0.19cm);
            \draw[thick] (A) -- (B) -- (C);
            \fill[black] (A) circle (2pt);
            \fill[black] (B) circle (2pt);
            \fill[black] (C) circle (2pt);
        \end{tikzpicture} }
        + S_{ \begin{tikzpicture}[scale=0.7]
            \coordinate (A) at (0,0);
            \coordinate (B) at (1/2,0);
            \coordinate (C) at (1,0);
            \coordinate (D) at (3/2,0);
            \draw[thick, black] (0.75,0) ellipse (0.41cm and 0.19cm);
            \draw[thick] (A) -- (B) -- (C);
            \fill[black] (A) circle (2pt);
            \fill[black] (B) circle (2pt);
            \fill[black] (C) circle (2pt);
        \end{tikzpicture} })
    }{
        S_{ \begin{tikzpicture}[scale=0.7]
            \coordinate (A) at (0,0);
            \coordinate (B) at (1/2,0);
            \coordinate (C) at (1,0);
            \coordinate (D) at (3/2,0);
            \draw[thick] (A) -- (B) -- (C);
            \draw[black,thick] (A) circle (5pt);
            \fill[black] (A) circle (2pt);
            \fill[black] (B) circle (2pt);
            \fill[black] (C) circle (2pt);
        \end{tikzpicture} }
        S_{ \begin{tikzpicture}[scale=0.7]
            \coordinate (A) at (0,0);
            \coordinate (B) at (1/2,0);
            \coordinate (C) at (1,0);
            \coordinate (D) at (3/2,0);
            \draw[thick] (A) -- (B) -- (C);
            \draw[black,thick] (B) circle (5pt);
            \fill[black] (A) circle (2pt);
            \fill[black] (B) circle (2pt);
            \fill[black] (C) circle (2pt);
        \end{tikzpicture} }
        S_{ \begin{tikzpicture}[scale=0.7]
            \coordinate (A) at (0,0);
            \coordinate (B) at (1/2,0);
            \coordinate (C) at (1,0);
            \coordinate (D) at (3/2,0);
            \draw[thick] (A) -- (B) -- (C);
            \draw[black,thick] (C) circle (5pt);
            \fill[black] (A) circle (2pt);
            \fill[black] (B) circle (2pt);
            \fill[black] (C) circle (2pt);
        \end{tikzpicture} }
        S_{ \begin{tikzpicture}[scale=0.7]
            \coordinate (A) at (0,0);
            \coordinate (B) at (1/2,0);
            \coordinate (C) at (1,0);
            \coordinate (D) at (3/2,0);
            \draw[thick, black] (0.25,0) ellipse (0.41cm and 0.19cm);
            \draw[thick] (A) -- (B) -- (C);
            \fill[black] (A) circle (2pt);
            \fill[black] (B) circle (2pt);
            \fill[black] (C) circle (2pt);
        \end{tikzpicture} }
        S_{ \begin{tikzpicture}[scale=0.7]
            \coordinate (A) at (0,0);
            \coordinate (B) at (1/2,0);
            \coordinate (C) at (1,0);
            \coordinate (D) at (3/2,0);
            \draw[thick, black] (0.75,0) ellipse (0.41cm and 0.19cm);
            \draw[thick] (A) -- (B) -- (C);
            \fill[black] (A) circle (2pt);
            \fill[black] (B) circle (2pt);
            \fill[black] (C) circle (2pt);
        \end{tikzpicture} }
        S_{ \begin{tikzpicture}[scale=0.7]
            \coordinate (A) at (0,0);
            \coordinate (B) at (1/2,0);
            \coordinate (C) at (1,0);
            \coordinate (D) at (3/2,0);
            \draw[thick, black] (0.5,0) ellipse (0.75cm and 0.25cm);
            \draw[thick] (A) -- (B) -- (C);
            \fill[black] (A) circle (2pt);
            \fill[black] (B) circle (2pt);
            \fill[black] (C) circle (2pt);
        \end{tikzpicture} }
    }
    ,
\label{eq:p3}
\end{aligned}\end{align}
where 
\begin{align}\begin{aligned}
    S_{ \begin{tikzpicture}[scale=0.8]
        \coordinate (A) at (0,0);
        \coordinate (B) at (1/2,0);
        \coordinate (C) at (1,0);
        \coordinate (D) at (3/2,0);
        \draw[thick] (A) -- (B) -- (C);
        \draw[black,thick] (A) circle (5pt);
        \fill[black] (A) circle (2pt);
        \fill[black] (B) circle (2pt);
        \fill[black] (C) circle (2pt);
    \end{tikzpicture} } 
    &=X_1 + Y_{12},
    &
    S_{ \begin{tikzpicture}[scale=0.8]
        \coordinate (A) at (0,0);
        \coordinate (B) at (1/2,0);
        \coordinate (C) at (1,0);
        \coordinate (D) at (3/2,0);
        \draw[thick] (A) -- (B) -- (C);
        \draw[black,thick] (B) circle (5pt);
        \fill[black] (A) circle (2pt);
        \fill[black] (B) circle (2pt);
        \fill[black] (C) circle (2pt);
    \end{tikzpicture} } 
    &=X_2+Y_{12}+Y_{23},  
    & 
    S_{ \begin{tikzpicture}[scale=0.8]
        \coordinate (A) at (0,0);
        \coordinate (B) at (1/2,0);
        \coordinate (C) at (1,0);
        \coordinate (D) at (3/2,0);
        \draw[thick] (A) -- (B) -- (C);
        \draw[black,thick] (C) circle (5pt);
        \fill[black] (A) circle (2pt);
        \fill[black] (B) circle (2pt);
        \fill[black] (C) circle (2pt);
    \end{tikzpicture} } &= X_{3}+Y_{23}, 
    \\
    S_{ \begin{tikzpicture}[scale=0.8]
        \coordinate (A) at (0,0);
        \coordinate (B) at (1/2,0);
        \coordinate (C) at (1,0);
        \coordinate (D) at (3/2,0);
        \draw[thick, black] (0.5,0) ellipse (0.75cm and 0.25cm);
        \draw[thick] (A) -- (B) -- (C);
        \fill[black] (A) circle (2pt);
        \fill[black] (B) circle (2pt);
        \fill[black] (C) circle (2pt);
    \end{tikzpicture} } 
    &=X_1+X_2+X_3, 
    &
    S_{ \begin{tikzpicture}[scale=0.8]
        \coordinate (A) at (0,0);
        \coordinate (B) at (1/2,0);
        \coordinate (C) at (1,0);
        \coordinate (D) at (3/2,0);
        \draw[thick, black] (0.25,0) ellipse (0.41cm and 0.19cm);
        \draw[thick] (A) -- (B) -- (C);
        \fill[black] (A) circle (2pt);
        \fill[black] (B) circle (2pt);
        \fill[black] (C) circle (2pt);
    \end{tikzpicture} } 
    &=X_1+X_2+Y_{23}, 
    &
    S_{ \begin{tikzpicture}[scale=0.8]
        \coordinate (A) at (0,0);
        \coordinate (B) at (1/2,0);
        \coordinate (C) at (1,0);
        \coordinate (D) at (3/2,0);
        \draw[thick, black] (0.75,0) ellipse (0.41cm and 0.19cm);
        \draw[thick] (A) -- (B) -- (C);
        \fill[black] (A) circle (2pt);
        \fill[black] (B) circle (2pt);
        \fill[black] (C) circle (2pt);
    \end{tikzpicture} } 
    &= X_2+X_3+Y_{12}.
\end{aligned}\end{align}
Similarly, the flat-space wavefunction coefficient for the one-loop bubble (figure~\ref{fig:feynman} right) is 
\begin{align}
\hat{\Omega}_{
\begin{tikzpicture}[scale=0.6]
\fill[black] (0,-1) circle (2pt);
\fill[black] (1,-1) circle (2pt);
\draw[thick] (0,-1) to[out=90,in=90] (1,-1);
\draw[thick] (0,-1) to[out=-90,in=180+90] (1,-1);
\end{tikzpicture}}
&=4 Y_{12} \tilde{Y}_{12}\frac{   S_{
\begin{tikzpicture}[scale=0.55]
\fill[black] (0,-1) circle (2pt);
\fill[black] (1,-1) circle (2pt);
\draw[thick] (0,-1) to[out=90,in=90] (1,-1);
\draw[thick] (0,-1) to[out=-90,in=180+90] (1,-1);
\draw[thick] (0-0.2,-1) to[out=90,in=90] (1+0.2,-1);
\draw[thick] (0+0.2,-1) to[out=90,in=90] (1-0.2,-1);
\draw[thick] (1-0.2,-1) to[out=90+180,in=180] (1,-1.2) to[out=0,in=90+180] (1+0.2,-1);
\draw[thick] (0-0.2,-1) to[out=90+180,in=180] (0,-1.2) to[out=0,in=90+180] (0+0.2,-1);
\end{tikzpicture}}+ S_{
\begin{tikzpicture}[scale=0.55]
\fill[black] (0,-1) circle (2pt);
\fill[black] (1,-1) circle (2pt);
\draw[thick] (0,-1) to[out=90,in=90] (1,-1);
\draw[thick] (0,-1) to[out=-90,in=180+90] (1,-1);
\draw[thick] (0-0.2,-1) to[out=-90,in=180+90] (1+0.2,-1);
\draw[thick] (0+0.2,-1) to[out=-90,in=180+90] (1-0.2,-1);
\draw[thick] (1-0.2,-1) to[out=90,in=180] (1,-0.8) to[out=0,in=90] (1+0.2,-1);
\draw[thick] (0-0.2,-1) to[out=90,in=180] (0,-0.8) to[out=0,in=90] (0+0.2,-1);
\end{tikzpicture}} }{S_{
\begin{tikzpicture}[scale=0.55]
\fill[black] (0,-1) circle (2pt);
\fill[black] (1,-1) circle (2pt);
\draw[black,thick] (0,-1) circle (5pt);
\draw[thick] (0,-1) to[out=90,in=90] (1,-1);
\draw[thick] (0,-1) to[out=-90,in=180+90] (1,-1);
\end{tikzpicture}}S_{
\begin{tikzpicture}[scale=0.55]
\fill[black] (0,-1) circle (2pt);
\fill[black] (1,-1) circle (2pt);
\draw[black,thick] (1,-1) circle (5pt);
\draw[thick] (0,-1) to[out=90,in=90] (1,-1);
\draw[thick] (0,-1) to[out=-90,in=180+90] (1,-1);
\end{tikzpicture}}    S_{
\begin{tikzpicture}[scale=0.55]
\fill[black] (0,-1) circle (2pt);
\fill[black] (1,-1) circle (2pt);
\draw[thick] (0,-1) to[out=90,in=90] (1,-1);
\draw[thick] (0,-1) to[out=-90,in=180+90] (1,-1);
\draw[thick] (0-0.2,-1) to[out=90,in=90] (1+0.2,-1);
\draw[thick] (0+0.2,-1) to[out=90,in=90] (1-0.2,-1);
\draw[thick] (1-0.2,-1) to[out=90+180,in=180] (1,-1.2) to[out=0,in=90+180] (1+0.2,-1);
\draw[thick] (0-0.2,-1) to[out=90+180,in=180] (0,-1.2) to[out=0,in=90+180] (0+0.2,-1);
\end{tikzpicture}} S_{
\begin{tikzpicture}[scale=0.55]
\fill[black] (0,-1) circle (2pt);
\fill[black] (1,-1) circle (2pt);
\draw[thick] (0,-1) to[out=90,in=90] (1,-1);
\draw[thick] (0,-1) to[out=-90,in=180+90] (1,-1);
\draw[thick] (0-0.2,-1) to[out=-90,in=180+90] (1+0.2,-1);
\draw[thick] (0+0.2,-1) to[out=-90,in=180+90] (1-0.2,-1);
\draw[thick] (1-0.2,-1) to[out=90,in=180] (1,-0.8) to[out=0,in=90] (1+0.2,-1);
\draw[thick] (0-0.2,-1) to[out=90,in=180] (0,-0.8) to[out=0,in=90] (0+0.2,-1);
\end{tikzpicture}}S_{
\begin{tikzpicture}[scale=0.5]
\fill[black] (0,-1) circle (2pt);
\fill[black] (1,-1) circle (2pt);
\draw[thick] (0,-1) to[out=90,in=90] (1,-1);
\draw[thick] (0,-1) to[out=-90,in=180+90] (1,-1);
\draw[thick, black] (0.5,-1) ellipse (0.8cm and 0.5cm);
\end{tikzpicture}}},
\label{eq:oneloop}
\end{align}
where
\begin{align}
&S_{
\begin{tikzpicture}[scale=0.6]
\fill[black] (0,-1) circle (2pt);
\fill[black] (1,-1) circle (2pt);
\draw[black,thick] (0,-1) circle (5pt);
\draw[thick] (0,-1) to[out=90,in=90] (1,-1);
\draw[thick] (0,-1) to[out=-90,in=180+90] (1,-1);
\end{tikzpicture}}=X_1 + Y_{12}+ \tilde{Y}_{12}, 
&&S_{
\begin{tikzpicture}[scale=0.6]
\fill[black] (0,-1) circle (2pt);
\fill[black] (1,-1) circle (2pt);
\draw[black,thick] (1,-1) circle (5pt);
\draw[thick] (0,-1) to[out=90,in=90] (1,-1);
\draw[thick] (0,-1) to[out=-90,in=180+90] (1,-1);
\end{tikzpicture}}=X_2 + Y_{12}+ \tilde{Y}_{12},&
&&S_{
\begin{tikzpicture}[scale=0.55]
\fill[black] (0,-1) circle (2pt);
\fill[black] (1,-1) circle (2pt);
\draw[thick] (0,-1) to[out=90,in=90] (1,-1);
\draw[thick] (0,-1) to[out=-90,in=180+90] (1,-1);
\draw[thick, black] (0.5,-1) ellipse (0.8cm and 0.5cm);
\end{tikzpicture}} = X_1+X_2, \notag \\
&S_{
\begin{tikzpicture}[scale=0.6]
\fill[black] (0,-1) circle (2pt);
\fill[black] (1,-1) circle (2pt);
\draw[thick] (0,-1) to[out=90,in=90] (1,-1);
\draw[thick] (0,-1) to[out=-90,in=180+90] (1,-1);
\draw[thick] (0-0.2,-1) to[out=90,in=90] (1+0.2,-1);
\draw[thick] (0+0.2,-1) to[out=90,in=90] (1-0.2,-1);
\draw[thick] (1-0.2,-1) to[out=90+180,in=180] (1,-1.2) to[out=0,in=90+180] (1+0.2,-1);
\draw[thick] (0-0.2,-1) to[out=90+180,in=180] (0,-1.2) to[out=0,in=90+180] (0+0.2,-1);
\end{tikzpicture}}=X_1+X_2 + 2 \tilde{Y}_{12}, 
&&S_{
\begin{tikzpicture}[scale=0.6]
\fill[black] (0,-1) circle (2pt);
\fill[black] (1,-1) circle (2pt);
\draw[thick] (0,-1) to[out=90,in=90] (1,-1);
\draw[thick] (0,-1) to[out=-90,in=180+90] (1,-1);
\draw[thick] (0-0.2,-1) to[out=-90,in=180+90] (1+0.2,-1);
\draw[thick] (0+0.2,-1) to[out=-90,in=180+90] (1-0.2,-1);
\draw[thick] (1-0.2,-1) to[out=90,in=180] (1,-0.8) to[out=0,in=90] (1+0.2,-1);
\draw[thick] (0-0.2,-1) to[out=90,in=180] (0,-0.8) to[out=0,in=90] (0+0.2,-1);
\end{tikzpicture}}=X_1+X_2 + 2 Y_{12}. &&&
\end{align}

\subsection{From flat-space to cosmology}

The flat-space wavefunction coefficients can be recycled to produce FRW wavefunction coefficients; they act as a universal integrand from which the corresponding wavefunctions in any power-law FRW space-time can be computed. To obtain the cosmological wavefunction from its flat-space counterpart we simply shift the kinematic variables associated to the graphs vertices and integrate against the so called {\it twist} factor $u$:
\begin{align} \label{eq:cosPsi}
\psi_{G} = \int_{0}^\infty u \Psi_G , \ \ \text{ where } \ \  \Psi_G=\hat{\Omega}_G({\bf x}+{\bf X},{\bf Y}) \ \d^{|V_G|}{\bf x}.
\end{align}
Here, the twist is defined as $u=\prod_{v \in V_G} x_v^{\alpha_v}$
where 
\begin{align}
    \alpha_v = d + \epsilon (d + 1) + \frac{1}{2} p_v (1 + \epsilon)(1 - d)
    \,,
\end{align} 
$d$ is the number of spatial dimensions $d$, and $p_v$ is the valency of the vertex $v$. 
It is a multi-valued power function with $\alpha_i \in \mathbb{C} \setminus \mathbb{Z}$ (since $\epsilon\notin\mathbb{C}\setminus\mathbb{Z}$) that vanishes on the coordinate hyperplanes $x_v=0$. Due to the twist $u$, the coordinate hyperplanes are branch surfaces of the integrand in \eqref{eq:cosPsi}. 
We call the coordinate hyperplanes $\mathcal{T}_i:= \Vsf(x_i)$ twisted hyperplanes
and define the variety $\mathcal{T} = \Vsf(x_1 \cdots x_n)$, which encodes the singular loci of the integrand in \eqref{eq:cosPsi} associated to the twist $u$.

The integrand in \eqref{eq:cosPsi} also inherits singularities from the the shifted flat-space wavefunction coefficient $\hat{\Omega}_G(\mbf{x}+\mbf{X},\mbf{Y})$.
The shifted flat-space canonical function is obtained by replacing all $S_t$ in \eqref{eq:lin_form} by 
\begin{align}
    B_t = \sum_{v\in t} x_v + S_t \,.
\end{align} 
The integrand in \eqref{eq:cosPsi} has potential singularities on the vanishing loci of the $B_t$: $\mathcal{B} := \Vsf\left( \prod_t B_t \right)$. 
We call the hyperplanes $\mathcal{B}_t := \Vsf(B_t)$ untwisted as they are not components of $\mathcal{T}$.
The nature of singularities associated to $\mathcal{B}$ are very different from those associated to $\mathcal{T}$: $\mathcal{B}$ is the loci where $u \Psi_G$ has potential poles (not branch cuts).
In more traditional quantum field theory language, the $B_i$'s should be thought of as propagators. 
Due to the distinct nature of the singularities associated to $\mathcal{T}$ and $\mathcal{B}$, we will have to treat these varieties differently when constructing the relevant cohomology theory in the next section.

\subsection{Structural overview of FRW-forms}

In this section, we provide an intuitive overview of the relevant aspects of twisted cohomology and intersection theory that motivates the cut-first perspective of our basis.

\paragraph{FRW-cohomology.} 
Due to the presence of the twist $u$ in \eqref{eq:cosPsi}, the cohomology theory describing such integrals is a twisted cohomology. 
Moreover, since the twist $u$ only regulates the twisted singular loci of $u \Psi_G$ ($\mathcal{T}$), the cohomology associated to \eqref{eq:cosPsi} can be thought of as a specialization of the twisted cohomology where the $B_t$ are included in the twist $u_\text{deformed}:=\prod_i x_i^{\alpha_i} \prod_{t} B_t^{\beta_i}$ by setting $\beta_t =0$: $u = u_\text{deformed}\vert_{\beta_\bullet=0}$. 
Explicitly, that is 
\be \label{eq:H}
    H^n(M\setminus\B;\nabla)
    := \frac{
        \{\text{$\nabla$-closed $n$-forms}\}
    }{
        \{\text{$\nabla$-exact $n$-forms}\}
    },
\ee
where $M:=\mathbb{C}^{|V_G|} \setminus \mathcal{T}$, $\nabla := \d + \omega \wedge$, and $\omega := \dlog u$ a flat connection.%
\footnote{A differential $n$-form $\vphi$ is \emph{$\nabla$-closed} if $\nabla\vphi = 0$ and \emph{$\nabla$-exact} if $\vphi = \nabla\psi$ for some ($n-1$)-form $\psi$.}
This is simply the de Rham cohomology of $M\setminus\B$ with respect to the covariant derivative $\nabla$.

While the above specialization changes the mathematical structure of the usual twisted cohomology (most notably in the dual cohomology introduced below), it preserves most of its key properties.  
In particular, the existence of a middle dimensional theorem:%
\footnote{%
    Note that the middle dimension theorem only holds when $\T$ satisfies certain generality assumptions that have been omitted. 
}
\footnote{%
    Top-dimensional holomorphic forms are middle dimensional since the real-dimension of a complex manifold or variety is twice that of the complex-dimension.
}
\be
    \text{dim} H^n(M\setminus\B;\nabla) 
    = \begin{cases}
        |\chi(M\setminus\B)|
        & n = \dim_\mathbb{C} (M\setminus \B) 
        = |V_G|
        \,,
        \\
        0 & \text{otherwise}
        \,,
    \end{cases}
\ee
where $\chi$ is the signed Euler characteristic of the variety $M \setminus \B$ (or hyperplane arrangement). 
This means that we expect the basis of cosmological integrals for a given graph to have size $|\chi(M\setminus\B)|$. 
In general, all FRW-forms can be expressed as a top-dimensional holomorphic form
\be \label{eq:generalFRWForm}
    \frac{\d^{|V_G|}\mbf{x}}{x_1^{\mu_1} \cdots x_n^{\mu_n} B_1^{\nu_1} \cdots B_m^{\nu_m}},
\ee
where $\mu_i, \nu_j \in \mathbb{Z}$. 
Note that all top-dimensional holomorphic forms are trivially $\nabla$-closed; one only needs to check exactness when building a basis for the FRW-cohomology. 

A natural organization for the FRW-cohomology is by the sets of propagators that appear in the denominator. 
Schematically, each basis element can be expressed as 
\be \label{eq:schematicFRWOrg}
        \underset{\text{can be cut}}{\underbrace{
            \left( \bigwedge_{t\in T} \dlog B_t \right)
        }}
        \wedge 
        \underset{\text{positive geometry}}{\underbrace{
            \left( 
                \sum_{i} \pm 
                \bigwedge_{j \in J_i} \dlog x_j 
            \right)
        }}
    \,,
\ee
where $T$ and the $J_i$ are multi-indices indexing which untwisted and twisted divisors appear such that $|T|+|J_i|=|V_G| \;\forall\; i$. 
The first factor contains only singularities in $\B$ $\dlog$'s (or possibly a linear combination of such forms) on which we can take generalized unitarity cuts. 
Moreover, when building our basis, we require the first factor to be linearly independent from from all other propagator $\dlog$-forms in the basis. 
This condition is non-trivial for the degenerate arrangements seen in cosmology (see section \ref{sec:indepCuts}).
The second factor is a linear combination of $\dlog$'s with poles on the twisted hyperplanes $\T$. 
The exact linear combinations that appear come from positive geometric considerations (sections \ref{sec:cutbasis} and \ref{sec:cutGeom}).

\begin{tcolorbox}[breakable, title=Residues and (generalized unitarity) cuts]
    Central to our analysis is the idea of a (generalized unitarity) cut. 
    Despite the fancy name, a cut is equivalent to taking the residue on the vanishing loci of a collection of $B_t$'s.%
    \footnote{These residues are linked to discontinuities which, in turn, are related to the notion of (generalized) unitarity in quantum field theory.}
    For example, cutting the ``propagators'' $B_{t_1}, B_{t_2}, \dots, B_{t_{|T|}}$ corresponds to the sequential residue $\res_T := \res_{\B_{j_{|T|}}} \circ \cdots \circ \res_{\B_{t_1}}$. 
    \\[1em]
    Importantly, the FRW-cohomology can be built from the twisted cohomology of \emph{simpler} cut spaces $M_T$ that are reached by taking sequential residues:
    \be \label{eq:MJ}
        M_T &:= M \cap \B_T 
        \,
        &
        \B_T &:= \bigcap_{t\in T} \Vsf(B_t)
        \,.
    \ee
    The $M_T$'s are positive geometries, and the (generalized unitarity) cuts $\res_T$ are induced by the dual forms (detailed below) in the intersection number.
    \\[1em]
    Additionally, the \emph{physical cuts}---those that do not annihilate the physical FRW-form $\Psi_G$---serve as projectors onto a distinguished subspace of the full FRW-cohomology called the \emph{physical subspace}.
    The physical subspace contains the physical FRW-form $\Psi_G$ such that the differential equations for any chosen basis close. 
    Moreover, the dimension of the physical subspace is generically much less than $|\chi(M\setminus\B)|$---the dimension of the full FRW-cohomology. 
    This was first observed in \cite{Arkani-Hamed:2023kig} and then it was understood how to isolate this subspace using physical sequential residues in \cite{De:2024zic}.

\end{tcolorbox}

\paragraph{The dual cohomology.}
The FRW twisted cohomology is a vector space equipped with an inner product called the \emph{intersection number}. 
Like all inner products, the intersection number is a pairing between a vector space $H^n(M\setminus\B;\nabla)$ and the associated dual space $\check{H}^{|V_G|}$ (to be discussed)
\be
    \check{H}^{|V_G|} \times H^{|V_G|}(M\setminus\B;\nabla) \to \mathbb{C}
    \quad \text{via} \quad
    \la \check{\vphi} \vert \phi \ra 
    := \res_{\check{\vphi}}[\phi]
    \,. 
\ee
Here, $\check{\vphi}$ is a dual form in the dual vector space and $\phi \in H^{|V_G|}(M\setminus\B;\nabla)$ is any FRW-form. 
For our purposes, the intersection number is equivalent to the action of a specific residue operator $\res_{\check{\vphi}}$ generated by the dual form $\check{\vphi}$. 
This residue operator includes the residues associated to the cuts discussed above as well as additional residue operators that fully localize the FRW-form $\phi$.
The intersection number is an algebraic operation; no integration is required---an improvement over generalized unitarity which still requires one to compute discontinuities via integration.
Besides entering as a component of the inner product, the dual cohomology suggests a natural organizational structure for the FRW-cohomology based on cuts (c.f., \eqref{eq:schematicFRWOrg} and \eqref{eq:Hdual}). 
After constructing the relevant dual space, we will use the intersection number to derive a closed form formula for the differential equations without relying on input from bulk physics. 

The dual space is a relative twisted cohomology where the variety $\B$ acts as boundaries instead of the loci for possible poles: 
$
    \check{H}^{|V_G|} 
    := H^{|V_G|}(M, \B; \check{\nabla})
$. 
The swapping of poles and boundaries is needed to make the definition of the intersection number finite and well defined.%
\footnote{%
    In order to construct the residue operator $\res_{\check{\vphi}}$ from $\check{\vphi}$, one needs to be able to integrate $\check{\vphi}$ infinitesimally near the components of $\B$ or $\T$ using only rational functions. 
    This is not possible for any $\check{\vphi}$ that has a simple pole on a component of $\B$; a logarithm is needed to integrate an untwisted pole. 
    Therefore, we simply have to enforce that all dual forms are regular on all components of $\B$; i.e., that the dual space is a relative cohomology with boundaries $\B$. 
    The technical details can be found in \cite{Caron-Huot:2021xqj, Caron-Huot:2021iev}. 
}
Moreover, this requires dual forms to be regular on all untwisted hyperplanes $\B_t$; poles on the twisted hyperplanes $\T$ are allowed. 
Dual integrals also come with the inverse twist $\check{u} = u^{-1}$. 
Hence, the relative twisted cohomology is equipped with the covariant derivative $\check{\nabla}:=\d + \check{\omega} \wedge$ where $\check{\omega} = - \omega$.

\begin{figure}
    \centering
    \begin{minipage}{.5\textwidth}
        \centering
        \includegraphics[scale=2,align=c]{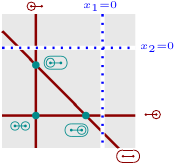}
    \end{minipage}
    \begin{minipage}{.4\textwidth}
        \centering
        \vspace{-2em}
        \begin{align*}
            M_{\includegraphics[scale=.25, align=c]{figures/2chain_B12.pdf}} 
            &=  \includegraphics[scale=2, align=c]{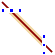}
            \\
            M_{\includegraphics[scale=.25, align=c]{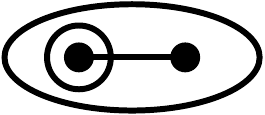}} 
            &=  \includegraphics[scale=2, align=c]{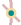}
            \\
            M_{\includegraphics[scale=.25, align=c]{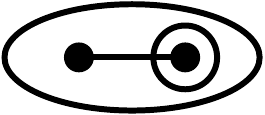}} 
            &=  \includegraphics[scale=2, align=c]{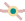}
            \\
            M_{\includegraphics[scale=.25, align=c]{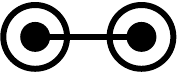}} 
            &=  \includegraphics[scale=2, align=c]{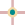}
        \end{align*}
    \end{minipage}
    \caption{%
        The two-chain dual hyperplane arrangement: $\mathbb{C}^2\setminus {\color{Blue}\Vsf(x_1x_2)}$ 
        with boundaries ${\color{BrickRed}\B}$ (left). 
        The geometry of the cut spaces $M_T$ (right). 
        Each of these geometries supports a 1-dimensional twisted cohomology; each has as single bounded chamber highlighted in light orange. 
        We always use the coboundary of the canonical forms associated to these bounded chambers as the basis for the boundary cohomology.
    }
    \label{fig:2chainHPA}
\end{figure}

The point of relative cohomology is to keep track of boundary integrals. 
To characterize the non-trivial differential forms that generate boundary integrals produced by Stokes' theorem when integrating over contours that have boundaries on the untwisted hyperplanes $\B_t$.
Since this is still an unfamiliar object for most of the physics community, we illustrate the formalism through an example instead of diving into mathematical definitions. 
To this end, consider the two-site chain graph $G=\includegraphics[scale=.3, align=c]{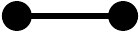}$: 
\be
    S_{\includegraphics[scale=.15, align=c]{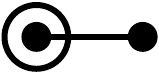}} &= X_1 + Y_{12}
    &
    S_{\includegraphics[scale=.15, align=c]{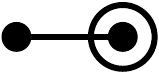}} &= X_2 + Y_{12}
    &
    S_{\includegraphics[scale=.15, align=c]{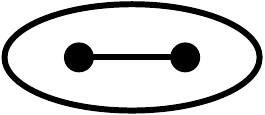}} &= X_1 + X_2
\ee
Here, $\B$ has three components: 
$
    \B = \B_{\includegraphics[scale=.15, align=c]{figures/2chain_B1.pdf}} 
    \cup \B_{\includegraphics[scale=.15, align=c]{figures/2chain_B2.pdf}}
    \cup \B_{\includegraphics[scale=.15, align=c]{figures/2chain_B12.pdf}}
$ where, as usual, $\B_t = \Vsf(B_\tau)$ and $B_t = \sum_{v \in V_t} x_v + S_t$. 
This hyperplane arrangement is depicted figure \ref{fig:2chainHPA} along with its cut spaces that have non-trivial cohomology. 

To illustrate the main features of the formalism, consider the dual integral 
\be \label{eq:2siteDualInt}
    \int_{\check{\gamma}} \check{u}\; \check{\vphi}\,,
    \qquad
    \check{\vphi} 
    = \frac{\d x_1 \wedge \d x_2}{x_1 x_2}\,,
\ee
where $\check{\gamma}$ is a generic relative twisted contour that can have boundaries on $\B$ (relative) and $\T$ (twisted). 
Note that while $\check{\vphi}$ is a good dual form (regular on $\B$ with poles on $T$), it is a total covariant derivative
\be
    \check{\vphi} 
    = \frac{\d x_1 \wedge \d x_2}{x_1 x_2}
    = \check{\nabla} \left(
        \frac{
            \alpha_1\; \dlog\,x_1
            - \alpha_2\; \dlog\,x_2
        }{2\alpha_1\alpha_2}
    \right)
    \,,
\ee
By Stokes theorem, the integral \eqref{eq:2siteDualInt} becomes 
\be
    \int_{\check{\gamma}} \check{u}\; \check{\nabla} 
    \left(
        \frac{
            \alpha_1\; \dlog\,x_1
            - \alpha_2\; \dlog\,x_2
        }{2\alpha_1\alpha_2}
    \right)
    = \sum_{t} \int_{\partial_{t}{\check{\gamma}}}
    \left.
        \frac{
            \alpha_1\; \dlog\,x_1
            - \alpha_2\; \dlog\,x_2
        }{2\alpha_1\alpha_2}
    \right\vert_{\B_t}
    \,,
\ee 
where $\partial_t\check{\gamma}$ computes the boundary of $\check{\gamma}$ that lies in $\B_t$.
Eliminating $x_1$ on the cut
$\B_{\includegraphics[scale=.15, align=c]{figures/2chain_B1.pdf}}$ 
and $x_2$ on the 
$\B_{\includegraphics[scale=.15, align=c]{figures/2chain_B2.pdf}}$ 
and $\B_{\includegraphics[scale=.15, align=c]{figures/2chain_B12.pdf}}$ 
cuts yields
\be \label{eq:2siteDualIntStokes1}
    \int_{\check{\gamma}} \check{u}\; \check{\vphi}
    &= \int_{\partial_{{\includegraphics[scale=.15, align=c]{figures/2chain_B1.pdf}}}{\check{\gamma}}} 
        (\check{u}\vert_{\B_{\includegraphics[scale=.15, align=b]{figures/2chain_B1.pdf}}})\; 
        \Big(
            -\frac{
                \alpha_2\; \dlog\, x_2
            }{2\alpha_1 \alpha_2} 
        \Big)
    + \int_{\partial_{{\includegraphics[scale=.15, align=c]{figures/2chain_B2.pdf}}}{\check{\gamma}}} 
        (\check{u}\vert_{\B_{\includegraphics[scale=.15, align=c]{figures/2chain_B2.pdf}}})\;
        \left(
            \frac{\alpha_1\; \dlog\, x_1}{2\alpha_1 \alpha_2} 
        \right)
    \\&\quad
    + \int_{\partial_{{\includegraphics[scale=.15, align=c]{figures/2chain_B12.pdf}}}{\check{\gamma}}} 
        (\check{u}\vert_{\B_{\includegraphics[scale=.15, align=c]{figures/2chain_B12.pdf}}})\;
        \frac{
            \alpha_1\; \dlog\,x_1
            - \alpha_2\; \dlog\,(x_1+S_3)
        }{2\alpha_1\alpha_2}
    \,,
\ee
where
\be
    \check{u}\vert_{\B_{\includegraphics[scale=.15, align=c]{figures/2chain_B1.pdf}}} &= (-S_1)^{-\alpha_1} x_2^{-\alpha_2}
    \,,
    & 
    \check{\omega}\vert_{\B_{\includegraphics[scale=.15, align=c]{figures/2chain_B1.pdf}}} &= -\alpha_2\; \dlog\, x_2 
    \,,
    \\
    \check{u}\vert_{\B_{\includegraphics[scale=.15, align=c]{figures/2chain_B2.pdf}}} &= (x_1)^{-\alpha_1} (-S_2)^{-\alpha_2}
    \,,
    & 
    \check{\omega}\vert_{\B_{\includegraphics[scale=.15, align=c]{figures/2chain_B2.pdf}}} &= -\alpha_1\; \dlog\, x_1 
    \,,
    \\
    \check{u}\vert_{\B_{\includegraphics[scale=.15, align=c]{figures/2chain_B12.pdf}}} &= (x_1)^{-\alpha_1} (-x_1-S_3)^{-\alpha_2}
    \,,
    &
    \check{\omega}\vert_{\B_{\includegraphics[scale=.15, align=c]{figures/2chain_B12.pdf}}} &= -\alpha_1\; \dlog\, x_1 - \alpha_2\; \dlog\, (x_1 + S_3) 
    \,.
\ee
Next, note that these lower dimensional integrals on cuts can also be simplified
\begin{align}
    \text{on } \B_{\includegraphics[scale=.15, align=c]{figures/2chain_B1.pdf}}:& 
        -\frac{
            \alpha_2\; \dlog\, x_2
        }{2\alpha_1 \alpha_2} 
        = \frac{
            \check{\omega}\vert_{
                \B_{\includegraphics[scale=.15, align=c]{figures/2chain_B1.pdf}}
            } 
        }{2\alpha_1 \alpha_2}
        = \check{\nabla}\vert_{
            \B_{\includegraphics[scale=.15, align=c]{figures/2chain_B1.pdf}}
        }
        \left(
            \frac{1}{2 \alpha_1 \alpha_2}
        \right)
        \,,
    \nn\\
    \text{on } \B_{\includegraphics[scale=.15, align=c]{figures/2chain_B2.pdf}}:& 
        \frac{\alpha_1\; \dlog\, x_1}{2\alpha_1 \alpha_2} 
        = -\frac{\check{\omega}\vert_{\B_{\includegraphics[scale=.15, align=c]{figures/2chain_B2.pdf}}}}{2 \alpha_1 \alpha_2} 
        = \check{\nabla}\vert_{\B_{\includegraphics[scale=.15, align=c]{figures/2chain_B2.pdf}}}\left(
            -\frac{1}{2 \alpha_1 \alpha_2}
        \right)    
        \,,
    \\
    \text{on } \B_{\includegraphics[scale=.15, align=c]{figures/2chain_B12.pdf}}:& 
        \frac{
            \alpha_1\; \dlog\,x_1
            {-} \alpha_2\; \dlog\,(x_1{+}S_3)
        }{2\alpha_1\alpha_2}
        {=} \frac{1}{\alpha_1\alpha_2} 
            \dlog\frac{x_1}{x_1+S_3}
        + \check{\nabla}\vert_{\B_{\includegraphics[scale=.15, align=c]{figures/2chain_B12.pdf}}} \left(
            \frac{
                \alpha_2{-}\alpha_1
            }{
                2 \alpha_1 \alpha_2
                (\alpha_1{+}\alpha_2)
            }
        \right)
        ,
    \nn
\end{align}
Substituting the above into \eqref{eq:2siteDualIntStokes1}, using Stokes theorem once again as well as $\partial_{t} \partial_{{t'}} = -\partial_{{t'}} \partial_{{t}}$, yields the fully reduced version of \eqref{eq:2siteDualInt}
\be \label{eq:2siteDualIntStokes2}
    \int_{\check{\gamma}} \check{u}\; \check{\vphi}
    &= \frac{1}{\alpha_1 \alpha_2}
        \int_{\partial_{{\includegraphics[scale=.15, align=c]{figures/2chain_B2.pdf}}}\partial_{{\includegraphics[scale=.15, align=c]{figures/2chain_B1.pdf}}}{\check{\gamma}}} 
        \check{u}\vert_{
            \B_{\includegraphics[scale=.15, align=c]{figures/2chain_B1B2.pdf}
            }
        }
    - \frac{1}{\alpha_2 (\alpha_1+\alpha_2)}
        \int_{
            \partial_{
                \includegraphics[scale=.15, align=c]{figures/2chain_B1.pdf}
            } 
            \partial_{
                \includegraphics[scale=.15, align=c]{figures/2chain_B12.pdf}
            } 
            \check{\gamma}
        } 
        \check{u}\vert_{
            \B_{
                \includegraphics[scale=.15, align=c]{figures/2chain_B1B12.pdf}
            }
        }
    \\&    
    + \frac{1}{\alpha_1 (\alpha_1+\alpha_2)}
        \int_{
            \partial_{
                \includegraphics[scale=.15, align=c]{figures/2chain_B2.pdf}
            }
            \partial_{
                \includegraphics[scale=.15, align=c]{figures/2chain_B12.pdf}
            }
            \check{\gamma}
        } 
        \check{u}\vert_{\B_{
            \includegraphics[scale=.15, align=c]{figures/2chain_B2B12.pdf}
        }}
    + \frac{1}{\alpha_1\alpha_2}
        \int_{\partial_{\B_{\includegraphics[scale=.15, align=c]{figures/2chain_B12.pdf}}}{\check{\gamma}}} 
        (\check{u}\vert_{\B_{\includegraphics[scale=.15, align=c]{figures/2chain_B12.pdf}}})\; 
        \dlog\frac{x_1}{x_1+S_3}
\ee
We have succeeded in simplifying the two dimensional integral \eqref{eq:2siteDualInt} into three zero-dimensional integrals, and one one-dimensional integral.
The integrands are restricted to either the one-dimensional cut space $M_{\includegraphics[scale=.15, align=c]{figures/2chain_B12.pdf}}$ or the zero-dimensional cut spaces $M_{\includegraphics[scale=.15, align=c]{figures/2chain_B1B12.pdf}}$, $M_{\includegraphics[scale=.15, align=c]{figures/2chain_B2B12.pdf}}$, and $M_{\includegraphics[scale=.15, align=c]{figures/2chain_B1B2.pdf}}$. 
Moreover, on each cut, the integrands cannot be simplified further. 
That is, $\dlog\frac{x_1}{x_1+S_3}$ is a basis for $H^1\left(M_{\includegraphics[scale=.15, align=c]{figures/2chain_B12.pdf}};\check{\nabla}\vert_{\B_{\includegraphics[scale=.15, align=c]{figures/2chain_B12.pdf}}}\right)$ and the constant function, $1$, is a basis for   $H^0\left(M_{\includegraphics[scale=.15, align=c]{figures/2chain_B1B12.pdf}}\right)$, 
$H^0\left(M_{\includegraphics[scale=.15, align=c]{figures/2chain_B2B12.pdf}}\right)$,
and 
$H^0\left(M_{\includegraphics[scale=.15, align=c]{figures/2chain_B1B12.pdf}}\right)$. 



The discussion so far has relied on the interplay between forms and contours. 
The relative twisted cohomology repackages everything so that we only need to talk only about differential forms on the cuts $M_T$ and how the cuts/boundaries are  coupled. 
To accomplish this, we introduce the (sequential) coboundary symbol $\delta_T := \delta_{t_1} \circ \cdots \circ \delta_{t_{|T|}}$.
The coboundary, is defined so that  
\be
    \int_{{\check{\gamma}}} \check{u}\; 
    \delta_{T} (\check{\phi})
    = \int_{
        \partial_{t_{|T|}}
        \cdots 
        \partial_{t_{1}}
        {\check{\gamma}}
    } (\check{u}\vert_{\B_{T}})\;
    \check{\phi}
    \,,
\ee
where $T=(t_{1}, \cdots, t_{|T|})$.
For generic hyperplane arrangements, the coboundary is anti-symmetric in its indices  $\delta_{t t'} = - \delta_{t' t}$. 
This mirrors the anti-symmetry of the partial boundary operators $\partial_{t} \partial_{t'} = - \partial_{t'} \partial_{t}$. 

Using this notation, \eqref{eq:2siteDualIntStokes2} becomes 
\be 
    \int_{\check{\gamma}} \check{u}\; \check{\vphi}
    = \int_{{\check{\gamma}}} \check{u}\; 
    \bigg[&
        \frac{1}{\alpha_1 \alpha_2}
            \delta_{
                \includegraphics[scale=.15, align=c]{figures/2chain_B1.pdf},
                \includegraphics[scale=.15, align=c]{figures/2chain_B2.pdf}
            } (1)
        - \frac{1}{\alpha_2 (\alpha_1+\alpha_2)} 
            \delta_{
                \includegraphics[scale=.15, align=c]{figures/2chain_B12.pdf},
                \includegraphics[scale=.15, align=c]{figures/2chain_B1.pdf}
            } (1)
    \\&\quad
        + \frac{1}{\alpha_1 (\alpha_1+\alpha_2)}
            \delta_{
                \includegraphics[scale=.15, align=c]{figures/2chain_B12.pdf},
                \includegraphics[scale=.15, align=c]{figures/2chain_B2.pdf}
            } (1)
        + \frac{1}{\alpha_1\alpha_2}
            \delta_{
                \includegraphics[scale=.15, align=c]{figures/2chain_B12.pdf}
            } \left(
                \dlog\frac{x_1}{x_1+S_3}
            \right) 
    \bigg]
    \,,
\ee
where 
\be \label{eq:2siteDualBasis}
    \left\{
        \delta_{
            \includegraphics[scale=.15, align=c]{figures/2chain_B12.pdf},
        } \left(
            \dlog\frac{x_1}{x_1+S_3}
        \right)
        \delta_{
            \includegraphics[scale=.15, align=c]{figures/2chain_B1.pdf},
            \includegraphics[scale=.15, align=c]{figures/2chain_B2.pdf}
        }(1), 
        \delta_{
            \includegraphics[scale=.15, align=c]{figures/2chain_B12.pdf},
            \includegraphics[scale=.15, align=c]{figures/2chain_B1.pdf}
        }(1), 
        \delta_{
            \includegraphics[scale=.15, align=c]{figures/2chain_B12.pdf},
            \includegraphics[scale=.15, align=c]{figures/2chain_B2.pdf}
        }(1)
    \right\}
    \,,
\ee
is a basis of differential forms for the relative twisted cohomology $H^2(M,\B;\check{\nabla})$. 
Note that $\delta_t$ has form-degree one; all elements of \eqref{eq:2siteDualBasis} are to be thought of as relative twisted 2-forms. 
By definition, the dual covariant derivative treats each $\delta_t$ as a 1-form.
Moreover, whenever the derivative hits a coboundary symbol, it produces the corresponding boundary terms
\be
    \check{\nabla} \delta_T(\check{\phi}) = (-1)^{|T|} \left[
        \delta_{T}(\check{\nabla}\vert_{\B_T} \check{\phi})
        + \sum_{t \notin T} \delta_{T \cup t}(\check{\phi}\vert_{\B_{T \cup t}})
    \right]
    \,.
\ee 
Also, the absence of a coboundary symbol is equivalent to $\delta_\emptyset$. 
In this way, boundary terms are still generated when acting with the covariant derivative on forms without any $\delta_\bullet$'s. 

We can now declare that $\check{\vphi}$ is cohomologous to the combination of forms in \eqref{eq:2siteDualIntStokes2}.
Explicitly, 
\be
    \frac{\d x_1 \wedge \d x_2}{x_1 x_2} 
    - \check{\nabla}\text{IBP}
    &= \frac{1}{\alpha_1 \alpha_2}
            \delta_{
                \includegraphics[scale=.15, align=c]{figures/2chain_B1.pdf},
                \includegraphics[scale=.15, align=c]{figures/2chain_B2.pdf}
            } (1)
        + \frac{1}{\alpha_2 (\alpha_1+\alpha_2)} 
            \delta_{
                \includegraphics[scale=.15, align=c]{figures/2chain_B1.pdf},
                \includegraphics[scale=.15, align=c]{figures/2chain_B12.pdf}
            } (1)
    \\&\quad
        - \frac{1}{\alpha_1 (\alpha_1+\alpha_2)}
            \delta_{
                \includegraphics[scale=.15, align=c]{figures/2chain_B2.pdf},
                \includegraphics[scale=.15, align=c]{figures/2chain_B12.pdf}
            } (1)
        + \frac{1}{\alpha_1\alpha_2}
            \delta_{
                \includegraphics[scale=.15, align=c]{figures/2chain_B12.pdf}
            } \left(
                \dlog\frac{x_1}{x_1+S_3}
            \right) 
    \,,
\ee
where 
\be
    \text{IBP} 
    {=} \frac{
            \alpha_1\, \dlog\,x_1
            {-} \alpha_2\, \dlog\,x_2
        }{2\alpha_1\alpha_2}
    {-} \frac{1}{2 \alpha_1 \alpha_2} 
        \delta_{
            \includegraphics[scale=.15, align=c]{figures/2chain_B1.pdf}
        }(1)
    {+} \frac{1}{2 \alpha_1 \alpha_2}
        \delta_{
            \includegraphics[scale=.15, align=c]{figures/2chain_B2.pdf}
        }(1)
    {-} \frac{
            \alpha_2{-}\alpha_1
        }{
            2 \alpha_1 \alpha_2
            (\alpha_1{+}\alpha_2)
        }
        \delta_{
            \includegraphics[scale=.15, align=c]{figures/2chain_B12.pdf}
        }(1)
    .
\ee

A general feature of the relative twisted cohomology is that it has a decomposition into the direct sum of the twisted cohomologies on each cut (for generic arrangements)
\be \label{eq:Hdual}
    H^{|V_G|}(M, \B; \check{\nabla})
    := 
        \bigoplus_{T \text{ (cuts)}}
        \delta_T \circ H^{|V_G|-|T|}(M_{T}; \check{\nabla}\vert_{\B_T})
    \,.
\ee
That is, relative twisted $|V_G|$-forms are vectors whose components (labeled by the $\delta_\bullet$) are twisted $(|V_G|-|T|)$-forms on $M_T$. 
Like the FRW-cohomology, there is a middle dimensional theorem and only $H^{|V_G|}\neq0$. 
For the two-site chain, this decomposition is 
\be 
    H^2(M, \B; \check{\nabla})
    &= \delta_{
         \includegraphics[scale=.15, align=c]{figures/2chain_B12.pdf}
    } \circ H^{1}\left(M_{\includegraphics[scale=.15, align=c]{figures/2chain_B12.pdf}}; \check{\nabla}\vert_{\B_{ \includegraphics[scale=.15, align=c]{figures/2chain_B12.pdf}}}\right)
    \oplus \delta_{
         \includegraphics[scale=.15, align=c]{figures/2chain_B1.pdf},
          \includegraphics[scale=.15, align=c]{figures/2chain_B2.pdf}
    } \circ H^{0}(M_{\includegraphics[scale=.15, align=c]{figures/2chain_B1B2.pdf}})
    \\&\qquad\qquad
    \oplus \delta_{
         \includegraphics[scale=.15, align=c]{figures/2chain_B12.pdf},
          \includegraphics[scale=.15, align=c]{figures/2chain_B1.pdf}
    } \circ H^{0}(M_{ \includegraphics[scale=.15, align=c]{figures/2chain_B2B12.pdf}})
    \oplus \delta_{
         \includegraphics[scale=.15, align=c]{figures/2chain_B12.pdf},
          \includegraphics[scale=.15, align=c]{figures/2chain_B2.pdf}
    } \circ H^{0}(M_{\includegraphics[scale=.15, align=c]{figures/2chain_B1B12.pdf}})
    \,,
\ee
where 
$
    H^2(M; \check{\nabla}) 
    = H^1\left(
        M_{
            \includegraphics[scale=.15, align=c]{figures/2chain_B1.pdf}
        }; \check{\nabla} \vert_{
            \B_{\includegraphics[scale=.15, align=c]{figures/2chain_B1.pdf}
        }}
    \right)
    = H^1\left(
        M_{
            \includegraphics[scale=.15, align=c]{figures/2chain_B1.pdf}
        }; \check{\nabla} \vert_{
            \B_{\includegraphics[scale=.15, align=c]{figures/2chain_B2.pdf}
        }}
    \right)
    = 0
$
since there are no bounded chambers on $M$, 
$M_{\includegraphics[scale=.15, align=c]{figures/2chain_B1.pdf}}$, 
and $M_{\includegraphics[scale=.15, align=c]{figures/2chain_B2.pdf}}$.
Moreover, each $M_T$ in \eqref{eq:Hdual} is a (twisted) hyperplane and therefore a positive geometry. 
To find a basis for the relative twisted cohomology, we use the coboundary of the canonical forms on each $M_T$. 

\begin{tcolorbox}[breakable, title=Sum of \emph{positive geometries!}]
    Equation \eqref{eq:Hdual} says that the dual cohomology is built up from the cohomology of the cuts $M_T$. 
    On each cut there is a (twisted) hyperplane arrangement which is a positive geometry.
    Hence, the dual cohomology is simply the direct sum of the cohomology associated to these positive geometries!
    \\[1em]
    We will build all FRW- and dual-forms from the positive geometry of the cuts \eqref{eq:phiCut} and show that this is equivalent to a simple combinatorial formula \eqref{eq:phiComb}.
    \\[1em]
    Via the intersection duality, the FRW-cohomology can be organized by cuts as schematically shown in \eqref{eq:schematicFRWOrg}. 
\end{tcolorbox}

\paragraph{Cuts from the coboundary and the intersection number.}
The last but most important part property of the coboundary is that it induces a (generalized unitarity) cut in the intersection number
\be \label{eq:coboundaryLocalization}
    \la \delta_T( \check{\phi} ) \vert \phi \ra 
    = \la \check{\phi} \vert \res_T[\phi] \ra 
    \,,
\ee
where $\res_T := \res_{B_{t_{|T|}}} \circ \cdots \circ \res_{\B_{t_1}}$. 
Note that when $\phi$ has higher order poles, the above formula must be modified slightly \cite{Caron-Huot:2021xqj, Caron-Huot:2021iev}.%
\footnote{When there are higher order poles, $\res_T[ \bullet ] \to \res_T[\frac{u\vert_{\B_T}}{u}\bullet]$.}
The intersection number also turns the remaining part of the dual form $\check{\phi}$ in \eqref{eq:coboundaryLocalization} into another residue operator acting on $\res_T[\phi]$
\be \label{eq:intNum0}
    \la \delta_T(\check{\phi}) \vert \phi \ra
    = \la \check{\phi} \vert \res_T[\phi] \ra
    = \res_{\check{\phi}} \circ \res_T[\phi]. 
\ee
Here, $\res_{\check{\phi}}$ localizes to where $\check{\phi}$ is maximally singular. 
We avoid the details to construct $\res_{\check{\phi}}$ (the interested reader is encouraged to look at \cite{Caron-Huot:2021iev, Caron-Huot:2021xqj} and references therein). 
Instead, we provide an explicit formula for $\res_{\check{\phi}}$ in the context of our basis forms in section \ref{sec:flowOfCuts}. 

In this paper, one should think of dual forms simply as an an instruction for how to construct residue operators that project FRW-integrals onto a basis. 
Given a basis $\{\phi_a\}$ for the FRW-cohomology and a basis $\{\check{\phi}_a\}$ for the dual cohomology, the matrix elements of the differential equation are 
\be \label{eq:Amat}
    A_{ab} = 
    C_{bc}^{-1} 
    \la \check{\phi}_c \vert \d_\text{kin} \phi_a \ra
\ee
where $C_{ab} = \la \check{\phi}_a \vert \phi_b \ra$ is the intersection matrix. 
Closed formula for both the intersection matrix $\mat{C}$ and the connection $\mat{A}$ are located in section \ref{sec:flowOfCuts}.

\section{Constructing the cut basis}
\label{sec:cutBasis}

In section \ref{sec:indepCuts}, we identify the degeneracy in the cosmological hyperplane arrangement $\B$ and deduce its implications on sequential residues and logarithmic forms. 
Then, we introduce graphical/combinatorial gadgets (acyclic minors) to help organize the structure of the basis and resulting differential equations in section \ref{sec:acyclicMinors}. 
Next, the cut basis is constructed directly from the acyclic minors (section \ref{sec:cutbasis}) and related to the canonical form of the positive geometry on each physical cut (\ref{sec:cutGeom}). 
Lastly, in section \ref{sec:reltaionToTimeInt}, we discuss how the cut basis is related to the time integral basis of \cite{Baumann:2025qjx,He:2024olr}.

\subsection{Independent cuts of the physical FRW-form}
\label{sec:indepCuts}

It is well-known that the kinematic derivative of a Feynman integral can be written as a linear combination of itself and other Feynman integrals where one propagator (pole in the integrand) is removed. 
Usually, a family of Feynman integrals is classified according to sectors that are defined by which propagators (poles) are present in the integrand. A similar approach carries over to the FRW-integrals we consider here. The kinematic differential of an FRW-integral cannot couple to another FRW integral unless the set of propagators of the second is a subset of the first. 
This is a direct consequence of equations \eqref{eq:coboundaryLocalization}, \eqref{eq:intNum0} and \eqref{eq:Amat}, and the fact that a kinematic derivative cannot introduce new poles that were not already present.  
Therefore, the structure of the differential equation is controlled by the independent cuts/residues of physical FRW-form.

Recall from \eqref{eq:lin_form} and \eqref{eq:cosPsi} that the physical FRW-form is given simply by
\begin{align}
    \Psi_G := \hat{\Omega}_G(\mbf{x} + \mbf{X},{\bf Y}) 
    \; \d^{|V_G|}\mbf{x}
    =  \prod_{e\in E_G} 2 Y_e 
    \sum_{ \tau \in \mathcal{T}_G^{\text{max}} }  
    \frac{1}{B_\tau}\d^{|V_G|}\mbf{x}
    \,. 
\label{eq:phys_form}
\end{align}
It is important to emphasize that $|\T^\max_G| = |V_G|+|E_G|$, which is larger than $|V_G|$, the number of integration variables $x_i$. 
Therefore, it is possible to use partial fractions such that each term in $\Psi_G$ has at most $|V_G|$ factors in the denominator. 
Graphical/combinatorial algorithms for such a decomposition are discussed in \cite{Fevola:2024nzj,Glew:2025arc, Glew:2025ugf}. 
For example, performing partial fractions on the shifted version of the flat-space wavefunction coefficient associated to the 3-site chain, \eqref{eq:p3}, yields
\begin{align} \begin{aligned} \label{eq:p3fractioned}
    \hat\Omega_{\includegraphics[scale=0.6]{figures/pt3}}({\bf x}+{\bf X},{\bf Y}) 
    &= \frac{1}{B_{\includegraphics[scale=0.6]{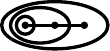}}}
    + 
    \textcolor{NavyBlue}{
        \frac{1}{B_{\includegraphics[scale=0.55]{figures/pt3_t2}}}
    }
    + \textcolor{NavyBlue}{
        \frac{1}{B_{\includegraphics[scale=0.55]{figures/pt3_t3}}}
    }
    + \frac{1}{B_{\includegraphics[scale=0.55]{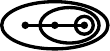}}}
    + \frac{1}{B_{\includegraphics[scale=0.55]{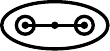}} }
    \\
    &- \frac{1}{B_{\includegraphics[scale=0.7]{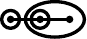}}}
    - \frac{1}{B_{\includegraphics[scale=0.7]{figures/pt3_line_t1_b}}}
    - \frac{1}{B_{\includegraphics[scale=0.7]{figures/pt3_line_t1_b}}}
    - \frac{1}{B_{\includegraphics[scale=0.7]{figures/pt3_line_t1_b}}}
    + \frac{1}{B_{\includegraphics[scale=0.7]{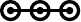}}}
    \,.
\end{aligned}\end{align}
Naively, one expects that after repeated use of partial fractions, so that each term in \eqref{eq:phys_form} has $|V_G|$ factors in the denominator, the resulting terms would expose all possible maximal cuts/residues of $\Psi_G$.
Continuing the example \eqref{eq:p3fractioned}, we expect to have $10$ distinct maximal cuts. 
However, this misses the fact that not all terms are linearly independent! 
Indeed, the {\color{RoyalBlue} blue} terms in \eqref{eq:p3fractioned} satisfy the linear relation
\begin{align} \label{eq:dependentPartialFractions}
    \textcolor{NavyBlue}{\frac{1}{B_{\includegraphics[scale=0.6]{figures/pt3_t2}}}}+\textcolor{NavyBlue}{\frac{1}{B_{\includegraphics[scale=0.6]{figures/pt3_t3}}}}
    = \frac{1}{B_{\includegraphics[scale=0.6]{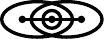}}} + \frac{1}{B_{\includegraphics[scale=0.6]{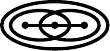}}}
    \,,
\end{align}
since 
\begin{align}
    B_{
        \begin{gathered}
        \begin{tikzpicture}[scale=.3]
            \coordinate (A0) at (-4,0);
            \coordinate (B0) at (-3,0);
            \coordinate (C0) at (-2,0);
            \draw[thick] (A0) -- (B0) -- (C0);
            \draw[thick, black] (-3.5,0) ellipse (0.8cm and 0.4cm);
            \fill[black] (A0) circle (3pt);
            \fill[black] (B0) circle (3pt);
            \fill[black] (C0) circle (3pt);
        \end{tikzpicture}
        \end{gathered}
    }
    +
    B_{
        \begin{gathered}
        \begin{tikzpicture}[scale=.3]
            \coordinate (A0) at (-4,0);
            \coordinate (B0) at (-3,0);
            \coordinate (C0) at (-2,0);
            \draw[thick] (A0) -- (B0) -- (C0);
            \draw[thick, black] (-2.5,0) ellipse (0.8cm and 0.4cm);
            \fill[black] (A0) circle (3pt);
            \fill[black] (B0) circle (3pt);
            \fill[black] (C0) circle (3pt);
        \end{tikzpicture}
        \end{gathered}
    }
    = 
    B_{
        \begin{gathered}
        \begin{tikzpicture}[scale=.3]
            \coordinate (A0) at (-4,0);
            \coordinate (B0) at (-3,0);
            \coordinate (C0) at (-2,0);
            \draw[thick] (A0) -- (B0) -- (C0);
            \draw[thick, black] (-3,0) ellipse (1.2cm and 0.4cm);
            \fill[black] (A0) circle (3pt);
            \fill[black] (B0) circle (3pt);
            \fill[black] (C0) circle (3pt);
        \end{tikzpicture}
        \end{gathered}
    }
    +
    B_{
        \begin{gathered}
        \begin{tikzpicture}[scale=.3]
            \coordinate (A0) at (-4,0);
            \coordinate (B0) at (-3,0);
            \coordinate (C0) at (-2,0);
            \draw[thick] (A0) -- (B0) -- (C0);
            \draw[thick, black] (-3,0) circle (0.3cm);
            \fill[black] (A0) circle (3pt);
            \fill[black] (B0) circle (3pt);
            \fill[black] (C0) circle (3pt);
        \end{tikzpicture}
        \end{gathered}
    }
    \,.
    \label{eq:lin_dep_examp}
\end{align}

The origin of this linear dependence is geometric.  
The untwisted hyperplane arrangement $\B$ is degenerate for $|V_G|>2$: there are codimension-$k$ surfaces in $\B$ where more than $k$ hyperplanes intersect. 
This is due to the hyperplane polynomials $B_\tau$ satisfying $(n\geq4)$-term identities\footnote{Note that $\tau_1 \cap \tau_2$ may correspond to more than one tube; in such cases, $B_{\tau_1 \cap \tau_2} := \sum_{\tau \in \tau_1 \cap \tau_2} B_\tau$.} 
\be \label{eq:linRel}
    B_{\tau_1} + B_{\tau_2} = B_{\tau_1 \cap \tau_2} + B_{\tau_1 \cup \tau_2}
    \,. 
\ee
Consequently, there are relations between sequential residue operators since setting any three $B_t$'s in \eqref{eq:lin_dep_examp} to zero automatically sets the fourth to zero. 
In terms of the cut spaces, which do not depend on an ordering, 
\begin{align}
    M_{
        \begin{gathered}
        \begin{tikzpicture}[scale=.5]
            \coordinate (A0) at (-4,0);
            \coordinate (B0) at (-3,0);
            \coordinate (C0) at (-2,0);
            \draw[thick] (A0) -- (B0) -- (C0);
            \draw[thick, black] (-3,0) ellipse (1.4cm and 0.6cm);
            \draw[thick, black] (-3.5,0) ellipse (0.8cm and 0.4cm);
            \draw[thick, black] (-3,0) circle (0.2cm);
            \fill[black] (A0) circle (3pt);
            \fill[black] (B0) circle (3pt);
            \fill[black] (C0) circle (3pt);
        \end{tikzpicture}
        \end{gathered}
    }
    =
    M_{
        \begin{gathered}
        \begin{tikzpicture}[scale=.5]
            \coordinate (A0) at (-4,0);
            \coordinate (B0) at (-3,0);
            \coordinate (C0) at (-2,0);
            \draw[thick] (A0) -- (B0) -- (C0);
            \draw[thick, black] (-3,0) ellipse (1.4cm and 0.6cm);
            \draw[thick, black] (-2.5,0) ellipse (0.8cm and 0.4cm);
            \draw[thick, black] (-3,0) circle (0.2cm);
            \fill[black] (A0) circle (3pt);
            \fill[black] (B0) circle (3pt);
            \fill[black] (C0) circle (3pt);
        \end{tikzpicture}
        \end{gathered}
    }
    =
    M_{
        \begin{gathered}
        \begin{tikzpicture}[scale=.5]
            \coordinate (A0) at (-4,0);
            \coordinate (B0) at (-3,0);
            \coordinate (C0) at (-2,0);
            \draw[thick] (A0) -- (B0) -- (C0);
            \draw[thick, black] (-3.5,0) ellipse (0.8cm and 0.4cm);
            \draw[thick, black] (-2.5,0) ellipse (0.8cm and 0.4cm);
            \draw[thick, black] (-3,0) circle (0.2cm);
            \fill[black] (A0) circle (3pt);
            \fill[black] (B0) circle (3pt);
            \fill[black] (C0) circle (3pt);
        \end{tikzpicture}
        \end{gathered}
    }
    =
    M_{
        \begin{gathered}
        \begin{tikzpicture}[scale=.5]
            \coordinate (A0) at (-4,0);
            \coordinate (B0) at (-3,0);
            \coordinate (C0) at (-2,0);
            \draw[thick] (A0) -- (B0) -- (C0);
            \draw[thick, black] (-3,0) ellipse (1.4cm and 0.6cm);
            \draw[thick, black] (-3.5,0) ellipse (.8cm and 0.4cm);
            \draw[thick, black] (-2.5,0) ellipse (0.8cm and 0.4cm);
            \fill[black] (A0) circle (3pt);
            \fill[black] (B0) circle (3pt);
            \fill[black] (C0) circle (3pt);
        \end{tikzpicture}
        \end{gathered}
    }
    \,.
\end{align}
To derive relations between residue operators, one needs to fix an ordering for the degenerate tubes in \eqref{eq:linRel} \cite{dupont2015orliksolomonmodelhypersurfacearrangements,  orlik1980combinatorics}.  
If $\{b_k\}_{k=1}^m$ is the set of tubes that appear in \eqref{eq:linRel} and they are ordered $(b_1, \dots, b_m)$, we have 
\begin{align}
    \res_{b_{m-1}=0} \circ 
    \res_{b_{m-2}=0} \circ
    \cdots \circ \res_{b_{1}=0} 
    = \res_{b_{m}=0} \circ 
    \res_{b_{m-2}=0} \circ
    \cdots \circ \res_{b_{1}=0}
    \,.
\end{align}
We also have the following relation among logarithmic forms
\begin{align} \label{eq:formLinRel}
    \sum_{i=1}^{m} 
    (-1)^{m-i} \bigwedge_{j\neq i } \dlog b_j = 0
    \,.
\end{align}
Importantly, the coboundary symbols $\delta_\bullet$ in the dual-forms also satisfy the same relations as the corresponding residues operators.

Returning to the 3-site chain example, and ordering the tubes via
\begin{align} \label{eq:3chainDegenTubeList}
    \left(
    \includegraphics[align=c,scale=.4]{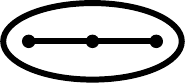},
    \includegraphics[align=c,scale=.4]{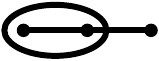},
    \includegraphics[align=c,scale=.4]{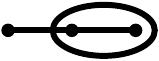},
    \includegraphics[align=c,scale=.4]{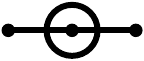}
    \right)
\end{align} 
reveals that there are only three linearly independent sequential residues that respect this tube ordering
\begin{align}\begin{aligned}
    \{ 
    \res_{\includegraphics[align=c,scale=.3]{figures/3chain_B2.pdf}}
    \circ \res_{\includegraphics[align=c,scale=.3]{figures/3chain_B12.pdf}}
    \circ \res_{\includegraphics[align=c,scale=.3]{figures/3chain_B123.pdf}}
    &=
    \res_{\includegraphics[align=c,scale=.3]{figures/3chain_B23.pdf}}
    \circ \res_{\includegraphics[align=c,scale=.3]{figures/3chain_B12.pdf}}
    \circ \res_{\includegraphics[align=c,scale=.3]{figures/3chain_B123.pdf}}
    \,,
    \\
    \res_{\includegraphics[align=c,scale=.3]{figures/3chain_B2.pdf}}
    \circ \res_{\includegraphics[align=c,scale=.3]{figures/3chain_B23.pdf}}
    \circ \res_{\includegraphics[align=c,scale=.3]{figures/3chain_B123.pdf}}
    & \,, 
    \res_{\includegraphics[align=c,scale=.3]{figures/3chain_B2.pdf}}
    \circ \res_{\includegraphics[align=c,scale=.3]{figures/3chain_B23.pdf}}
    \circ \res_{\includegraphics[align=c,scale=.3]{figures/3chain_B12.pdf}}
    \}
    \,.
\end{aligned}\end{align}
Moreover, equation \eqref{eq:dependentPartialFractions} follows form  \eqref{eq:formLinRel}
\begin{align}\begin{aligned}
    &\dlog\, \includegraphics[align=c,scale=.3]{figures/3chain_B123.pdf}
    \wedge \dlog\, 
    \includegraphics[align=c,scale=.3]{figures/3chain_B12.pdf}
    \wedge \dlog\,
    \includegraphics[align=c,scale=.3]{figures/3chain_B23.pdf}
    -\dlog\, \includegraphics[align=c,scale=.3]{figures/3chain_B123.pdf}
    \wedge \dlog\, 
    \includegraphics[align=c,scale=.3]{figures/3chain_B12.pdf}
    \wedge \dlog\,
    \includegraphics[align=c,scale=.3]{figures/3chain_B2.pdf}
    \\&\quad
    + \dlog\, \includegraphics[align=c,scale=.3]{figures/3chain_B123.pdf}
    \wedge \dlog\, 
    \includegraphics[align=c,scale=.3]{figures/3chain_B23.pdf}
    \wedge \dlog\,
    \includegraphics[align=c,scale=.3]{figures/3chain_B2.pdf}
    -\dlog\, \includegraphics[align=c,scale=.3]{figures/3chain_B12.pdf}
    \wedge \dlog\, 
    \includegraphics[align=c,scale=.3]{figures/3chain_B23.pdf}
    \wedge \dlog\,
    \includegraphics[align=c,scale=.3]{figures/3chain_B2.pdf}
    = 0 
    \,,
\end{aligned}\end{align}
after factoring out the volume form $\d^3\mbf{x}$.

\begin{tcolorbox}[breakable, title=\hypertarget{box:cutOrd}{Cut ordering}]
The cut multi-index $J$ in $\delta_J$, $\res_J$ and $M_J$ is \emph{always} ordered such that tubes with the most vertices comes first. 
For tubes of the same size, we proceed in lexicographical order.
\\[1em]
This choice is natural in the sense that sequential residues with crossed tubes vanish respecting the compatibility of tubes established earlier. 
In particular, all residues with crossed tubes annihilate the physical FRW-form when ordered this way. 
\end{tcolorbox}

An algorithm for enumerating all independent cuts of the physical FRW-form $\Psi_G$ using {\it cut tubings} was provided in \cite{De:2024zic}. However, this approach requires careful bookkeeping of degeneracies between cuts arising from linear relations such as \eqref{eq:linRel}. In this work, we improve the algorithm by demonstrating how the {\it acyclic minors} introduced in \cite{Glew:2025ugf, He:2024olr} capture the set of independent cuts of $\Psi_G$, without encountering the issue of degeneracy.


\subsection{Acyclic minors}
\label{sec:acyclicMinors}
\begin{figure}[]
\centering
 \includegraphics[width=\textwidth]{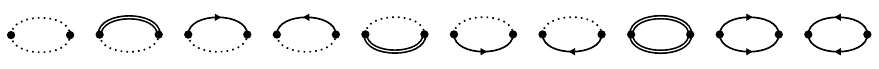}
\caption{The $10$ acyclic minors of the two-cycle.}
\label{fig:sunrise_ado}
\end{figure}

To label the set of independent cuts of the physical FRW-form we introduce the notion of an {\it acyclic minor}. An acyclic minor of a graph $G$ is a decorated graph $\g$, obtained by assigning to each edge $e \in E_G$ one of three decorations: $\begin{tikzpicture}[scale=1]
        \coordinate (A) at (0,0);
        \coordinate (B) at (1/2,0);
        \coordinate (C) at (1,0);
        \coordinate (D) at (3/2,0);
        \draw[thick,double] (A) -- (B);
        \fill[black] (A) circle (2pt);
        \fill[black] (B) circle (2pt);
    \end{tikzpicture}$ (solid), $\begin{tikzpicture}[scale=1]
        \coordinate (A) at (0,0);
        \coordinate (B) at (1/2,0);
        \coordinate (C) at (1,0);
        \coordinate (D) at (3/2,0);
        \draw[ultra thick,dotted] (A) -- (B);
        \fill[black] (A) circle (2pt);
        \fill[black] (B) circle (2pt);
    \end{tikzpicture}$ (broken) or $\raisebox{-0.11cm}{\begin{tikzpicture}[scale=1]
        \coordinate (A) at (0,0);
        \coordinate (B) at (1/2,0);
        \coordinate (C) at (1,0);
        \coordinate (D) at (3/2,0);
        \draw[ultra thick] (A) --  node {\arb} (B);
        \fill[black] (A) circle (2pt);
        \fill[black] (B) circle (2pt);
    \end{tikzpicture}}$ 
(oriented), such that the oriented graph resulting from contracting all solid edges and deleting all broken edges is acyclic in the usual sense. For example, the complete set of acyclic minors for the three-site chain and two-cycle are illustrated in figure~\ref{fig:all_mins} and figure~\ref{fig:sunrise_ado}. 
We use $\Vso$ to  denote the set of solid edges of $\g$, and similarly $\Vbr$ for the broken edges. The set of all acyclic minors is denoted by $\mathcal{A}(G)$. Given a subset of edges $E \subset E_G$, it will prove useful to refine the set of acyclic minors further and define  
\begin{align}
    \mathcal{A}(G) \supset \mathcal{A}_{E}(G)= \{ \g \in \mathcal{A}(G) : \Vbr =E \} \supset \mathcal{A}^{\emptyset}_{E}(G)= \{ \g \in \mathcal{A}_E(G) : \Vso = \emptyset \}.
\end{align}
In this context, $\mathcal{A}_E(G)$ denotes the set of acyclic minors with fixed broken edge set $E$ and $\mathcal{A}_E^\emptyset(G)$ is the subset with fixed broken edge set $E$ and no solid edges.

Unlike the cut tubings of \cite{De:2024zic}, the set of acyclic minors is in one-to-one correspondence with the independent cuts of the physical FRW-form. Each acyclic minor $\g$ determines a set $\mathcal{C}_{\mathfrak{g}}$ of cut tubings. 
If the cut is degenerate, then $|\mathcal{C}_{\mathfrak{g}}| > 1$. Otherwise, the acyclic minor $\mathfrak{g}$ corresponds to a unique cut tubing.

Given an acyclic minor $\g$, the set of associated cut tubings $\mathcal{C}_\g$ is obtained by the following simple recipe:
\begin{enumerate}
    \item Contract all solid edges and delete all broken edges of the decorated graph $\g$.
    \item On the resulting oriented graph, identify all vertex-induced\footnote{A vertex-induced tube of a graph $G$ is a connected subgraph formed by a subset of vertices $V \subset V_G$, along with all edges of $G$ whose endpoints are both contained in $V$. } tubes that have no incoming oriented edges from vertices outside the tube.
    \item Uplift each tube back to the original decorated graph by reversing the contractions and reinstating the deleted edges.
    \item Form all maximal sets of compatible tubes from this collection to obtain the set of cut tubings $\mathcal{C}_\g$.
\end{enumerate}
To illustrate this definition, consider the following examples of cut tubings
\begin{align} \label{eq:acyclicMinorExample}
\mathcal{C}_{\raisebox{-0.08cm}{\begin{tikzpicture}[scale=0.8]
        \coordinate (A) at (0,0);
        \coordinate (B) at (1/2,0);
        \coordinate (C) at (1,0);
        \coordinate (D) at (3/2,0);
        \coordinate (E) at (2,0);
        \draw[thick] (B) --  node {\ar} (C);
        \draw[thick,double] (A) --  (B);
        \fill[black] (A) circle (2pt);
        \fill[black] (B) circle (2pt);
        \fill[black] (C) circle (2pt);
    \end{tikzpicture}}} &=  \{ \raisebox{-0.2cm}{\begin{tikzpicture}[scale=0.8]
        \coordinate (A) at (0,0);
        \coordinate (B) at (0.6,0);
        \coordinate (C) at (1.2,0);
        \draw[thick] (B) -- (C);
        \draw[thick,double] (A) --  (B);
        \draw node at (0.95,0) {\ar};
        \draw[thick, black] (0.6,0) ellipse (0.9cm and 0.35cm);
        \draw[thick, black] (0.3,0) ellipse (0.5cm and 0.25cm);
        \fill[black] (B) circle (2pt);
        \fill[black] (A) circle (2pt);
        \fill[black] (C) circle (2pt);
    \end{tikzpicture}}  \},  &&
\mathcal{C}_{\raisebox{-0.08cm}{\begin{tikzpicture}[scale=0.8]
        \coordinate (A) at (0,0);
        \coordinate (B) at (1/2,0);
        \coordinate (C) at (1,0);
        \coordinate (D) at (3/2,0);
        \coordinate (E) at (2,0);
        \draw[thick] (B) --  node {\ar} (C);
        \draw[thick] (A) --  node {\al} (B);
        \fill[black] (A) circle (2pt);
        \fill[black] (B) circle (2pt);
        \fill[black] (C) circle (2pt);
    \end{tikzpicture}}} =  \{ \raisebox{-0.2cm}{\begin{tikzpicture}[scale=0.8]
        \coordinate (A) at (0,0);
        \coordinate (B) at (0.6,0);
        \coordinate (C) at (1.2,0);
        \draw[thick] (B) -- (C);
        \draw[thick] (A) --  (B);
        \draw node at (0.95,0) {\ar};
        \draw node at (0.6-0.35,0) {\al};
        \draw[thick, black] (0.6,0) ellipse (0.9cm and 0.35cm);
        \draw[thick, black] (0.3,0) ellipse (0.51cm and 0.26cm);
        \draw[black,thick] (B) circle (4pt);
        \fill[black] (B) circle (2pt);
        \fill[black] (A) circle (2pt);
        \fill[black] (C) circle (2pt);
    \end{tikzpicture}},  \raisebox{-0.2cm}{\begin{tikzpicture}[scale=0.8]
        \coordinate (A) at (0,0);
        \coordinate (B) at (0.6,0);
        \coordinate (C) at (1.2,0);
        \draw[thick] (B) -- (C);
        \draw[thick] (A) --  (B);
        \draw node at (0.95,0) {\ar};
        \draw node at (0.6-0.35,0) {\al};
        \draw[thick, black] (0.6,0) ellipse (0.9cm and 0.35cm);
        \draw[thick, black] (0.9,0) ellipse (0.51cm and 0.26cm);
        \draw[black,thick] (B) circle (4pt);
        \fill[black] (B) circle (2pt);
        \fill[black] (A) circle (2pt);
        \fill[black] (C) circle (2pt);
    \end{tikzpicture}}   \}.
\end{align}
The resulting set of tubes contain no broken edges, do not cross any solid edges and have all oriented edges `flowing outwards' from the tubes. 

\paragraph{Letters.} 

A letter of $\mathfrak{g}$ is defined to be any connected subgraph consisting of only solid edges, together with all incident incoming (half) oriented edges. We denote by $\mathbb{L}_\g$ the set of letters associated to $\g$, and represent each letter by a coloring on the graph, as illustrated in the following examples
\begin{align} \label{eq:letterExample}
    \mathbb{L}_{
 \raisebox{-0.1cm}{\begin{tikzpicture}[scale=0.8]
        \coordinate (A) at (0,0);
        \coordinate (B) at (1/2,0);
        \coordinate (C) at (1,0);
        \coordinate (D) at (3/2,0);
        \coordinate (E) at (2,0);
        \draw[thick] (B) -- node {\ar} (C);
        \draw[thick,double] (A) --  (B);
        \fill[black] (B) circle (2pt);
        \fill[black] (A) circle (2pt);
        \fill[black] (C) circle (2pt);
    \end{tikzpicture}}} &= \{ \raisebox{-0.1 cm}{\begin{tikzpicture}[scale=1]
        \coordinate (A) at (0,0);
        \coordinate (B) at (1/2,0);
        \coordinate (C) at (1,0);
        \coordinate (D) at (3/2,0);
        \coordinate (E) at (2,0);
        \draw[thick] (B) -- node {\ar} (C);
        \draw[ultra thick,double,Magenta] (A) -- (B);
        \fill[black,Magenta] (B) circle (2.3pt);
        \fill[black,Magenta] (A) circle (2.3pt);
        \fill[black] (C) circle (2pt);
    \end{tikzpicture}}
    ,
    \raisebox{-0.12 cm}
    {\begin{tikzpicture}[scale=1]
        \coordinate (A) at (0,0);
        \coordinate (B) at (1/2,0);
        \coordinate (C) at (1,0);
        \coordinate (D) at (3/2,0);
        \coordinate (E) at (2,0);
        \draw[ultra thick,Green] (B) -- node {\arb}  (C);
        \draw[thick,double] (A) --  (B);
        \fill[black] (B) circle (2pt);
        \fill[black] (A) circle (2pt);
        \fill[black,Green] (C) circle (2.3pt);
    \end{tikzpicture}} \},  &&
\mathbb{L}_{
 \raisebox{-0.1cm}{\begin{tikzpicture}[scale=0.8]
        \coordinate (A) at (0,0);
        \coordinate (B) at (1/2,0);
        \coordinate (C) at (1,0);
        \coordinate (D) at (3/2,0);
        \coordinate (E) at (2,0);
        \draw[thick] (B) -- node {\ar} (C);
        \draw[thick] (A) --  node {\al} (B);
        \fill[black] (B) circle (2pt);
        \fill[black] (A) circle (2pt);
        \fill[black] (C) circle (2pt);
    \end{tikzpicture}}} = \{ \raisebox{-0.1 cm}{\begin{tikzpicture}[scale=1]
        \coordinate (A) at (0,0);
        \coordinate (B) at (1/2,0);
        \coordinate (C) at (1,0);
        \coordinate (D) at (3/2,0);
        \coordinate (E) at (2,0);
        \draw[thick] (B) -- node {\ar} (C);
        \draw[ultra thick,NavyBlue] (A) -- node {\alb} (B);
        \fill[black] (B) circle (2pt);
        \fill[black,NavyBlue] (A) circle (2.3pt);
        \fill[black] (C) circle (2pt);
    \end{tikzpicture}}
    ,
    \raisebox{-0.075 cm}{\begin{tikzpicture}[scale=1]
        \coordinate (A) at (0,0);
        \coordinate (B) at (1/2,0);
        \coordinate (C) at (1,0);
        \coordinate (D) at (3/2,0);
        \coordinate (E) at (2,0);
        \draw[thick] (B) -- node {\ar}  (C);
        \draw[thick] (A) --  node {\al} (B);
        \fill[black,Red] (B) circle (2.3pt);
        \fill[black] (A) circle (2pt);
        \fill[black] (C) circle (2pt);
    \end{tikzpicture}}
    ,
    \raisebox{-0.1 cm}
    {\begin{tikzpicture}[scale=1]
        \coordinate (A) at (0,0);
        \coordinate (B) at (1/2,0);
        \coordinate (C) at (1,0);
        \coordinate (D) at (3/2,0);
        \coordinate (E) at (2,0);
        \draw[ultra thick,Green] (B) -- node {\arb}  (C);
        \draw[thick] (A) -- node {\al}  (B);
        \fill[black] (B) circle (2pt);
        \fill[black] (A) circle (2pt);
        \fill[black,Green] (C) circle (2.3pt);
    \end{tikzpicture}} \},
\end{align}
For each letter $\mathfrak{l} \in \mathbb{L}_{\g}$ we denote the associated vertex set by $V_\mathfrak{l}$, which is assumed to be in ascending order. 

Additionally, the set of letters $\mathbb{L}_\g$ is ordered such that $\lfrak_i < \lfrak_j$ whenever $\overline{V}_{\lfrak_i} < \overline{V}_{\lfrak_j}$ where $\overline{V}_\mathfrak{l} := \min V_\mathfrak{l}$. 
This ordering is introduced here for future convenience where it is used to keep track of signs generated by the wedge product.

Crucially, each letter defines a function of the graph variables and we will use the terminology letters interchangeably for both the colored subgraph and their associated functions. To read off the letter associated to a colored subgraph we simply: sum over $X_v$ for all colored vertices, subtract $Y_e$ for each colored half edge, and sum over $Y_e$ for each uncolored edge incident to a colored vertex, schematically that is\footnote{Here we emphasize the second sum is over only those uncolored edges incident to a colored vertex.}
\begin{align}
    f_{\mathfrak{l}} (\mbf{X},\mbf{Y})
    = \sum_{v\in V_{\mathfrak{l}}} X_v 
    + \sum_{e:\text{uncolored}} Y_e
    - \sum_{e:\text{colored}} Y_e
    \,.
\end{align}
As an example of the above formula, the colored subgraphs appearing in \eqref{eq:letterExample} correspond to the following functions of graph variables
\begin{align}\label{eq:lettersExample}
    \mathbb{L}_{
 \raisebox{-0.1cm}{\begin{tikzpicture}[scale=0.8]
        \coordinate (A) at (0,0);
        \coordinate (B) at (1/2,0);
        \coordinate (C) at (1,0);
        \coordinate (D) at (3/2,0);
        \coordinate (E) at (2,0);
        \draw[thick] (B) -- node {\ar} (C);
        \draw[thick,double] (A) -- (B);
        \fill[black] (B) circle (2pt);
        \fill[black] (A) circle (2pt);
        \fill[black] (C) circle (2pt);
    \end{tikzpicture}}} &= \{ \textcolor{Magenta}{X_1+X_2+Y_{23}},\textcolor{Green}{X_3-Y_{23}} \}, \notag \\
\mathbb{L}_{
 \raisebox{-0.1cm}{\begin{tikzpicture}[scale=0.8]
        \coordinate (A) at (0,0);
        \coordinate (B) at (1/2,0);
        \coordinate (C) at (1,0);
        \coordinate (D) at (3/2,0);
        \coordinate (E) at (2,0);
        \draw[thick] (B) -- node {\ar} (C);
        \draw[thick] (A) --  node {\al} (B);
        \fill[black] (B) circle (2pt);
        \fill[black] (A) circle (2pt);
        \fill[black] (C) circle (2pt);
    \end{tikzpicture}}} &= \{ \textcolor{NavyBlue}{X_1-Y_{12}}, \textcolor{Red}{X_2+Y_{12}+Y_{23}},\textcolor{Green}{X_3-Y_{23}} \}.
\end{align}
These letters serve two important purposes: they appear as arguments of the kinematic $\dlog$'s in the differential equations, and, their shifted versions control the geometry of the physical cuts, as detailed in section \ref{sec:cutGeom}.

\paragraph{Mergers.} 

We say a pair of acyclic minors $(\g,\g')$ are related by a {\it merger} if $\g'$ can be obtained from $\g$ by replacing all oriented edges between two letters of $\g$ with solid edges, this follows the terminology introduced in \cite{Arkani-Hamed:2023bsv,Baumann:2025qjx}. Moreover, for any pair $(\g,\g^\prime)$ related by a merger, their set of letters can be organized as:
\begin{align}
    \mathbb{L}_{\g'}    
    &=  
        \mathbb{L}_{\g'}^\downarrow \cup 
        \{ \underline{\lfrak} \} \cup
        \mathbb{L}_{\g'}^\uparrow,
    &&
    \mathbb{L}_{\g'}^\downarrow = \{ \lfrak' \in \mathbb{L}_{\g'} : \overline{V}_{\lfrak'} < \overline{V}_{\underline{\lfrak}} \}
    \,, \notag 
    \\
    \mathbb{L}_{\g}
    &=  
        \mathbb{L}_{\g'}^\downarrow \cup 
        \{ \underline{\lfrak}_\downarrow,
        \underline{\lfrak}_\uparrow \} \cup 
        \mathbb{L}_{\g'}^\uparrow 
    \,,
    &&
    \mathbb{L}_{\g'}^\uparrow 
    = \{ \lfrak' \in \mathbb{L}_{\g'} 
    : \overline{V}_{\lfrak'} > \overline{V}_{\underline{\lfrak}} \}
    \,.
\end{align}
Whenever this is the case, it follows that for each $\tau' \in \mathcal{C}_\g$, there exists a unique $\tau \in \mathcal{C}_{\g}$ such that $\tau' \subset \tau$. 

In deriving the off-diagonal components of the differential equation (section \eqref{sec:DEQs} and appendix \ref{app:AmatOffDiag}), it will be useful to define the set of such pairs of cut tubings by
\begin{align}\label{eq:Cgg}
\mathcal{C}_{\g\g'} = \{(\tau,\tau') \in \mathcal{C}_\g \times \mathcal{C}_{\g'} : \tau' \subset \tau \}.
\end{align}
For example, the acyclic minors appearing in \eqref{eq:lettersExample} are related by a merger, and we have 
\begin{align} 
&\mathcal{C}_{
    \raisebox{-0.09cm}{\begin{tikzpicture}[scale=0.8]
        \coordinate (A) at (0,0);
        \coordinate (B) at (1/2,0);
        \coordinate (C) at (1,0);
        \coordinate (D) at (3/2,0);
        \coordinate (E) at (2,0);
        \draw[thick] (B) --  node {\ar} (C);
        \draw[thick] (A) -- node {\al} (B);
        \fill[black] (A) circle (2pt);
        \fill[black] (B) circle (2pt);
        \fill[black] (C) circle (2pt);
    \end{tikzpicture}}
    ,
    \raisebox{-0.09cm}{\begin{tikzpicture}[scale=0.8]
        \coordinate (A) at (0,0);
        \coordinate (B) at (1/2,0);
        \coordinate (C) at (1,0);
        \coordinate (D) at (3/2,0);
        \coordinate (E) at (2,0);
        \draw[thick] (B) --  node {\ar} (C);
        \draw[thick,double] (A) -- (B);
        \fill[black] (A) circle (2pt);
        \fill[black] (B) circle (2pt);
        \fill[black] (C) circle (2pt);
    \end{tikzpicture}}
    } =  \{( 
    \raisebox{-0.25cm}{\begin{tikzpicture}[scale=0.8]
        \coordinate (A) at (0,0);
        \coordinate (B) at (1/2,0);
        \coordinate (C) at (1,0);
        \coordinate (D) at (3/2,0);
        \coordinate (E) at (2,0);
        \draw[thick] (B) -- (C);
        \draw[thick] (A) --  (B);
        \draw[thick, black] (0.5,0) ellipse (0.9cm and 0.35cm);
        \draw[thick, black] (0.25,0) ellipse (0.5cm and 0.25cm);
        \draw[black,thick] (B) circle (4pt);
        \fill[black] (B) circle (2pt);
        \fill[black] (A) circle (2pt);
        \fill[black] (C) circle (2pt);
    \end{tikzpicture}},\raisebox{-0.25cm}{\begin{tikzpicture}[scale=0.8]
        \coordinate (A) at (0,0);
        \coordinate (B) at (1/2,0);
        \coordinate (C) at (1,0);
        \coordinate (D) at (3/2,0);
        \coordinate (E) at (2,0);
        \draw[thick] (B) -- (C);
        \draw[thick] (A) --  (B);
        \draw[thick, black] (0.5,0) ellipse (0.9cm and 0.35cm);
        \draw[thick, black] (0.25,0) ellipse (0.5cm and 0.25cm);
        \fill[black] (B) circle (2pt);
        \fill[black] (A) circle (2pt);
        \fill[black] (C) circle (2pt);
    \end{tikzpicture}}  
    )\}.
\end{align}
In this case, the letters are organized as: $\mathbb{L}^{\downarrow}_{\g}=\emptyset$, $\underline{\mathfrak{l}} = \textcolor{Magenta}{X_1+X_2+Y_{23}}$, $\underline{\mathfrak{l}}_\downarrow = \textcolor{NavyBlue}{X_1-Y_{12}}$, $\underline{\mathfrak{l}}_\uparrow = \textcolor{Red}{X_2+Y_{12}+Y_{23}}$ and $\mathbb{L}^{\uparrow}_{\g}=\{ \textcolor{Green}{X_3-Y_{23}} \}$.

\subsection{Cut basis}
\label{sec:cutbasis}
In this section we explain how to associate an FRW-form to each acyclic minor, then in section \ref{sec:cutGeom}, we explain the positive geometric origin of this basis. 

Given an acyclic minor $\g$, we define the form
\begin{tcolorbox}
\vspace{-1em}
\begin{align} \label{eq:phiComb}
    \phi_{\mathfrak{g}} &:= \sum_{\tau \in \C_{\g}}\dlog_\tau, 
    &\text{ where }& &
    \dlog_{\tau} &:= \sum_{r \in \mathcal{R}_{\g}} \bigwedge_{v\in V_G} \dlog\begin{cases} B_{\langle v \rangle_\tau} \text{ if } v \in r,\\  x_v \text{ otherwise,}\end{cases}
\end{align}
\end{tcolorbox}
\noindent
where the sum appearing on the right is over the cartesian product of the vertex sets of letters $$\mathcal{R}_\g = V_{\lfrak_1} \times \ldots \times V_{\lfrak_{|\mathbb{L}_\g|}}.$$
In addition, given a vertex $v \in V_G$ and tube $\tau \in \mathcal{C}_\g$, we define $\langle v \rangle_\tau$ to be the smallest tube $t \in \tau$ such that $v \in V_t$. We refer to the set of forms in \eqref{eq:phiComb} as the cut basis, which is closely related to the cut basis introduced in \cite{De:2024zic}, differing only by overall signs.  
Continuing our running examples for the three-site chain we find the following elements of the cut basis
\begin{align}
\phi_{\raisebox{-0.08cm}{\begin{tikzpicture}[scale=0.8]
        \coordinate (A) at (0,0);
        \coordinate (B) at (1/2,0);
        \coordinate (C) at (1,0);
        \coordinate (D) at (3/2,0);
        \coordinate (E) at (2,0);
        \draw[thick] (B) --  node {\ar} (C);
        \draw[thick,double] (A) --  (B);
        \fill[black] (A) circle (2pt);
        \fill[black] (B) circle (2pt);
        \fill[black] (C) circle (2pt);
    \end{tikzpicture}}} = \dlog_{\raisebox{0cm}{\begin{tikzpicture}[scale=0.8]
        \coordinate (A) at (0,0);
        \coordinate (B) at (1/2,0);
        \coordinate (C) at (1,0);
        \coordinate (D) at (3/2,0);
        \draw[thick] (A) -- (B) --(C);
        \draw[thick, black] (0.27,0) ellipse (0.44cm and 0.25cm);
        \draw[thick, black] (0.5,0) ellipse (0.78cm and 0.34cm);
        \fill[black] (B) circle (2pt);
        \fill[black] (A) circle (2pt);
        \fill[black] (C) circle (2pt);
    \end{tikzpicture}}}
    &= \dlog \ x_1 \wedge  \dlog  \raisebox{-0.1cm}{\begin{tikzpicture}[scale=0.7]
        \coordinate (A) at (0,0);
        \coordinate (B) at (1/2,0);
        \coordinate (C) at (1,0);
        \draw[thick] (A) -- (B) --(C);
        \draw[thick, black] (0.25,0) ellipse (0.44cm and 0.22cm);
        \fill[black] (B) circle (2pt);
        \fill[black] (A) circle (2pt);
        \fill[black] (C) circle (2pt);
    \end{tikzpicture}}  \wedge \dlog \  \raisebox{-0.1cm}{\begin{tikzpicture}[scale=0.7]
        \coordinate (A) at (0,0);
        \coordinate (B) at (1/2,0);
        \coordinate (C) at (1,0);
        \coordinate (D) at (3/2,0);
        \draw[thick] (A) -- (B) --(C);
        \draw[thick, black] (0.5,0) ellipse (0.75cm and 0.22cm);
        \fill[black] (B) circle (2pt);
        \fill[black] (A) circle (2pt);
        \fill[black] (C) circle (2pt);
    \end{tikzpicture}} \notag \\
    &+\dlog  \raisebox{-0.1cm}{\begin{tikzpicture}[scale=0.7]
        \coordinate (A) at (0,0);
        \coordinate (B) at (1/2,0);
        \coordinate (C) at (1,0);
        \draw[thick] (A) -- (B) --(C);
        \draw[thick, black] (0.25,0) ellipse (0.44cm and 0.22cm);
        \fill[black] (B) circle (2pt);
        \fill[black] (A) circle (2pt);
        \fill[black] (C) circle (2pt);
    \end{tikzpicture}}  \wedge \dlog \ x_2 \wedge \dlog \  \raisebox{-0.1cm}{\begin{tikzpicture}[scale=0.7]
        \coordinate (A) at (0,0);
        \coordinate (B) at (1/2,0);
        \coordinate (C) at (1,0);
        \coordinate (D) at (3/2,0);
        \draw[thick] (A) -- (B) --(C);
        \draw[thick, black] (0.5,0) ellipse (0.75cm and 0.22cm);
        \fill[black] (B) circle (2pt);
        \fill[black] (A) circle (2pt);
        \fill[black] (C) circle (2pt);
    \end{tikzpicture}}.
\end{align}
together with
\begin{align}
\phi_{\begin{tikzpicture}[scale=0.8]
        \coordinate (A) at (0,0);
        \coordinate (B) at (1/2,0);
        \coordinate (C) at (1,0);
        \draw[thick] (B) --  node {\ar} (C);
        \draw[thick] (A) --  node {\al} (B);
        \fill[black] (A) circle (2pt);
        \fill[black] (B) circle (2pt);
        \fill[black] (C) circle (2pt);
    \end{tikzpicture}}  =  \dlog_{\raisebox{0cm}{\begin{tikzpicture}[scale=0.8]
        \coordinate (A) at (0,0);
        \coordinate (B) at (1/2,0);
        \coordinate (C) at (1,0);
        \coordinate (D) at (3/2,0);
        \draw[thick] (A) -- (B) --(C);
        \draw[thick, black] (0.27,0) ellipse (0.44cm and 0.25cm);
        \draw[thick, black] (0.5,0) ellipse (0.78cm and 0.34cm);
        \draw[black,thick] (B) circle (4pt);
        \fill[black] (B) circle (2pt);
        \fill[black] (A) circle (2pt);
        \fill[black] (C) circle (2pt);
    \end{tikzpicture}}}+\dlog_{\raisebox{0cm}{\begin{tikzpicture}[scale=0.8]
        \coordinate (A) at (0,0);
        \coordinate (B) at (1/2,0);
        \coordinate (C) at (1,0);
        \coordinate (D) at (3/2,0);
        \draw[thick] (A) -- (B) --(C);
        \draw[thick, black] (0.73,0) ellipse (0.44cm and 0.25cm);
        \draw[thick, black] (0.5,0) ellipse (0.78cm and 0.34cm);
        \draw[black,thick] (B) circle (4pt);
        \fill[black] (B) circle (2pt);
        \fill[black] (A) circle (2pt);
        \fill[black] (C) circle (2pt);
    \end{tikzpicture}}} &= \dlog  \raisebox{-0.1cm}{\begin{tikzpicture}[scale=0.7]
        \coordinate (A) at (0,0);
        \coordinate (B) at (1/2,0);
        \coordinate (C) at (1,0);
        \draw[thick] (A) -- (B) --(C);
        \draw[thick, black] (0.25,0) ellipse (0.44cm and 0.22cm);
        \fill[black] (B) circle (2pt);
        \fill[black] (A) circle (2pt);
        \fill[black] (C) circle (2pt);
    \end{tikzpicture}}  \wedge \dlog \  \raisebox{-0.02cm}{\begin{tikzpicture}[scale=0.7]
        \coordinate (A) at (0,0);
        \coordinate (B) at (1/2,0);
        \coordinate (C) at (1,0);
        \coordinate (D) at (3/2,0);
        \draw[thick] (A) -- (B) --(C);
        \draw[black,thick] (B) circle (4pt);
        \fill[black] (B) circle (2pt);
        \fill[black] (A) circle (2pt);
        \fill[black] (C) circle (2pt);
    \end{tikzpicture}} \wedge \dlog \  \raisebox{-0.1cm}{\begin{tikzpicture}[scale=0.7]
        \coordinate (A) at (0,0);
        \coordinate (B) at (1/2,0);
        \coordinate (C) at (1,0);
        \coordinate (D) at (3/2,0);
        \draw[thick] (A) -- (B) --(C);
        \draw[thick, black] (0.5,0) ellipse (0.75cm and 0.22cm);
        \fill[black] (B) circle (2pt);
        \fill[black] (A) circle (2pt);
        \fill[black] (C) circle (2pt);
    \end{tikzpicture}} \notag \\
    &+\dlog  \ \raisebox{-0.1cm}{\begin{tikzpicture}[scale=0.7]
        \coordinate (A) at (0,0);
        \coordinate (B) at (1/2,0);
        \coordinate (C) at (1,0);
        \coordinate (D) at (3/2,0);
        \draw[thick] (A) -- (B) --(C);
        \draw[thick, black] (0.5,0) ellipse (0.75cm and 0.22cm);
        \fill[black] (B) circle (2pt);
        \fill[black] (A) circle (2pt);
        \fill[black] (C) circle (2pt);
    \end{tikzpicture}} \wedge \dlog \ \raisebox{-0.02cm}{\begin{tikzpicture}[scale=0.7]
        \coordinate (A) at (0,0);
        \coordinate (B) at (1/2,0);
        \coordinate (C) at (1,0);
        \coordinate (D) at (3/2,0);
        \draw[thick] (A) -- (B) --(C);
        \draw[black,thick] (B) circle (4pt);
        \fill[black] (B) circle (2pt);
        \fill[black] (A) circle (2pt);
        \fill[black] (C) circle (2pt);
    \end{tikzpicture}} \wedge \dlog \ \raisebox{-0.1cm}{\begin{tikzpicture}[scale=0.7]
        \coordinate (A) at (0,0);
        \coordinate (B) at (1/2,0);
        \coordinate (C) at (1,0);
        \draw[thick] (A) -- (B) --(C);
        \draw[thick, black] (0.75,0) ellipse (0.44cm and 0.22cm);
        \fill[black] (B) circle (2pt);
        \fill[black] (A) circle (2pt);
        \fill[black] (C) circle (2pt);
    \end{tikzpicture}}.
\end{align}

Following the results of \cite{Glew:2025ugf,Glew:2025arc} it is clear that the physical FRW-form has a simple expansion in terms of the cut basis
\begin{tcolorbox}
\vspace{-1em}
\begin{align} \label{eq:psiPhysCutBasis}
    \Psi_G = \sum_{E \subset E_G} (-1)^{|E|} \sum_{\mathfrak{g} \in \mathcal{A}^{\emptyset}_E(G)} \ \phi_{\mathfrak{g}}
    ,
\end{align}
\end{tcolorbox}
\noindent
As a simple example, the wavefunction for the path graph on three vertices is given by  
\begin{align}\begin{aligned}\label{eq:WavefunctCutBasis}
    \Psi_{\includegraphics[]{figures/amp_t1}} =&
    \left( 
        \phi_{\includegraphics[]{figures/amp_t2}}
        +\phi_{\includegraphics[]{figures/amp_t3}}
        +\phi_{\includegraphics[]{figures/amp_t4}} 
        +\phi_{\includegraphics[]{figures/amp_t5}}
    \right)
    \\
    -&\left(
        \phi_{\includegraphics[]{figures/amp_t8}}
        +\phi_{\includegraphics[]{figures/amp_t9}}
    \right)
    \\
    -&\left(
        \phi_{\includegraphics[]{figures/amp_t6}} 
        +\phi_{\includegraphics[]{figures/amp_t7}} 
    \right)
    \\
    +& \left( \phi_{\includegraphics[]{figures/amp_t10}} \right)
    \,.
\end{aligned}\end{align}
An alternative set of forms $\{ \tilde{\phi}_G \}$, sharing similar properties with the cut basis, was recently proposed in \cite{Capuano:2025ehm}. However, the degeneracy of the hyperplane arrangement was not taken into account, so that in general the cardinality $|\{\tilde{\phi}_G \}|$ is greater than that of the cut basis.

\paragraph{Organizing the wedge product.}

When deriving the differential equations it will prove useful to rewrite \eqref{eq:phiComb} using the notation of \cite{De:2024zic}. 
To do this, note that $\phi_{\g}$ can be re-written as
\begin{align} \label{eq:phiCutAlt}
    \phi_\g
    &:= \sgn_\g
    \left(
        \sum_{\tau \in \mathcal{C}_\g}
        \bigwedge_{\lfrak\in\mathbb{L}_\g}
        B_{\la\overline{V}_\lfrak\ra_\tau}
    \right) 
    \wedge
    \overset{\tilde\Omega_\g}{\overbrace{
        \left(
            \bigwedge_{\lfrak\in \mathbb{L}_\g}
            \tilde\Omega_{\lfrak}
        \right)
    }}
    \,,
    \qquad
    \tilde{\Omega}_{\mathfrak{l}}
    = \bigwedge_{v \in V_\mathfrak{l}\setminus \overline{V}_\mathfrak{l}} 
    \dlog \frac{x_v}{x_{ \overline{V}_\mathfrak{l}}}
    \,,
\end{align}
where $
    \sgn_\g := \mathrm{sig}(
        \overline{V}_{\lfrak_1},
        \dots, 
        \overline{V}_{\lfrak_{|\mathbb{L}_\g|}},
        V_{\lfrak_1}\setminus\overline{V}_{\lfrak_1},
        \dots,
        V_{\lfrak_{|\mathbb{L}_\g|}} \setminus \overline{V}_{\lfrak_{|\mathbb{L}_\g|}}
    )
$ 
and the two wedge products are ordered such that $\min V_{\lfrak_i} < \min V_{\lfrak_{i+1}}$ for all $\lfrak_i \in \mathbb{L}_\g$. 
Note, when $|V_\mathfrak{l}|=1$ we have 
$\tilde{\Omega}_\lfrak = 1$.
Moreover, as discussed in the next section, $\tilde{\Omega}_\g$ and $\tilde{\Omega}_\lfrak$ are closely related the canonical forms of positive geometries. 

Already in \eqref{eq:phiCutAlt}, the cut structure is 
made manifest by the two terms in brackets; the first has the potential to be cut while the second becomes the canonical form on the maximal cut. 
However, to isolate the physical cuts, we had to choose a global ordering for the $B_\bullet$. 
Implementing this ordering on the $B_\bullet$ introduces an additional compensating sign
\begin{tcolorbox}
    \vspace{-1em}
    \begin{align} \label{eq:phiCut}
        \phi_{\mathfrak{g}} 
        &:= \text{sgn}_\g  \dlog\mathcal{C_\g} 
            \wedge \tilde\Omega_\g \,,  
        & 
        \dlog\mathcal{C}_\g 
        &:= \sum_{\tau \in \mathcal{C}_\g} \text{sgn}_\tau  
        \bigwedge_{\lfrak\in\mathbb{L}_\g} 
        \dlog B_{\sigma(\la\overline{V}_{\lfrak}\ra_\tau)},
    \end{align}
\end{tcolorbox}
\noindent
where $\sigma$ is the permutation that implements the global ordering and 
$
    \sgn_\tau := \mathrm{sig}(
        \la \overline{V}_{\lfrak_1}\ra_\tau, 
        \dots,  \allowbreak
        \la \overline{V}_{\lfrak_{|\mathbb{L}_\g|}} \ra_\tau
    ) 
    /
    \mathrm{sig}(
        \sigma(\la \overline{V}_{\lfrak_1}\ra_\tau), 
        \dots, 
        \sigma(\la \overline{V}_{\lfrak_{|\mathbb{L}_\g|}} \ra_\tau)
    )
$.
Note that $\sgn_\g$ and $\sgn_\tau$ are there to undo signs generated by reordering the wedge product in \eqref{eq:phiComb} and will drop out of the final expression for the differential equations.

\subsection{Positive geometry of \emph{physical} cuts}
\label{sec:cutGeom}

We now seek to motivate the definition of the cut basis in \eqref{eq:phiComb} by examining the positive geometry associated to physical cuts. Since each acyclic minor corresponds uniquely to a physical cut, it is convenient to introduce the notation $\mathcal{B}_\g = \mathcal{B}_\tau$ for $\tau\in \mathcal{C_\g}$, where, by construction, we have $\mathcal{B}_\tau=\mathcal{B_{\tau'}}$ for all $\tau,\tau'\in\mathcal{C}_\g$. 
Each cut space, 
$$
    M_{\C_\g} = (\mathbb{CP}^{|V_G|} \setminus \T) \cap \B_{\g}, 
$$
is the complement of a hyperplane arrangement with a {\it single bounded chamber}  carved out by the restriction of the twisted divisors to $\B_{\g}$. This bounded chamber $\Gamma_\g$ is an example of a positive geometry which turns out to have a product structure given by $\Gamma_{\g} = \bigtimes_{\mathfrak{l} \in \mathbb{L}_\g} \Gamma_\mathfrak{l}$ where each component is a simplex: 
\begin{align}
    \label{eq:GammaFactor}
    \Gamma_\lfrak &:= \{ (x_{v_1}, \cdots, x_{v_{|V_\lfrak|}}) \in \mathbb{R}^{|V_\lfrak|} : x_v \leq 0 \; \text{for all } v \in V_\lfrak \text{ and } f_\lfrak({\bf x} + {\bf X},{\bf Y})=0\}
    \,.
\end{align}
Moreover, the maximal cut of $\phi_\g$ is exactly the canonical form $\Omega_\g$ of the bounded chamber $\Gamma_\g$, that is 
\begin{align}
    \res_{\C_\g} \phi_\g
    = \Omega_\g
    \,,
\end{align}
where 
\begin{align} \label{eq:res_Cg}
     \res_{\C_\g} &:= \sum_{\tau \in \mathcal{C}_\g} 
     \text{sgn}_{\tau}\,
     \res_{\tau}
    \,,
    &
    \res_\tau &:= \res_{\B_{t_{|\tau|}}} \circ \cdots \circ \res_{\B_{t_1}}
    \,,
\end{align}
is the residue operator for both non-degenerate and degenerate cuts.

To see why each cut has a geometry described by \eqref{eq:GammaFactor},  consider the following acyclic minor of the double box graph together with its cut conditions 
\begin{center}
    \begin{minipage}{.3\textwidth}
    $\g=\includegraphics[align=c,scale=1.2]{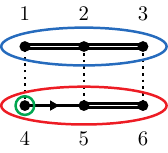}$
    \end{minipage}
    \begin{minipage}{.69\textwidth}
        \be 
            -(x_1+x_2+x_3)=& \textcolor{NavyBlue}{X_1+X_2+X_3+Y_{14}+Y_{25}+Y_{36}}
            \,,
            \\
            -(x_4+x_5+x_6)=& \textcolor{Red}{X_4 + X_5 + X_6 + Y_{14}+ Y_{25}+Y_{36}}
            \,,
            \\
            -x_4=& \textcolor{Green}{X_4 + Y_{14}+Y_{45}}
            \,.
            \label{eq:double_examp1}
        \ee
    \end{minipage}
\end{center}
The bounded chamber $\Gamma_\g$, is 
\begin{align}
    \Gamma_\g &= \{
        \mbf{x} \in \mathbb{R}^6
        : x_i \leq 0 
        \text{ and } 
        \eqref{eq:double_examp1}
        \text{ is satisfied}
    \}
    \,.
\end{align}
To see the Minkowski decomposition of this region, combine the \textcolor{Green}{green} cut condition with the \textcolor{Red}{red} cut condition. 
This yields a new but equivalent linear system in terms of letters
\be 
    f_{
        \adjustbox{align=c}{ 
            \includegraphics[align=c,scale=.45]{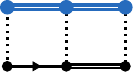}
        }
    }(\mbf{x}+\mbf{X},\mbf{Y})&= \textcolor{NavyBlue}{x_1+x_2+x_3+X_1+X_2+X_3+Y_{14}+Y_{25}+Y_{36}}
     =0
    \,,
    \\
    f_{
        \adjustbox{align=c}{ 
            \includegraphics[align=c,scale=.45]{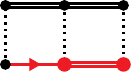}
        }
    }(\mbf{x}+\mbf{X},\mbf{Y})&= \textcolor{Red}{x_5+x_6+X_5 +X_6 + Y_{25}+Y_{36}- Y_{45}}
    =0 
    \,,
    \\
    f_{
        \adjustbox{align=c}{ 
            \includegraphics[align=c,scale=.45]{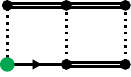}
        }
    }(\mbf{x}+\mbf{X},\mbf{Y})&= \textcolor{Green}{x_4+X_4 + Y_{14}+Y_{45}}
    =0
    \,.
\label{eq:double_examp2}
\ee
Each equation above places a constraint on a subset of the integration variables, which together partition $(x_1, \dots, x_6)$. 
Therefore, we can interpret each line of equation \eqref{eq:double_examp2} as a 2-, 1- and 0-simplex, respectively
\begin{align}
    \Gamma_\g  &= \Gamma_{
        \adjustbox{align=c}{ 
            \includegraphics[align=c,scale=.45]{figures/double_box_exam_alt.pdf}
        }
    } \times \Gamma_{
        \adjustbox{align=c}{ 
            \includegraphics[align=c,scale=.45]{figures/double_box_exam_alt_1.pdf}
        }
    } \times \Gamma_{
        \adjustbox{align=c}{ 
            \includegraphics[align=c,scale=.45]{figures/double_box_exam_alt_copy_2.pdf}
        }
    }
    \,.
\end{align}
Moreover, since the left-hand-sides of \eqref{eq:double_examp2} partition the integration variables, each simplex is orthogonal to the others. 
Thus, the Minkowski sum of these simplices is equivalent to the Cartesian product. 
In the context of the above example, the Cartesian product makes a triangular prism:
\begin{align} \label{eq:toblerone}
    \includegraphics[align=c, scale=.6]{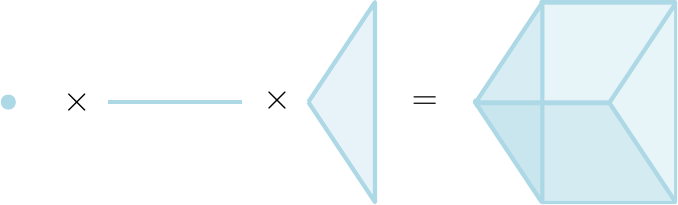} \ .
\end{align}

\subsection{Relation to the time integral basis}
\label{sec:reltaionToTimeInt}
It should be noted that the basis of forms presented here is closely related to the {\it time integral basis} recently proposed in \cite{Baumann:2025qjx}. As an example, for the path graph on two vertices, the cut basis is given by  
\begin{align}
\phi_{\includegraphics[]{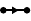}} & = \int \d x_1 \wedge \d x_2 \frac{ x_1^{\alpha_1}x_2^{\alpha_2}}{(X_1+x_1+Y_{12})(X_1+X_2 +x_1+x_2)}, \notag \\
\phi_{\includegraphics[]{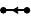}} & = \int \d x_1 \wedge \d x_2  \frac{x_1^{\alpha_1}x_2^{\alpha_2}}{(X_1+X_2+x_1+x_2)(X_2+x_2+Y_{12})}, \notag \\ 
\phi_{\includegraphics[]{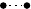}} & = \int \d x_1 \wedge \d x_2 \frac{x_1^{\alpha_1}x_2^{\alpha_2}}{(X_1+x_1+Y_{12})(X_2+x_2+Y_{12})}, \notag \\
\phi_{\includegraphics[]{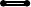}} & = \int \d x_1 \wedge \d x_2 \frac{x_1+x_2}{x_1 x_2} \frac{x_1^{\alpha_1}x_2^{\alpha_2}}{(X_1+X_2+x_1+x_2)}.
\end{align}
The time integral basis on the other hand includes the first three forms above together with
\begin{align}
\phi'_{\includegraphics[]{figures/amp_2_t1}} & = \int \d x_1 \wedge \d x_2 \frac{x_1^{\alpha_1}x_2^{\alpha_2}}{(X_1+X_2+x_1+x_2)^2}.
\end{align}
However, it is easy to show that these two forms differ by a total derivative given explicitly by
\begin{align}
\frac{\alpha_1 +\alpha_2}{\alpha_1 \alpha_2}\phi'_{\includegraphics[]{figures/amp_2_t1}}-\phi_{\includegraphics[]{figures/amp_2_t1}}  = 
 \nabla\left(  \frac{ \frac{1}{\alpha_2} \d x_1-\frac{1}{\alpha_1} \d x_2}{x_1+x_2+X_1+X_2}
\right).
\end{align}
As demonstrated by the above example, when the acyclic minor contains no solid edges, the cut basis and time integral basis forms coincide. 
More generally, by comparing the structure of our differential equations (c.f., \eqref{eq:AmatDiag} and  \eqref{eq:AmatOffDiag}, or \eqref{eq:kin_flow}) to those of \cite{Baumann:2025qjx}, suggests that the cut basis and time integral basis are equivalent up to total derivatives and a rescaling.

\section{Zonotopes and the flow of cuts}
\label{sec:zono}
\begin{figure}
    \centering
    \includegraphics[width=.55\textwidth,align=c]{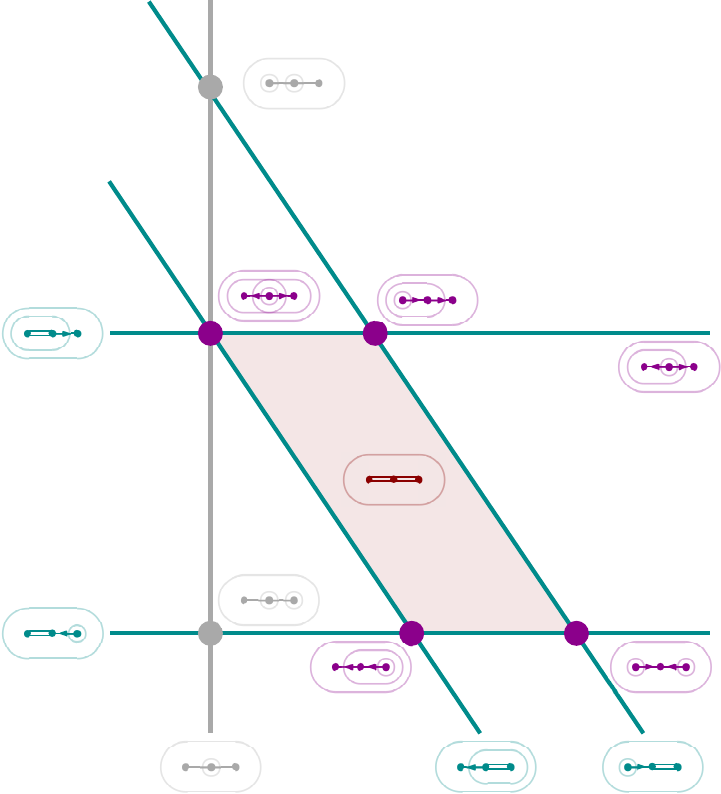}
    \qquad
    \includegraphics[width=.3\textwidth,align=c]{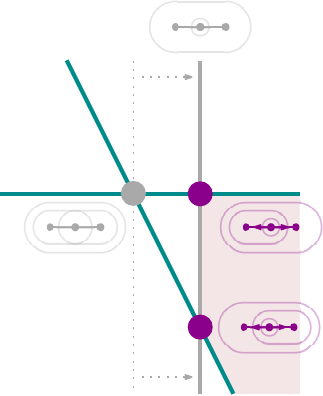}
    \caption{%
    The arrangement of untwisted divisors on the total energy plane of the 3-site chain graph     
    (not the cut space $M_{
            \includegraphics[scale=.6,align=c]{figures/delta_cc.pdf}
    }$). 
    The zonotope is the {\color{BrickRed}red} shaded region, the {\color{darkCyan}cyan} lines are the intersection of the total energy plane and the remaining untwisted planes.
    The dots in are the intersection of the total energy plane with at least two other untwisted planes. 
    The {\color{darkMagenta} magenta} dots are vertices of the zonotope where the physical form has non-trivial triple residues while at the {\color{gray} gray} dots, the physical form has trivial residue. 
    On the total energy cut, the physical form $\Psi_{\includegraphics[scale=0.6]{figures/pt3.pdf}}$ is proportional to the canonical form of the zonotope. 
    }
    \label{fig:3chainTotEnergyZono}
\end{figure}

It has recently been observed that {\it graphical zonotopes} play an important role in characterising the structure of differential equations for the wavefunction \cite{Baumann:2025qjx}. These zonotopes (defined as Minkowski sums of line segments) naturally arise by considering the representation of the (flat-space) wavefunction in terms of cut tubings. To illustrate this, consider the following subset of terms in the wavefunction for the three-site chain
\begin{align}
\hat{\Omega}_{\includegraphics[scale=0.6]{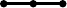}} \supset \textcolor{Green}{\frac{1}{S_{\includegraphics[scale=0.6]{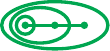}}}}+\textcolor{NavyBlue}{\frac{1}{S_{\includegraphics[scale=0.6]{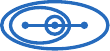}}}}+\textcolor{NavyBlue}{\frac{1}{S_{\includegraphics[scale=0.6]{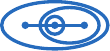}}}}+\textcolor{Red}{\frac{1}{S_{\includegraphics[scale=0.6]{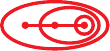}}}}+\textcolor{Orange}{\frac{1}{S_{\includegraphics[scale=0.6]{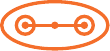}}}}.
\label{eq:SubsetCutTube}
\end{align}
The acyclic minors 
$
    \g \in \{
        \includegraphics[scale=0.8, align=c]{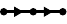},
        \includegraphics[scale=0.8, align=c]{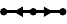},
        \includegraphics[scale=0.8, align=c]{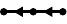},
        \includegraphics[scale=0.8, align=c]{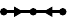}
    \}
$ which label each set of (degenerate) cut tubings appearing in \eqref{eq:SubsetCutTube} have the property $|\mathbb{L}_\g|=|V_G|$. As a result, solving for ${\bf X}$ in the equations $\mathbb{L}_\g=0$ yields a unique point ${\bf X}^*_\g \in \mathbb{R}^{|V_G|}$. In this example, by solving $\mathbb{L}_\g=0$ for each acyclic minor, we produce the four points 
\begin{align}
&-{\bf X}^*_{\includegraphics[scale=0.8]{figures/pt3_X_g.pdf}}=f_{\raisebox{-0.1cm}{\includegraphics[scale=0.8]{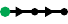}}}({\bf 0},{\bf Y}) \ \hat{e}_1+f_{\raisebox{-0.1cm}{\includegraphics[scale=0.8]{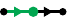}}}({\bf 0},{\bf Y}) \ \hat{e}_2+f_{\raisebox{-0.1cm}{\includegraphics[scale=0.8]{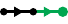}}}({\bf 0},{\bf Y}) \ \hat{e}_3, \notag \\  
&-{\bf X}^*_{\includegraphics[scale=0.8]{figures/pt3_X_b.pdf}}=f_{\raisebox{-0.1cm}{\includegraphics[scale=0.8]{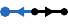}}}({\bf 0},{\bf Y}) \ \hat{e}_1+f_{\raisebox{-0.1cm}{\includegraphics[scale=0.8]{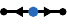}}}({\bf 0},{\bf Y}) \ \hat{e}_2+f_{\raisebox{-0.1cm}{\includegraphics[scale=0.8]{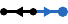}}}({\bf 0},{\bf Y}) \ \hat{e}_3, \notag \\
&-{\bf X}^*_{\includegraphics[scale=0.8]{figures/pt3_X_r.pdf}}=f_{\raisebox{-0.1cm}{\includegraphics[scale=0.8]{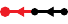}}}({\bf 0},{\bf Y}) \ \hat{e}_1+f_{\raisebox{-0.1cm}{\includegraphics[scale=0.8]{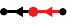}}}({\bf 0},{\bf Y}) \ \hat{e}_2+f_{\raisebox{-0.1cm}{\includegraphics[scale=0.8]{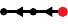}}}({\bf 0},{\bf Y}) \ \hat{e}_3, \notag \\
& -{\bf X}^*_{\includegraphics[scale=0.8]{figures/pt3_X_o.pdf}}=f_{\raisebox{-0.1cm}{\includegraphics[scale=0.8]{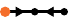}}}({\bf 0},{\bf Y}) \ \hat{e}_1+f_{\raisebox{-0.1cm}{\includegraphics[scale=0.8]{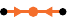}}}({\bf 0},{\bf Y}) \ \hat{e}_2+f_{\raisebox{-0.1cm}{\includegraphics[scale=0.8]{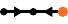}}}({\bf 0},{\bf Y}) \ \hat{e}_3. 
\end{align}
The unit vector $\hat{e}_i$, in the $i$-th vertex direction, appears above multiplied by the unique letter whose vertex set contains $v_i$. By taking the convex hull of these four points we generate a zonotope (parallelogram) which resides on the total energy facet within the cosmological hyperplane arrangement (see figure \ref{fig:3chainTotEnergyZono}). 

\begin{figure}
\center
\label{fig:zonoExample}
\includegraphics[align=c,width=\textwidth]{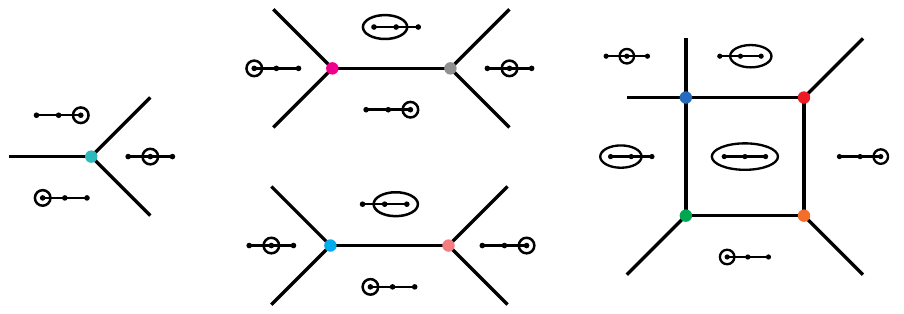}
\caption{The four zonotopes for the three-site chain. 
These are certain projections of the positive region of the hyperplane arrangement.  Each line/point represents the intersection of the hyperplanes labeled by the adjacent tubes. 
The right most zonotope resides on the total energy facet $X_1+X_2+X_3=0$ and is bounded by five planes. Due to degeneracy, the five planes intersect the total energy facet in a square rather than a pentagon. The two middle zonotopes live on the cut surfaces $X_1+X_2+Y_{23}=X_{3}+Y_{23}=0$ and $X_1+Y_{12}=X_2+X_3+Y_{12}=0$ respectively, both producing line segments. The final zonotope is a point at $X_1+Y_{12}=X_2+Y_{12}+Y_{23}=X_3+Y_{23}=0$.} 
\end{figure}

This procedure can be performed for each subset of terms in the partial fraction expression for the wavefunction that contain the same set of maximal (by inclusion) tubes. The four zonotopes generated in this way for the three-site chain are illustrated in figure~\ref{fig:zonoExample}. For example, taking the points generated by either of the subsets 
\begin{align} \begin{aligned} 
    \hat\Omega_{\includegraphics[scale=0.6]{figures/pt3.pdf}} \supset
    &-\left(\textcolor{Magenta}{\frac{1}{S_{\includegraphics[scale=0.6]{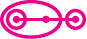}}}
    }+ 
    \textcolor{Gray}{\frac{1}{S_{\includegraphics[scale=0.6]{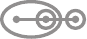}}}}\right)\quad \text{ or } \quad \hat\Omega_{\includegraphics[scale=0.6]{figures/pt3.pdf}} \supset
    &-\left(\textcolor{Cyan}{\frac{1}{S_{\includegraphics[scale=0.6]{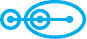}}}}
    + 
    \textcolor{Salmon}{\frac{1}{S_{\includegraphics[scale=0.5]{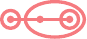}}}}\right)
    \,,
\end{aligned}\end{align}
we obtain two line segments defined as the convex hull of the pairs of points
\begin{align}
&-{\bf X}^*_{\includegraphics[scale=0.8]{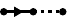}}=f_{\raisebox{-0.1cm}{\includegraphics[scale=0.8]{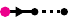}}}({\bf 0},{\bf Y}) \ \hat{e}_1+f_{\raisebox{-0.1cm}{\includegraphics[scale=0.8]{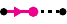}}}({\bf 0},{\bf Y}) \ \hat{e}_2+f_{\raisebox{-0.1cm}{\includegraphics[scale=0.8]{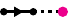}}}({\bf 0},{\bf Y}) \ \hat{e}_3, \notag \\  
&-{\bf X}^*_{\includegraphics[scale=0.8]{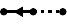}}=f_{\raisebox{-0.1cm}{\includegraphics[scale=0.8]{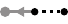}}}({\bf 0},{\bf Y}) \ \hat{e}_1+f_{\raisebox{-0.1cm}{\includegraphics[scale=0.8]{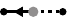}}}({\bf 0},{\bf Y}) \ \hat{e}_2+f_{\raisebox{-0.1cm}{\includegraphics[scale=0.8]{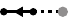}}}({\bf 0},{\bf Y}) \ \hat{e}_3, \notag \\
& \hspace{5cm} \text{ or } \notag \\
&-{\bf X}^*_{\includegraphics[scale=0.8]{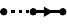}}=f_{\raisebox{-0.1cm}{\includegraphics[scale=0.8]{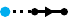}}}({\bf 0},{\bf Y}) \ \hat{e}_1+f_{\raisebox{-0.1cm}{\includegraphics[scale=0.8]{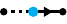}}}({\bf 0},{\bf Y}) \ \hat{e}_2+f_{\raisebox{-0.1cm}{\includegraphics[scale=0.8]{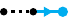}}}({\bf 0},{\bf Y}) \ \hat{e}_3, \notag \\
& -{\bf X}^*_{\includegraphics[scale=0.8]{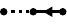}}=f_{\raisebox{-0.1cm}{\includegraphics[scale=0.8]{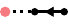}}}({\bf 0},{\bf Y}) \ \hat{e}_1+f_{\raisebox{-0.1cm}{\includegraphics[scale=0.8]{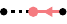}}}({\bf 0},{\bf Y}) \ \hat{e}_2+f_{\raisebox{-0.1cm}{\includegraphics[scale=0.8]{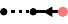}}}({\bf 0},{\bf Y}) \ \hat{e}_3.
\end{align}
Finally, for the remaining term
$
1/S_{\includegraphics[scale=0.5]{figures/pt3_point_1.pdf}}
$, we generate a single point 
\begin{align}
-{\bf X}^*_{\includegraphics[scale=0.8]{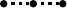}}=f_{\raisebox{-0.1cm}{\includegraphics[scale=0.8]{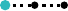}}}({\bf 0},{\bf Y}) \ \hat{e}_1+f_{\raisebox{-0.1cm}{\includegraphics[scale=0.8]{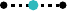}}}({\bf 0},{\bf Y}) \ \hat{e}_2+f_{\raisebox{-0.1cm}{\includegraphics[scale=0.8]{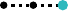}}}({\bf 0},{\bf Y}) \ \hat{e}_3.
\end{align}

The above discussion can be generalized to arbitrary graphs. For each acyclic minor labeling a set of terms appearing in \eqref{eq:psiPhysCutBasis} we define the point 
\begin{align}
{\bf X}^*_{\g} = \sum_{v \in V_G} -f_{\mathfrak{l}_v}({\bf 0}, {\bf Y}) \hat{e}_v,
\end{align}
where $\mathfrak{l}_v \in \mathbb{L}_\g$ is the unique letter containing $v$ in its vertex set. This allows us to define a family of zonotopes for each graph $G$: one for each subset of edges $E \subset E_G$
\begin{align}
\mathcal{Z}_{\g(E)}= \text{conv} \left\{ {\bf X}^*_{\g} :\g \in \mathcal{A}_E^\emptyset(G) \right\},
\label{eq:Zonotope}
\end{align}
where $\g(E)$ is the acyclic minor associated to the bulk of the polytope---the acyclic minor with broken edges $E$ and all remaining edges solid. For instance, the four zonotopes depicted in figure~\ref{fig:zonoExample} for the three-site chain are denoted by
\be
\mathcal{Z}_{\includegraphics[scale=.8,align=c]{figures/delta_cc.pdf}} 
	&= \text{conv} \{ {\bf X}^*_{\includegraphics[scale=0.75]{figures/amp_t2}}
        ,{\bf X}^*_{\includegraphics[scale=0.75]{figures/amp_t3}}
        ,{\bf X}^*_{\includegraphics[scale=0.75]{figures/amp_t4}} 
        ,{\bf X}^*_{\includegraphics[scale=0.75]{figures/amp_t5}}
         \}, 
        && \text{{(square in figure \ref{fig:zonoExample}),}}
        \\
	\mathcal{Z}_{\includegraphics[scale=.8,align=c]{figures/delta_cb.pdf}} 
	&= \text{conv} \{ {\bf X}^*_{\includegraphics[scale=0.75]{figures/amp_t8}}
        ,{\bf X}^*_{\includegraphics[scale=0.75]{figures/amp_t9}}
        \}, 
        && \text{{(line in figure \ref{fig:zonoExample}),}}
        \\
	\mathcal{Z}_{\includegraphics[scale=.8,align=c]{figures/delta_bc.pdf}} 
	&= \text{conv} \{ {\bf X}^*_{\includegraphics[scale=0.75]{figures/amp_t6}}
        ,{\bf X}^*_{\includegraphics[scale=0.75]{figures/amp_t7}}
         \}, 
         && \text{{(line in figure \ref{fig:zonoExample}),}}
         \\
        \mathcal{Z}_{\includegraphics[scale=.8,align=c]{figures/delta_bb.pdf}} &= \text{conv} \{ {\bf X}^*_{\includegraphics[scale=0.75]{figures/amp_t10}} \},
        && \text{{(point in figure \ref{fig:zonoExample})}}. 
\ee
Alternatively, given an edge $e \in E_G$ with end points $v_i$ and $v_j$, we define the {\it point} $\Delta_{ij}^+$ and {\it line segment} $\Delta_{ij}^-$ as 
\begin{align}
\Delta^+_e = -Y_e(\hat{e}_i+\hat{e}_j), \quad \quad \Delta^-_e =\text{conv} \left\{ Y_{e}(\hat{e}_i-\hat{e}_j),Y_{e}(\hat{e}_j- \hat e_i) \right\}.
\end{align}
With this notation the zonotope \eqref{eq:Zonotope} can be written as the following (translated) Minkowski sum of line segments
\begin{align}
\mathcal{Z}_{\g(E)} = \bigoplus_{e \in E_G} \begin{cases}
\Delta^+_e & \text{if } e\in E,\\
\Delta^-_e & \text{otherwise}.
\end{cases}
\end{align}
Remarkably, the set of faces associated to this family of zonotopes are in one-to-one correspondence with the independent cuts of the physical FRW-form and (by construction) with the elements of the cut basis used to formulate our differential equations. This provides a geometric origin for the appearance of zonotopes in the differential equations of \cite{Baumann:2025qjx}. 

\paragraph{Flow of cuts/sequential residues.}

The zonotope structure observed for the flat-space wavefunction $\hat{\Omega}_G$ straightforwardly carries over to the physical FRW-form $\Psi_G := \hat{\Omega}_G(\mbf{x}+\mbf{X},\mbf{Y})\d^{|V_G|}\mbf{x}$ and the arrangement of untwisted hyperplanes $\B$. 
With the coordinates that specify the vertices of the zonotopes now given by 
\begin{align}
{\bf x}^*_{\g} = \sum_{v \in V_G} -f_{\mathfrak{l}_v}({\bf X},{\bf Y}) \hat{e}_v.
\end{align}
Moreover, the residues of the physical FRW-form $\Psi_G$ are the canonical form of the appropriate zonotope (or its boundaries). 
To see this, we recall equation~\eqref{eq:psiPhysCutBasis} here for convenience 
\begin{align}
    \Psi_G = \sum_{E \subset E_G} (-1)^{|E|} \sum_{\mathfrak{g} \in \mathcal{A}^{\emptyset}_E(G)} \ \phi_{\mathfrak{g}}
    .
\end{align}
Each set of acyclic minors $\mathcal{A}_E^\emptyset(G)$ with fixed broken edges and no solid edges is compatible with the cut $\C_{g(E)}$ . Therefore, taking the sequential residue $\res_{\C_{\g(E)}}$ of the physical FRW-form annihilates all terms that do not have the same set of broken edges. 
The resulting sum is the canonical form of the zonotope $\mathcal{Z}_{\g(E)}$:%
\footnote{%
    We note that a similar observation was made by Harry Goodhew during the online Universe+ seminar \cite{U+HarryTalk}.
}
\begin{align} \label{eq:physResZono}
    \res_{\C_{\g(E)}}[\Psi_G]
    = \sum_{
        \g \in \mathcal{A}^\emptyset_E(G)
    } \res_{\C_{\g(E)}}[\phi_\g ]
    = \Omega[\mathcal{Z}_{\g(E)}]
    \,. 
\end{align}
If the hyperplane arrangement of the untwisted diviors $\B$ were generic, this statement would be obvious since each $\g \in \mathcal{A}^\emptyset_E(G)$ corresponds to a vertex of $\mathcal{Z}_{g(E)}$. 
However, since $\B$ is degenerate, we must check that the cut basis correctly assigns $\dlog$-forms to the degenerate vertices of the zonotopes. Choosing the correct prescription for the $\phi_\g$ where $\g$ corresponds to a set of degenerate tubings was already solved in section \ref{sec:cutBasis} and \cite{De:2024zic} with the prescription for $\dlog\C_\g$ in \eqref{eq:phiCut}  (we will revisit this below in the context of three-site chain).

\begin{tcolorbox}[breakable, title=\hypertarget{box:cutFlow}{Flow of cuts}]
    From \eqref{eq:physResZono} it is clear that the combinatorics of further sequential residues of $\Psi_G$ is determined by the geometry of the zonotopes through the canonical form map. 
    The sequential residue $\res_{\mathcal{C}_{\g'}}$ can flow to $\res_{\mathcal{C}_{\g}}$ if and only if $\g$ labels a boundary of the facet labeled by $\g'$; equivalently when $(\g,\g')$ are related by a merger (c.f., section \ref{sec:acyclicMinors}). 
    We call this combinatorics the \emph{flow of cuts}.
\end{tcolorbox}

To make the above discussion concrete, consider the three-site chain as an example. The arrangement of the untwisted divisors on the total energy plane is depicted in figure \ref{fig:3chainTotEnergyZono} (left). 
The {\color{BrickRed}red} shaded region is identified with a zonotope whose top-dimensional facet is labeled by the acyclic minor with only solid edges. 
The codimension-one boundaries ({\color{darkCyan}cyan}) are also labeled by acyclic minors with one solid edge; they correspond to the intersection of the planes associated to tubes attached to the acyclic minor.
Lastly, the vertices ({\color{darkMagenta}magenta}) of the zonotope are labeled by acyclic minors with only directed edges; again, these correspond to the intersection of the planes associated to tubes attached to the acyclic minor.

To compute the canonical form of a region with degenerate facets, the arrangement must be perturbed in order to become generic. This is illustrated in figure \ref{fig:3chainTotEnergyZono} (right) where we shift the line 
$\Vsf (B_{\includegraphics[scale=.3]{figures/3chain_B2.pdf}}) \cap \Vsf (B_{\includegraphics[scale=.3]{figures/3chain_B123.pdf}})$. The canonical form of the degenerate arrangement is then obtained as the limit of the canonical form of the generic arrangement. 
Note that in the generic arrangement, there are two vertices labeled by the cut tubings associated to the acyclic minor ${\includegraphics[align=c]{figures/amp_t3}}$. The contribution of these two vertices to the canonical form is given by
\begin{align}
\res_{
        \includegraphics[scale=.3,align=c]{figures/3chain_B123.pdf}
    }[\phi_{\includegraphics[]{figures/amp_t3}}]
    =
    \left. \left[ 
        \textcolor{NavyBlue}{
            \frac{
                1
            }{
                B_{
                    \includegraphics[scale=0.3]{figures/3chain_B12.pdf}
                }
                B_{
                    \includegraphics[scale=0.3]{figures/3chain_B2.pdf}
                }
            }
        }
        +
        \textcolor{NavyBlue}{
            \frac{
                1
            }{
                B_{
                    \includegraphics[scale=0.3]{figures/3chain_B23.pdf}
                }
                B_{
                    \includegraphics[scale=0.3]{figures/3chain_B2.pdf}
                }
            }
        }
    \right] \right\vert_{
        B_{\includegraphics[scale=.3,align=c]{figures/3chain_B123.pdf}}=0
    }
    \d^2\mbf{x}_{
        \parallel 
    }
    \,,
\end{align}
where 
$
    \mbf{x}_{
        \parallel 
    }
$ are coordinates on the total energy plane 
$ \B_{\includegraphics[scale=.3,align=c]{figures/3chain_B123.pdf}}$ and we recall that 
\begin{align}
    \phi_{\includegraphics[]{figures/amp_t3}}
    = \dlog\C_{
        \includegraphics[]{figures/amp_t3}
    }
    &= \dlog B_{
        \includegraphics[scale=.3,align=c]{figures/3chain_B123.pdf}
    }
    \wedge \dlog\frac{
        B_{
            \includegraphics[scale=.3,align=c]{figures/3chain_B23.pdf}
        }
    }{
        B_{
            \includegraphics[scale=.3,align=c]{figures/3chain_B12.pdf}
        }
    }
    \wedge \dlog B_{
        \includegraphics[scale=.3,align=c]{figures/3chain_B2.pdf}
    }
    \,.
\end{align}
Therefore, 
$
    \res_{
        \C_{\includegraphics[scale=.6]{figures/delta_cc.pdf}}
    }[\Psi_{\includegraphics[scale=0.5]{figures/pt3.pdf}}]
    = \res_{
        \includegraphics[scale=.3,align=c]{figures/3chain_B123.pdf}
    }[\Psi_{\includegraphics[scale=0.5]{figures/pt3.pdf}}]
$, is the canonical form of the zonotope 
$ \mathcal{Z}_{\includegraphics[scale=.6,align=c]{figures/delta_cc.pdf}}
$ 
even though the point ${\bf x}^*_{\includegraphics[align=c,scale=0.8]{figures/amp_t3}}$ is degenerate. We conclude that the definition of $\dlog\C_\g$ in \eqref{eq:phiCut} is compatible with the interpretation of the canonical form for zonotopes with degenerate facets.

\begin{figure}
    \centering
    \includegraphics[width=.45\textwidth,align=c]{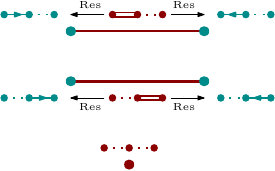}
    \qquad
    \includegraphics[width=.45\textwidth,align=c]{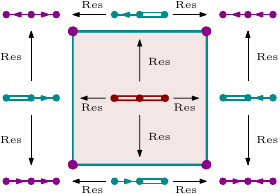}
    \caption{%
    Flow of cuts for the 3-site chain graph. 
    The top-dimensional facets of the zonotope are {\color{BrickRed}red}, the codimension-1 facets are {\color{darkCyan}cyan} and the codimension-2 facets are {\color{darkMagenta}magenta}. 
    }
    \label{fig:3chainZonos}
\end{figure}

The remaining zonotopes are generic and it is easily verified that 
\begin{align}\begin{aligned}
    \res_{
        \C_{\includegraphics[scale=.8,align=c]{figures/delta_cb.pdf}}
    }[\Psi_{\includegraphics[]{figures/amp_t1}}]
    &= \res_{
        \C_{\includegraphics[scale=.8,align=c]{figures/delta_cb.pdf}}
    }[
       \phi_{\includegraphics[]{figures/amp_t8}}
        +\phi_{\includegraphics[]{figures/amp_t9}}
    ]
    = \Omega[\mathcal{Z}_{
            \includegraphics[scale=.8,align=c]{figures/delta_cb.pdf}
        }]
    \,,
    \\
    \res_{
        \C_{\includegraphics[scale=.8,align=c]{figures/delta_bc.pdf}}
    }[\Psi_{\includegraphics[]{figures/amp_t1}}]
    &= \res_{
        \C_{\includegraphics[scale=.8,align=c]{figures/delta_bc.pdf}}
    }[
       \phi_{\includegraphics[]{figures/amp_t6}}
        +\phi_{\includegraphics[]{figures/amp_t7}}
    ]
    = \Omega[\mathcal{Z}_{
            \includegraphics[scale=.8,align=c]{figures/delta_bc.pdf}
        }]
    \,,
    \\
    \res_{
        \C_{\includegraphics[scale=.8,align=c]{figures/delta_bb.pdf}}
    }[\Psi_{\includegraphics[]{figures/amp_t1}}]
    &= \res_{
        \C_{\includegraphics[scale=.8,align=c]{figures/delta_bb.pdf}}
    }[
       \phi_{\includegraphics[]{figures/amp_t10}}
    ]
    = \Omega[
        \mathcal{Z}_{
            \includegraphics[scale=.8,align=c]{figures/delta_bb.pdf}
        }
    ]
    \,.
\end{aligned}\end{align}
The zonotopes of the 3-site chain are presented in figure \ref{fig:3chainZonos} ({\color{BrickRed}red}).
Further residues of the physical FRW-form are governed by the combinatorics of these zonotopes leading the the flow of (physical) cuts in figure \ref{fig:3chainZonos}. 

\section{Deriving the kinematic flow}
\label{sec:flowOfCuts}

In section \ref{sec:DEQs}, we derive the differential equations for the cut basis, without input from bulk physics, using intersection theory and simple residue calculus. 
Since the flow of physical sequential residues/cuts is tied to the combinatorics of zonotopes (c.f., section \ref{sec:zono}), the differential equations inherit the \emph{reverse} of this flow.\footnote{%
The flow is reversed for the differential equations because discontinuities (computed by changing the contour to one that has a residue component) and the differential are correlated by the coproduct but act on opposite ends of the tensor product \cite{Abreu:2019wzk, McLeodPokrakaRen}. 
Another way to see this reversal is by noting that the differential equations of the FRW-cohomology and dual-cohomology are related my minus transpose \cite{Caron-Huot:2021xqj, De:2023xue} and because the coboundary symbols $\delta_\bullet$ are directly related to sequential residues.
}
In section \ref{sec:DEQexamples}, we provide worked examples for the master formula for kinematic flow. 
\texttt{Mathematica} code for many of the worked examples and more (such as the five-site chain and the two-loop/cycle kite) can be found at the github repository \github\;  or in the ancillary files. 

\subsection{The differential equations from intersection theory}
\label{sec:DEQs}

In this section, we sketch the intersection theory computation for the DEQs. 
We provide the basis of forms dual to the those introduced in section \ref{sec:cutbasis}. 
Then using these dual forms, we provide a residue formula for the matrix elements of the DEQ. 
More details can be found in appendix \ref{app:AmatDets}. 

The basis of forms dual to $\phi_\g$ is 
\be
    \check{\phi}_{\mathfrak{g}}
    &:= \sgn_\g \,
    \delta_{\C_\mathfrak{g}} \left( 
        \Omega_{\mathfrak{g}}
    \right)
    \,,
    &\qquad
    \delta_{\g} &:= \sum_{\tau\in \C_\g} 
    \text{sgn}_\tau
    \delta_\tau
    \,,
    \\
    \Omega_{\mathfrak{g}} &= \tilde{\Omega}_\g\vert_{\C_\g}
    = \bigwedge_{\lfrak\in\mathbb{L}_\g} \Omega_\lfrak
    \,,
    &\qquad
    \Omega_\lfrak &= \tilde{\Omega}_\lfrak\vert_{\C_\g}
\ee
where we recall that inside the intersection number $\delta_{\mathcal{C}_\g} \to \res_{\C_\g}$. 
Furthermore, the argument of $\delta_{\mathcal{C}_\g}$ must be a form restricted to the cut $M_{\C_\g}$.
In fact, we use the canonical form, $\Omega_\g = \tilde{\Omega}_\g\vert_{\C_\g}$, of the cut geometry, $\Gamma_\g$, as the argument for the $\delta_{\mathcal{C}_\g}$.\footnote{%
Note that we use $\vert_{\C_\mathfrak{g}}$ to denote the pull-back to the cut associated this cut. 
} 
These dual forms are designed to have a diagonal intersection matrix with the cut basis $\phi_\g$.
Note that in the wedge product, $\Omega_{\lfrak_1}$ comes before $\Omega_{\lfrak_2}$ if $\min(V_{\lfrak_1})<\min(V_{\lfrak_2})$.

Assuming that the FRW-form $\phi$ has only simple poles, the intersection number has a simple localization formula
\be\label{eq:intNum}
    \la \check{\phi}_{\mathfrak{g}} \vert \phi \ra
    = \la \delta_{\C_\g}(\Omega_\g) \vert \phi \ra
    &:= \sgn_\g
    \overset{\res_{\Omega_{\mathfrak{g}}}}{\overbrace{
        \res_{\lfrak_{|\mathbb{L}_\g|}} \circ \cdots \res_{\lfrak_1} 
    }}
    \circ \res_{\C_\mathfrak{g}}[\phi]
\ee
where $\res_{\Omega_{\mathfrak{g}}}$ is the (leading order term of the) residue operator generated by the canonical form $\Omega_\g$ and each
\be \label{eq:resl}
    \res_{\lfrak} &:= 
    \left(\frac{1}{\prod_{v\in V_\lfrak} \alpha_{v}}\right)
    \sum_{v \in V_\lfrak}
    \mathrm{sig}(v,V_\lfrak\setminus v) 
    \alpha_{v} 
    \underset{u \in V_\lfrak \setminus v}{\bigcirc}
    \res_{x_v=0}
\ee
is the (leading order term of the) residue operator generated by the factor $\Omega_\lfrak$ in the product \eqref{eq:GammaFactor}.
The residue operator $\res_\lfrak$ is a weighted sum of residues that localizes to the vertices of the simplex whose canonical form is $\Omega_\lfrak$.
Note that we order the $\res_\lfrak$ such that $\res_{\lfrak_1}$ comes before $\res_{\lfrak_2}$ if $\min(V_{\lfrak_1})<\min(V_{\lfrak_2})$.

Using \eqref{eq:intNum}, \eqref{eq:resl}, and \eqref{eq:res_Cg}, computing the intersection matrix is straightforward
\begin{align} \label{eq:intMat}
    C_{\g\g^\prime} := 
    \la \check{\phi}_{\mathfrak{g}} \vert \phi_{\mathfrak{g}^\prime} \ra 
    = |\C_\g| \delta_{\mathfrak{g},\mathfrak{g}^\prime} \prod_{
        \lfrak \in \mathbb{L}_{\mathfrak{g}} 
    } 
    \frac{
        \sum_{l\in V_\lfrak} \alpha_l 
    }{
        \prod_{l\in V_\lfrak} \alpha_l
    }
    \,.
\end{align}
This matrix is diagonal since the residue operator $\res_{\C_\g}$ in \eqref{eq:intNum} projects out a single element of our basis $\res_{\C_\g}[\phi_{\g^\prime}] = |\C_\g| \delta_{\g\g^\prime} \Omega_\g$. 
Then, each $\res_\lfrak$ act on each $\Omega_\lfrak$ that makes up $\Omega_\g$. 
Each $\Omega_\lfrak$ is the canonical form of a simplex in the product decomposition of $\Gamma_{\g}$. 
See appendix \ref{app:intMat} for a simple derivation of \eqref{eq:intMat}.

Having a diagonal intersection matrix \eqref{eq:intMat} makes it easy to derive the differential equations from \eqref{eq:Amat}. 
Moreover, since we have designed our basis of FRW- and dual-forms to have only one independent cut the resulting DEQs will have the minimal amount of mixing between different cuts.

To compute the differential equations, we project $\d_\kin \phi_\g$ back onto the cut basis. 
However, $\d_\kin \phi_\g$ has double poles and the simplified formulas for the intersection number (\eqref{eq:intNum}, \eqref{eq:resl}, and \eqref{eq:res_Cg}) cannot be used. 
To remedy this, we need a representation for $\d_\kin \phi_\g$ with only simple poles. 

To find a form cohomologous (differs by a total covariant derivative) to $\d_\kin \phi_\g$ with simple poles, it is useful to promote the exterior derivative on the integration variables to the exterior derivative on the combined  integration and kinematic space: 
\begin{align} \label{eq:dtot}
    \d \to 
    \sum_{v\in V_G} \d x_v\, \partial_{x_v}
    + \sum_{v\in V_G} \d X_v\, \partial_{X_v} 
    + \sum_{e\in E_G} \d Y_e\, \partial_{Y_e} 
    \,,
\end{align}
{(see \cite[section 3.3]{De:2023xue} for more on this technique).}
Then, any logarithmic differential form (all forms so far have been logarithmic differential forms: the FRW-forms, dual-forms and the connection $\omega:= \dlog u = \sum_{v\in V_G} \alpha_v \dlog x_v$) can be \emph{canonically} upgraded to forms on the larger integration plus kinematic space by interpreting $\d$ as in \eqref{eq:dtot}.
The total covariant derivative of the upgraded FRW-forms $\phi_\g$ is especially simple and leads to 
\begin{align} \label{eq:dkin}
    \d_\kin \phi_\g
    \simeq 
    \left(
        \omega \wedge \phi_\g 
    \right)_{\d \text{ as in \eqref{eq:dtot}}} \vert_{ 
    \cdots \wedge \d Z \wedge \cdots \wedge \d Z^\prime \wedge \cdots \to 0
    }
    \,,
\end{align}
where $Z,Z^\prime \in \{X_v, Y_e\}$ and $\simeq$ denotes equivalence in twisted cohomology in the usual sense (using $\nabla$ with $\d$ acting only on the integration variables).  
Equation \eqref{eq:dkin} allows us to use the simplified formulas for the intersection number: \eqref{eq:intNum}, \eqref{eq:resl}, and \eqref{eq:res_Cg}. 

For the remainder of this derivation we will interpret $\d$ as in \eqref{eq:dtot} and extract the appropriate component of the differential form by sending $\cdots \wedge \d Z \wedge \cdots \wedge \d Z^\prime \wedge \cdots \to 0$.
Moreover, note that the residue operators of the intersection number project out the appropriate component of the differential form $\omega\wedge\phi_\g$ interpreted on the integration plus kinematic space automatically. 
Therefore, we leave this projection implicit.

Using \eqref{eq:intNum}, the diagonal entries of the DEQ are simple to evaluate 
\begin{tcolorbox}
\be\label{eq:AmatDiag}
    A_{\mathfrak{g}\mathfrak{g}} 
    &= \frac{1}{\la \check{\phi}_{\mathfrak{g}} \vert \phi_{\mathfrak{g}} \ra}
    \res_{\lfrak_{|\mathbb{L}_\g|}} \circ \cdots \circ \res_{\lfrak_1} \circ \res_{\C_\mathfrak{g}}
    \left[
        \omega \wedge \dlog\C_{\mathfrak{g}} \wedge \tilde\Omega_{\mathfrak{g}}
    \right]
    \,,
    \\
    &= (-1)^{|V_G|} \sum_{\lfrak\in\mathbb{L}_\g} 
    \bigg(\sum_{l\in V_\lfrak}\alpha_{l}\bigg)
    \dlog f_\lfrak(\mbf{X},\mbf{Y})
    \,.
\ee
\end{tcolorbox}
\noindent
To pass from the first to second line in \eqref{eq:AmatDiag}, we move $\omega$ so that it appears after $\tilde{\Omega}_\g$ at the cost of a sign. 
Then, $\res_{\C_\g}$ is easy to perform. 
After this residue, the remaining differential form is simply $(-1)^{|V_G|} \Omega_\g\wedge\omega \vert_{\C_\g}$. 
Next, we perform the residues $\res_\lfrak$ by breaking $\Omega_\g\wedge\omega \vert_{\C_\g}$ into its constituent pieces by isolating the $x_{l\in V_\lfrak}$-dependence
\be
    \Omega_\g\wedge\omega \vert_{\C_\g}
    &= \sum_{i=1}^{|\mathbb{L}_\g|} 
    (-1)^\bullet\; 
    \Omega_{\lfrak_1} \wedge \cdots
    \wedge \left(\Omega_{\lfrak_i} \wedge \omega_{\lfrak_i}\right)
    \wedge \cdots \Omega_{\lfrak_{|\mathbb{L}_\g|}}
    \\
    \omega_\lfrak 
    &:= \sum_{l\in V_\lfrak} \alpha_l\;  \dlog x_l 
    \bigg\vert_{\sum_{i\in V_\lfrak} x_i + f_\lfrak(\mbf{X},\mbf{Y}) = 0}
\ee
where the exact form of the sign $(-1)^\bullet$ is not important for understanding gross features of the localization in \eqref{eq:AmatDiag}.
Since $\res_\lfrak[\Omega_{\lfrak^\prime}] \propto \delta_{\lfrak\lfrak^\prime}$,
\begin{align}\begin{aligned}
    &\res_{\lfrak_{|\mathbb{L}_\g|}} \circ \cdots \circ \res_{\lfrak_1}
    \left[
        \Omega_{\lfrak_1} \wedge \cdots
        \wedge \left(\Omega_{\lfrak_i} \wedge \omega_{\lfrak_i}\right)
        \wedge \cdots \Omega_{\lfrak_{|\mathbb{L}_\g|}}
    \right]
    \\&= 
    \res_{\lfrak_1}[\Omega_{\lfrak_1}]
    \cdots \res_{\lfrak_i}[\Omega_{\lfrak_i} \wedge \omega_{\lfrak_i}]
    \cdots \res_{\lfrak_{|\mathbb{L}_\g|}}[\Omega_{\lfrak_{|\mathbb{L}_\g|}}]
    \,.
\end{aligned}\end{align}
The factors $\res_{\lfrak}[\Omega_{\lfrak}]$ evaluate to constants while the $\res_{\lfrak}[\Omega_{\lfrak} \wedge \omega_{\lfrak}]$ is proportional to a combination of the $\dlog x_{l \in V_\lfrak}$ where each $\dlog x_{l \in V_\lfrak}$ is subject to the conditions $\sum_{i\in V_\lfrak} x_i+f_\lfrak(\mbf{X},\mbf{Y})=0$ (from the cut) and  $x_{k \in V_\lfrak \setminus \{l\}} = 0$ (from the residue $\res_\lfrak$). 
Fortunately, these conditions have the same solution for each $x_{l \in V_\lfrak}$ as seen in the second line of \eqref{eq:AmatDiag} where there is only one $\dlog$ for each letter $\lfrak$.
A detailed derivation of \eqref{eq:AmatDiag} can be found in appendix \ref{app:Adaig}. 

Next are the off-diagonal terms
\begin{align}\label{eq:AmatOffDiag1}
    A_{\mathfrak{g}\mathfrak{g}^\prime} 
    &= \frac{(-1)^{|V_G|}}{\la \check{\phi}_{\mathfrak{g}^\prime} \vert \phi_{\mathfrak{g}^\prime} \ra}
    \sgn_\g \sgn_{\g^\prime}
    \res_{\lfrakp_{|\mathbb{L}_\gp|}} \circ \cdots \circ \res_{\lfrakp_1} \circ \res_{\C_\gp}
    \left[
       \dlog\C_{\mathfrak{g}} \wedge \tilde\Omega_{\mathfrak{g}}
       \wedge \omega
    \right]
    \,.
\end{align}
Obviously, $A_{\mathfrak{g}\mathfrak{g}^\prime} \neq 0$ if and only if $\res_{\C_\gp}[\dlog\C_{\g}] \neq 0$. 
This only happens when all tubings in $\C_{\g}$ can be obtained from a tubing in $\C_{\gp}$ by adding additional compatible tubes. 
This makes it clear that forms associated to acyclic minors with different broken edges cannot couple in the differential equation since the tubings associated to these acyclic minors  necessarily contain an incompatible pair.
Moreover, the residues $\res_\lfrak$ localize to where the $x_i\vert_{\C_\gp} = 0$ and are non-trivial if and only if there are enough $\dlog x_i$'s in $\tilde\Omega_{\g}\wedge\omega$. 
This further restricts which matrix elements $A_{\mathfrak{g}\mathfrak{g}^\prime}$ are non-trivial. 
Indeed, $A_{\mathfrak{g}\mathfrak{g}^\prime} \neq 0$ if and only if for all $\tau^\prime \in \C_\gp$ there exists a tubing $\tau \in \C_{\g}$ such that $\tau^\prime$ and $\tau$ differ by exactly one tube: $|\tau\setminus\tau^\prime|=1$. 
This is because $\omega$ can only contribute a single factor of $\dlog x_i$. 
Translating this condition into a condition on acyclic minors implies that $A_{\mathfrak{g}\mathfrak{g}^\prime} \neq 0$ if and only if the pair $(\g,\gp)$ are related by a merger (c.f., \ref{sec:acyclicMinors}).
That is, the differential equations inherit the combinatorics of the zonotopes governing the cuts. 

Then, for a pair $(\g,\g')$ that are related by a merger, we have  
\begin{tcolorbox}
\be\label{eq:AmatOffDiag}
    A_{\mathfrak{g}\mathfrak{g}^\prime} 
    &= 
    \mathfrak{s}(\g,\g')
    \frac{
        \left(\sum_{i\in V_{\lfrak_\uparrow}} \alpha_i\right)
        \left(\sum_{j\in V_{\lfrak_\downarrow}} \alpha_j\right)
    }{
        \sum_{k\in  V_{\lfrak_\uparrow} \cup  V_{\lfrak_\downarrow}} \alpha_k
    }
    \dlog \frac{
        f_{\lfrak_\uparrow}(\mbf{X},\mbf{Y})
    }{
        f_{\lfrak_\downarrow}(\mbf{X},\mbf{Y})
    }
    \,,
    \\
    \mathfrak{s}(\g,\g') 
    &:= \begin{cases}
        \phantom{-} 1 
        & \text{if oriented edges connecting $\lfrak_\uparrow$ to $\lfrak_\downarrow$ point away from $\lfrak_\uparrow$}
        \\
        -1 
        & \text{if oriented edges connecting $\lfrak_\uparrow$ to $\lfrak_\downarrow$ point towards $\lfrak_\uparrow$}
    \end{cases}
    \,,
\ee
\end{tcolorbox}
\noindent
where we recall that $\overline{V}_{\lfrak_\uparrow} < \overline{V}_{\lfrak_\downarrow}$.
Up to the overall sign, the factors in this formula are easy to understand. 
After taking the cut, 
\be \label{eq:AmatOffDiagWithSgn}
    A_{\g\g^\prime}
    &{\propto} \frac{
        |\mathcal{C}_{\g'}|
    }{
        \la \check{\phi}_{\gp} 
        \vert \phi_{\gp} \ra
    }
    \res_{\lfrakp_{|\mathbb{L}_\gp|}} {\circ} 
    {\cdots} {\circ} \res_{\lfrakp_1} 
    \left[
       \dlog B_{\g \setminus \g^\prime} {\wedge} \tilde\Omega_{\mathfrak{g}}
       {\wedge} \omega
    \right]
    \bigg\vert_{B_{t'\in\tau^\prime}=0}
    \,,
\ee
and
\be \label{eq:usingCggp}
    \dlog B_{\g\setminus\gp}
    &:= \dlog B_{t\in\tau\setminus\tau^\prime}
    \bigg\vert_{B_{t'\in\tau^\prime}=0}
    \,\forall\, 
    (\tau,\tau^\prime)\in\C_{\g\g^\prime}
    \,,
\ee
where we recall the definition of $\C_{\g\g^\prime}$ in equation \eqref{eq:Cgg}. 
Then, ignoring the overall sign, the residue in $A_{\g\g'}$ factorizes since each $\tilde{\Omega}_\lfrak \vert_{B_{t\in\tau^\prime}=0}$ depends solely on the $x_{v\in V_\lfrak}$
\be
    A_{\g\g'} &\propto
    \underset{
        \frac{
            \prod_{v \in V_{\underline{\lfrak}} } \alpha_v
        }{
            \sum_{v \in V_{\underline{\lfrak}} } \alpha_v
        }
        = 
        \frac{
            \prod_{v \in V_{\lfrak_\uparrow} \cup  V_{\lfrak_\downarrow} } \alpha_v
        }{
            \sum_{v V_{\lfrak_\uparrow} \cup  V_{\lfrak_\downarrow} } \alpha_v
        }
    }{\underbrace{
        \frac{
            |\mathcal{C}_{\g'}|
        }{
            \la \check{\phi}_{\gp} 
            \vert \phi_{\gp} \ra
        }
        \left(
            \prod_{
                \lfrak \in 
                \mathbb{L}_{\g'}^\uparrow
                \cup 
                \mathbb{L}_{\g'}^\downarrow
            }
            \res_{\lfrak}\tilde{\Omega}_\lfrak
            \bigg\vert_{B_{t\in\tau^\prime}=0}
        \right)
    }}
    \res_{\underline{\lfrak}} 
    \left[
       \tilde\Omega_{\lfrak_\uparrow}
       {\wedge} 
       \tilde\Omega_{\lfrak_\downarrow}
       {\wedge}
       \omega_{\underline{\lfrak}}
        {\wedge}
       \dlog B_{\g \setminus \g^\prime}
    \right]
    \bigg\vert_{B_{t'\in\tau^\prime}=0}
    \,,
\ee
where $\res_{\lfrak}\tilde{\Omega}_\lfrak\vert_{B_{t'\in\tau^\prime}=0} = \sum_{v \in V_\lfrak}\alpha_v / \prod_{v\in V_\lfrak} \alpha_v$, $V_{\underline{\lfrak}} = V_{\lfrak_\uparrow} \cup V_{\lfrak_\downarrow}$ and the only part of $\omega$ that survives the restriction of the wedge product to the cut is $\omega_\lfrak = \sum_{v \in V_{\lfrak_\uparrow}} \alpha_v \dlog x_v + \sum_{v \in V_{\lfrak_\downarrow}} \alpha_v \dlog x_v$.
To get \eqref{eq:AmatOffDiag} up to the overall sign, use 
\be
    \res_{\underline{\lfrak}} 
    \left[
       \tilde\Omega_{\lfrak_\uparrow}
       {\wedge} 
       \tilde\Omega_{\lfrak_\downarrow}
       {\wedge}
       \omega_{\underline{\lfrak}}
        {\wedge}
       \dlog B_{\g \setminus \g^\prime}
    \right]
    \bigg\vert_{B_{t'\in\tau^\prime}=0}
    \propto 
    \frac{
        \left(\sum_{i\in V_{\lfrak_\uparrow}} \alpha_i\right)
        \left(\sum_{j\in V_{\lfrak_\downarrow}} \alpha_j\right)
    }{
        \prod_{k\in  V_{\lfrak_\uparrow} \cup  V_{\lfrak_\downarrow}} \alpha_k
    }
    \dlog \frac{
        f_{\lfrak_\uparrow}(\mbf{X},\mbf{Y})
    }{
        f_{\lfrak_\downarrow}(\mbf{X},\mbf{Y})
    }
\ee
where 
\be
    \dlog B_{\g\setminus\g'}\vert_{x_{i\in V_{\underline{\lfrak}_\uparrow}} = 0}
    &=  f_{\underline{\lfrak}_\uparrow}(\mbf{X},\mbf{Y})
    \,,
    &
    \dlog B_{\g\setminus\g'}\vert_{x_{i\in V_{\underline{\lfrak}_\downarrow}} = 0}
    &=  f_{\underline{\lfrak}_\downarrow}(\mbf{X},\mbf{Y})
    \,.
\ee
Getting the sign correct takes careful bookkeeping; a full derivation can be found in appendix \ref{app:AmatOffDiag}.

\subsection{Examples}
\label{sec:DEQexamples}
To summarize the results of the previous section, the action of the derivative on an arbitrary element of the cut basis is given by  
\begin{tcolorbox}
\vspace{-1em}
\begin{align}
(-1)^{|V_\g|} \ \d_\kin\phi_{\g} &\simeq  \sum_{\lfrak \in \mathbb{L}_\g}\sum_{v \in V_\lfrak} \alpha_v \ \dlog \left( f_{\mathfrak{l}} \right) \wedge  \phi_\g \notag\\ &+ \sum_{\lfrak \rightarrow \lfrak' }  \frac{\sum_{v \in V_{\lfrak}}\alpha_v\sum_{v' \in V_{\lfrak'}} \alpha_{v'}}{\sum_{v \in V_{\lfrak}}\alpha_v+\sum_{v' \in V_{\lfrak'}}  \alpha_{v'}}  \dlog  \left(\frac{ f_{\lfrak'}}{ f_{\lfrak}} \right) \wedge \phi_{\g_{\lfrak \cup \lfrak'}}.
    \label{eq:kin_flow}
\end{align}
\end{tcolorbox}
\noindent
The sum on the second line is over all pairs of letters $\{ \lfrak,\lfrak'\} \subset \mathbb{L}_\g$ for which there exists at least one oriented edge from $V_{\lfrak}$ to $V_{\lfrak'}$. Furthermore, we use $\g_{\lfrak \cup\lfrak'}$ to denote the acyclic minor obtained from $\g$ by replacing all oriented edges between $V_{\lfrak}$ and $V_{\lfrak'}$ with solid edges. 

As emphasized throughout the paper, there are two important consequences which follow from the fact that only acyclic minors related by a merger---defined as the exchange of oriented edges with solid edges---can couple to one another in the differential equations. First, the differential equations decouple into $2^{|E_G|}$ blocks, one for each choice of broken edges. Second, within each block, the acyclic minors with fixed broken edges label the boundary stratification of the graphical zonotopes introduced in section \ref{sec:zono}. As a result, the differential equations inherit this zonotopal structure as we now demonstrate through examples. 

Throughout the examples we abuse notation and use a letter $\mathfrak{l}$ to denote $\dlog(f_\mathfrak{l})$. 
For example, in the context of the three-site chain
\begin{align}
&\raisebox{0.05cm}{\begin{tikzpicture}[scale=0.8]
        \coordinate (A) at (0,0);
        \coordinate (B) at (1/2,0);
        \coordinate (C) at (1,0);
        \coordinate (D) at (3/2,0);
        \draw[thick,dotted] (A) -- (B) --(C) ;
        \fill[Red] (A) circle (2.4pt);
        \fill[black] (B) circle (2pt);
        \fill[black] (C) circle (2pt);
    \end{tikzpicture}} \equiv \dlog (X_1+Y_{12}), && \raisebox{-0.1cm}{\begin{tikzpicture}[scale=0.8]
        \coordinate (A) at (0,0);
        \coordinate (B) at (1/2,0);
        \coordinate (C) at (1,0);
        \coordinate (D) at (3/2,0);
        \draw[ultra thick,Red] (A) -- node {\arbb} (B);
        \draw[thick,dotted] (B) -- (C);
        \fill[black] (A) circle (2pt);
        \fill[Red] (B) circle (2.4pt);
        \fill[black] (C) circle (2pt);
    \end{tikzpicture}} \equiv \dlog (X_2-Y_{12}+Y_{23}),   
\end{align}
and the three-cycle
\begin{align}
&\raisebox{-0.23cm}{ \includegraphics[scale=1]{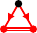}} \equiv \dlog(X_1+X_3-Y_{12}-Y_{23}), && \raisebox{-0.23cm}{ \includegraphics[scale=1]{figures/c3_bulk_let}} \equiv \dlog (X_1+X_2+X_3).
\end{align}

We also do not exemplify the differential equation for every element of the cut basis in each example since the cut basis is generally too large to fit into the main text of this work. 
However, the interested reader can find a the full differential equation for the one-loop bubble or two-cycle graph in appendix \ref{app:twoCycle}. 
The full results of the worked examples and more can be found at the github repository \github\; or in the ancillary files.

\subsubsection{Three-site chain}
\begin{figure}
    \centering
    \includegraphics[width=1\linewidth]{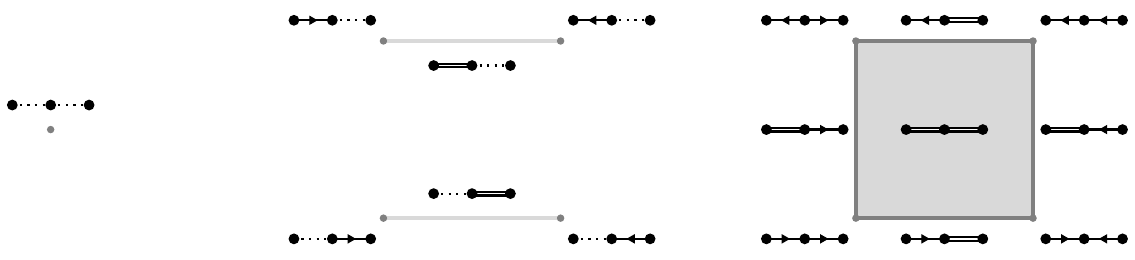}
    \caption{The differential equations for the path graph on three vertices decomposes into four blocks each described by the boundary stratification of a graphical zonotope.}
    \label{fig:path_zono_2}
\end{figure}

Consider the three-site chain. In the cut basis, the differential equations decouple into four blocks, each of which is described by a zonotope in figure \ref{fig:path_zono_2}. First, the point on the left-hand-side of figure \ref{fig:path_zono_2} encodes the following $1\times 1$ system of differential equations 
\begin{align}
-\d_\kin \phi_{\includegraphics[scale=0.9]{figures/delta_bb}}\simeq  \left( \alpha_1 \raisebox{0.05cm}{\begin{tikzpicture}[scale=0.8]
        \coordinate (A) at (0,0);
        \coordinate (B) at (1/2,0);
        \coordinate (C) at (1,0);
        \coordinate (D) at (3/2,0);
        \draw[thick,dotted] (A) -- (B) --(C) ;
        \fill[Red] (A) circle (2.4pt);
        \fill[black] (B) circle (2pt);
        \fill[black] (C) circle (2pt);
    \end{tikzpicture}}+\alpha_2 \raisebox{0.05cm}{\begin{tikzpicture}[scale=0.8]
        \coordinate (A) at (0,0);
        \coordinate (B) at (1/2,0);
        \coordinate (C) at (1,0);
        \coordinate (D) at (3/2,0);
        \draw[thick,dotted] (A) -- (B) --(C) ;
        \fill[black] (A) circle (2pt);
        \fill[Red] (B) circle (2.4pt);
        \fill[black] (C) circle (2pt);
    \end{tikzpicture}}+\alpha_3 \raisebox{0.05cm}{\begin{tikzpicture}[scale=0.8]
        \coordinate (A) at (0,0);
        \coordinate (B) at (1/2,0);
        \coordinate (C) at (1,0);
        \coordinate (D) at (3/2,0);
        \draw[thick,dotted] (A) -- (B) --(C) ;
        \fill[black] (A) circle (2pt);
        \fill[black] (B) circle (2pt);
        \fill[Red] (C) circle (2.4pt);
    \end{tikzpicture}}  \right) \phi_{\includegraphics[scale=0.9]{figures/delta_bb}}.
 \end{align}
Less trivially, the top middle line segment of figure \ref{fig:path_zono_2} is associated to the following $3 \times 3$ system of differential equations
\begin{align}
-\d_\kin \phi_{\begin{tikzpicture}[scale=0.8]
        \coordinate (A) at (0,0);
        \coordinate (B) at (1/2,0);
        \coordinate (C) at (1,0);
        \coordinate (D) at (3/2,0);
        \draw[thick,dotted] (B) -- (C);
        \draw[thick] (A) -- node {\ar} (B);
        \fill[black] (A) circle (2pt);
        \fill[black] (B) circle (2pt);
        \fill[black] (C) circle (2pt);
    \end{tikzpicture}} &\simeq \left( \alpha_1 \raisebox{-0.1cm}{\begin{tikzpicture}[scale=0.8]
        \coordinate (A) at (0,0);
        \coordinate (B) at (1/2,0);
        \coordinate (C) at (1,0);
        \coordinate (D) at (3/2,0);
        \draw[thick] (A) -- node {\ar} (B);
        \draw[thick,dotted] (B) -- (C);
        \fill[Red] (A) circle (2.4pt);
        \fill[black] (B) circle (2pt);
        \fill[black] (C) circle (2pt);
    \end{tikzpicture}}+\alpha_2\raisebox{-0.1cm}{\begin{tikzpicture}[scale=0.8]
        \coordinate (A) at (0,0);
        \coordinate (B) at (1/2,0);
        \coordinate (C) at (1,0);
        \coordinate (D) at (3/2,0);
        \draw[ultra thick,Red] (A) -- node {\arbb} (B);
        \draw[thick,dotted] (B) -- (C);
        \fill[black] (A) circle (2pt);
        \fill[Red] (B) circle (2.4pt);
        \fill[black] (C) circle (2pt);
        \end{tikzpicture}}+\alpha_3\raisebox{-0.1cm}{\begin{tikzpicture}[scale=0.8]
        \coordinate (A) at (0,0);
        \coordinate (B) at (1/2,0);
        \coordinate (C) at (1,0);
        \coordinate (D) at (3/2,0);
        \draw[thick] (A) -- node {\ar} (B);
        \draw[thick,dotted] (B) -- (C);
        \fill[black] (A) circle (2pt);
        \fill[black] (B) circle (2pt);
        \fill[Red] (C) circle (2.4pt);
    \end{tikzpicture}} \right) \phi_{\begin{tikzpicture}[scale=0.8]
        \coordinate (A) at (0,0);
        \coordinate (B) at (1/2,0);
        \coordinate (C) at (1,0);
        \coordinate (D) at (3/2,0);
        \draw[thick,dotted] (B) -- (C);
        \draw[thick] (A) -- node {\ar} (B);
        \fill[black] (A) circle (2pt);
        \fill[black] (B) circle (2pt);
        \fill[black] (C) circle (2pt);
    \end{tikzpicture}} + \frac{\alpha_1\alpha_2}{\alpha_1+\alpha_2}\left(\raisebox{-0.1cm}{\begin{tikzpicture}[scale=0.8]
        \coordinate (A) at (0,0);
        \coordinate (B) at (1/2,0);
        \coordinate (C) at (1,0);
        \coordinate (D) at (3/2,0);
        \draw[ultra thick,Red] (A) -- node {\arbb} (B);
        \draw[thick,dotted] (B) -- (C);
        \fill[black] (A) circle (2pt);
        \fill[Red] (B) circle (2.4pt);
        \fill[black] (C) circle (2pt);
    \end{tikzpicture}}- \raisebox{-0.1cm}{\begin{tikzpicture}[scale=0.8]
        \coordinate (A) at (0,0);
        \coordinate (B) at (1/2,0);
        \coordinate (C) at (1,0);
        \coordinate (D) at (3/2,0);
        \draw[thick] (A) -- node {\ar} (B);
        \draw[thick,dotted] (B) -- (C);
        \fill[Red] (A) circle (2.4pt);
        \fill[black] (B) circle (2pt);
        \fill[black] (C) circle (2pt);
    \end{tikzpicture}} \right) \phi_{\begin{tikzpicture}[scale=0.8]
        \coordinate (A) at (0,0);
        \coordinate (B) at (1/2,0);
        \coordinate (C) at (1,0);
        \coordinate (D) at (3/2,0);
        \draw[thick,dotted] (B) -- (C);
        \draw[thick,double] (A) -- (B);
        \fill[black] (A) circle (2pt);
        \fill[black] (B) circle (2pt);
        \fill[black] (C) circle (2pt);
    \end{tikzpicture}}, \notag \\
   -\d_\kin \phi_{\begin{tikzpicture}[scale=0.8]
        \coordinate (A) at (0,0);
        \coordinate (B) at (1/2,0);
        \coordinate (C) at (1,0);
        \coordinate (D) at (3/2,0);
        \draw[thick,dotted] (B) -- (C);
        \draw[thick] (A) -- node {\al} (B);
        \fill[black] (A) circle (2pt);
        \fill[black] (B) circle (2pt);
        \fill[black] (C) circle (2pt);
    \end{tikzpicture}}  &\simeq\left( \alpha_1 \raisebox{-0.1cm}{\begin{tikzpicture}[scale=0.8]
        \coordinate (A) at (0,0);
        \coordinate (B) at (1/2,0);
        \coordinate (C) at (1,0);
        \coordinate (D) at (3/2,0);
        \draw[ultra thick,Red] (A) -- node {\albb} (B);
        \draw[thick,dotted] (B) -- (C);
        \fill[Red] (A) circle (2.4pt);
        \fill[black] (B) circle (2pt);
        \fill[black] (C) circle (2pt);
    \end{tikzpicture}}+ \alpha_2 \raisebox{-0.1cm}{\begin{tikzpicture}[scale=0.8]
        \coordinate (A) at (0,0);
        \coordinate (B) at (1/2,0);
        \coordinate (C) at (1,0);
        \coordinate (D) at (3/2,0);
        \draw[thick] (A) -- node {\al} (B);
        \draw[thick,dotted] (B) -- (C);
        \fill[black] (A) circle (2pt);
        \fill[Red] (B) circle (2.4pt);
        \fill[black] (C) circle (2pt);
    \end{tikzpicture}}+ \alpha_3 \raisebox{-0.1cm}{\begin{tikzpicture}[scale=0.8]
        \coordinate (A) at (0,0);
        \coordinate (B) at (1/2,0);
        \coordinate (C) at (1,0);
        \coordinate (D) at (3/2,0);
        \draw[thick] (A) -- node {\al} (B);
        \draw[thick,dotted] (B) -- (C);
        \fill[black] (A) circle (2pt);
        \fill[black] (B) circle (2pt);
        \fill[Red] (C) circle (2.4pt);
    \end{tikzpicture}} \right) \phi_{\begin{tikzpicture}[scale=0.8]
        \coordinate (A) at (0,0);
        \coordinate (B) at (1/2,0);
        \coordinate (C) at (1,0);
        \coordinate (D) at (3/2,0);
        \draw[thick,dotted] (B) -- (C);
        \draw[thick] (A) -- node {\al} (B);
        \fill[black] (A) circle (2pt);
        \fill[black] (B) circle (2pt);
        \fill[black] (C) circle (2pt);
    \end{tikzpicture}}+\frac{\alpha_1\alpha_2}{\alpha_1+\alpha_2}\left(\raisebox{-0.1cm}{\begin{tikzpicture}[scale=0.8]
        \coordinate (A) at (0,0);
        \coordinate (B) at (1/2,0);
        \coordinate (C) at (1,0);
        \coordinate (D) at (3/2,0);
        \draw[ultra thick,Red] (A) -- node {\albb} (B);
        \draw[thick,dotted] (B) -- (C);
        \fill[Red] (A) circle (2.4pt);
        \fill[black] (B) circle (2pt);
        \fill[black] (C) circle (2pt);
    \end{tikzpicture}}- \raisebox{-0.1cm}{\begin{tikzpicture}[scale=0.8]
        \coordinate (A) at (0,0);
        \coordinate (B) at (1/2,0);
        \coordinate (C) at (1,0);
        \coordinate (D) at (3/2,0);
        \draw[thick] (A) -- node {\al} (B);
        \draw[thick,dotted] (B) -- (C);
        \fill[black] (A) circle (2pt);
        \fill[Red] (B) circle (2.4pt);
        \fill[black] (C) circle (2pt);
    \end{tikzpicture}} \right) \phi_{\begin{tikzpicture}[scale=0.8]
        \coordinate (A) at (0,0);
        \coordinate (B) at (1/2,0);
        \coordinate (C) at (1,0);
        \coordinate (D) at (3/2,0);
        \draw[thick,dotted] (B) -- (C);
        \draw[thick,double] (A) -- (B);
        \fill[black] (A) circle (2pt);
        \fill[black] (B) circle (2pt);
        \fill[black] (C) circle (2pt);
    \end{tikzpicture}}, \notag \\
    -\d_\kin \phi_{\begin{tikzpicture}[scale=0.8]
        \coordinate (A) at (0,0);
        \coordinate (B) at (1/2,0);
        \coordinate (C) at (1,0);
        \coordinate (D) at (3/2,0);
        \draw[thick,dotted] (B) -- (C);
        \draw[thick,double] (A) -- (B);
        \fill[black] (A) circle (2pt);
        \fill[black] (B) circle (2pt);
        \fill[black] (C) circle (2pt);
    \end{tikzpicture}} &\simeq \left((\alpha_1+\alpha_2) \ \raisebox{0.05cm}{\begin{tikzpicture}[scale=0.8]
        \coordinate (A) at (0,0);
        \coordinate (B) at (1/2,0);
        \coordinate (C) at (1,0);
        \coordinate (D) at (3/2,0);
        \draw[very thick,Red,double] (A) -- (B) ;
        \draw[thick,dotted] (B) -- (C);
        \fill[Red] (A) circle (2.4pt);
        \fill[Red] (B) circle (2.4pt);
        \fill[black] (C) circle (2pt);
    \end{tikzpicture}}+\alpha_3 \raisebox{0.05cm}{\begin{tikzpicture}[scale=0.8]
        \coordinate (A) at (0,0);
        \coordinate (B) at (1/2,0);
        \coordinate (C) at (1,0);
        \coordinate (D) at (3/2,0);
        \draw[thick,double] (A) -- (B);
        \draw[thick,dotted] (B) -- (C);
        \fill[black] (A) circle (2pt);
        \fill[black] (B) circle (2pt);
        \fill[Red] (C) circle (2.4pt);
    \end{tikzpicture}} \right) \phi_{\begin{tikzpicture}[scale=0.8]
        \coordinate (A) at (0,0);
        \coordinate (B) at (1/2,0);
        \coordinate (C) at (1,0);
        \coordinate (D) at (3/2,0);
        \draw[thick,dotted] (B) -- (C);
        \draw[thick,double] (A) -- (B);
        \fill[black] (A) circle (2pt);
        \fill[black] (B) circle (2pt);
        \fill[black] (C) circle (2pt);
    \end{tikzpicture}}.
\end{align}
Notably, the differential of each vertex produces a term proportional to the bulk of the line segment. 

Finally, the remaining square describes a $9 \times 9$ block in the differential equations. We provide a handful of examples. Consider the top left vertex of the square in figure \ref{fig:path_zono_2}, the differential of the associated function is given by 
\begin{align}\label{eq:ThreeChainDerivExam}
-\d_\kin \phi_{\begin{tikzpicture}[scale=0.8]
        \coordinate (A) at (0,0);
        \coordinate (B) at (1/2,0);
        \coordinate (C) at (1,0);
        \coordinate (D) at (3/2,0);
        \draw[thick] (B) -- node {\ar} (C);
        \draw[thick] (A) -- node {\al} (B);
        \fill[black] (A) circle (2pt);
        \fill[black] (B) circle (2pt);
        \fill[black] (C) circle (2pt);
    \end{tikzpicture}}  &\simeq\left( \alpha_1 \raisebox{-0.1cm}{\begin{tikzpicture}[scale=0.8]
        \coordinate (A) at (0,0);
        \coordinate (B) at (1/2,0);
        \coordinate (C) at (1,0);
        \coordinate (D) at (3/2,0);
        \draw[ultra thick,Red] (A) -- node {\albb} (B);
        \draw[thick] (B) -- node {\ar} (C);
        \fill[Red] (A) circle (2.4pt);
        \fill[black] (B) circle (2pt);
        \fill[black] (C) circle (2pt);
    \end{tikzpicture}}+ \alpha_2 \raisebox{-0.08cm}{\begin{tikzpicture}[scale=0.8]
        \coordinate (A) at (0,0);
        \coordinate (B) at (1/2,0);
        \coordinate (C) at (1,0);
        \coordinate (D) at (3/2,0);
        \draw[thick] (A) -- node {\al} (B);
        \draw[thick] (B) -- node {\ar} (C);
        \fill[black] (A) circle (2pt);
        \fill[Red] (B) circle (2.4pt);
        \fill[black] (C) circle (2pt);
    \end{tikzpicture}}+ \alpha_3 \raisebox{-0.1cm}{\begin{tikzpicture}[scale=0.8]
        \coordinate (A) at (0,0);
        \coordinate (B) at (1/2,0);
        \coordinate (C) at (1,0);
        \coordinate (D) at (3/2,0);
        \draw[thick] (A) -- node {\al} (B);
        \draw[very thick,Red] (B) -- node {\arbb} (C);
        \fill[black] (A) circle (2pt);
        \fill[black] (B) circle (2pt);
        \fill[Red] (C) circle (2.4pt);
    \end{tikzpicture}} \right) \wedge \phi_{\begin{tikzpicture}[scale=0.8]
        \coordinate (A) at (0,0);
        \coordinate (B) at (1/2,0);
        \coordinate (C) at (1,0);
        \coordinate (D) at (3/2,0);
        \draw[thick] (B) -- node {\ar} (C);
        \draw[thick] (A) -- node {\al} (B);
        \fill[black] (A) circle (2pt);
        \fill[black] (B) circle (2pt);
        \fill[black] (C) circle (2pt);
    \end{tikzpicture}}+\frac{\alpha_1\alpha_2}{\alpha_1+\alpha_2}\left(\raisebox{-0.1cm}{\begin{tikzpicture}[scale=0.8]
        \coordinate (A) at (0,0);
        \coordinate (B) at (1/2,0);
        \coordinate (C) at (1,0);
        \coordinate (D) at (3/2,0);
        \draw[ultra thick,Red] (A) -- node {\albb} (B);
        \draw[thick] (B) -- node {\ar} (C);
        \fill[Red] (A) circle (2.4pt);
        \fill[black] (B) circle (2pt);
        \fill[black] (C) circle (2pt);
    \end{tikzpicture}}- \raisebox{-0.08cm}{\begin{tikzpicture}[scale=0.8]
        \coordinate (A) at (0,0);
        \coordinate (B) at (1/2,0);
        \coordinate (C) at (1,0);
        \coordinate (D) at (3/2,0);
        \draw[thick] (A) -- node {\al} (B);
        \draw[thick] (B) -- node {\ar} (C);
        \fill[black] (A) circle (2pt);
        \fill[Red] (B) circle (2.4pt);
        \fill[black] (C) circle (2pt);
    \end{tikzpicture}} \right) \wedge\phi_{\begin{tikzpicture}[scale=0.8]
        \coordinate (A) at (0,0);
        \coordinate (B) at (1/2,0);
        \coordinate (C) at (1,0);
        \coordinate (D) at (3/2,0);
        \draw[thick] (B) -- node {\ar} (C);
        \draw[thick,double] (A) -- (B);
        \fill[black] (A) circle (2pt);
        \fill[black] (B) circle (2pt);
        \fill[black] (C) circle (2pt);
    \end{tikzpicture}} \notag \\
    &+ \frac{\alpha_2 \alpha_3}{\alpha_2 +\alpha_3}\left(\raisebox{-0.1cm}{\begin{tikzpicture}[scale=0.8]
        \coordinate (A) at (0,0);
        \coordinate (B) at (1/2,0);
        \coordinate (C) at (1,0);
        \coordinate (D) at (3/2,0);
        \draw[thick] (A) -- node {\al} (B);
        \draw[very thick,Red] (B) -- node {\arbb} (C);
        \fill[black] (A) circle (2pt);
        \fill[black] (B) circle (2pt);
        \fill[Red] (C) circle (2.4pt);
    \end{tikzpicture}}- \raisebox{-0.08cm}{\begin{tikzpicture}[scale=0.8]
        \coordinate (A) at (0,0);
        \coordinate (B) at (1/2,0);
        \coordinate (C) at (1,0);
        \coordinate (D) at (3/2,0);
        \draw[thick] (A) -- node {\al} (B);
        \draw[thick] (B) -- node {\ar} (C);
        \fill[black] (A) circle (2pt);
        \fill[Red] (B) circle (2.4pt);
        \fill[black] (C) circle (2pt);
    \end{tikzpicture}}  \right) \wedge \phi_{\begin{tikzpicture}[scale=0.8]
        \coordinate (A) at (0,0);
        \coordinate (B) at (1/2,0);
        \coordinate (C) at (1,0);
        \coordinate (D) at (3/2,0);
        \draw[thick,double] (B) -- (C);
        \draw[thick] (A) -- node {\al} (B);
        \fill[black] (A) circle (2pt);
        \fill[black] (B) circle (2pt);
        \fill[black] (C) circle (2pt);
    \end{tikzpicture}}.
\end{align}
On the right-hand side two new functions appear, each corresponding to one of the codimension-one faces which contain the specified vertex in their boundary. Next, consider the derivative acting on the first new function appearing in \eqref{eq:ThreeChainDerivExam}, given by
\begin{align}
-\d_\kin \phi_{\begin{tikzpicture}[scale=0.75]
        \coordinate (A) at (0,0);
        \coordinate (B) at (1/2,0);
        \coordinate (C) at (1,0);
        \coordinate (D) at (3/2,0);
        \draw[thick,double] (B) -- (C);
        \draw[thick] (A) -- node {\al} (B);
        \fill[black] (A) circle (2pt);
        \fill[black] (B) circle (2pt);
        \fill[black] (C) circle (2pt);
    \end{tikzpicture}}  &\simeq\left( \alpha_1 \raisebox{-0.1cm}{\begin{tikzpicture}[scale=0.75]
        \coordinate (A) at (0,0);
        \coordinate (B) at (1/2,0);
        \coordinate (C) at (1,0);
        \coordinate (D) at (3/2,0);
        \draw[ultra thick,Red] (A) -- node {\albb} (B);
        \draw[thick,dotted] (B) -- (C);
        \fill[Red] (A) circle (2.4pt);
        \fill[black] (B) circle (2pt);
        \fill[black] (C) circle (2pt);
    \end{tikzpicture}}+ (\alpha_2+\alpha_3) \raisebox{-0.1cm}{\begin{tikzpicture}[scale=0.75]
        \coordinate (A) at (0,0);
        \coordinate (B) at (1/2,0);
        \coordinate (C) at (1,0);
        \coordinate (D) at (3/2,0);
        \draw[thick] (A) -- node {\al} (B);
        \draw[very thick,double,Red] (B) -- (C);
        \fill[black] (A) circle (2pt);
        \fill[Red] (B) circle (2.4pt);
        \fill[Red] (C) circle (2.4pt);
    \end{tikzpicture}} \right)\wedge \phi_{\begin{tikzpicture}[scale=0.75]
        \coordinate (A) at (0,0);
        \coordinate (B) at (1/2,0);
        \coordinate (C) at (1,0);
        \coordinate (D) at (3/2,0);
        \draw[thick,double] (B) -- (C);
        \draw[thick] (A) -- node {\al} (B);
        \fill[black] (A) circle (2pt);
        \fill[black] (B) circle (2pt);
        \fill[black] (C) circle (2pt);
    \end{tikzpicture}}+\frac{\alpha_1(\alpha_2+\alpha_3)}{\alpha_1+\alpha_2+\alpha_3}\left(\raisebox{-0.1cm}{\begin{tikzpicture}[scale=0.75]
        \coordinate (A) at (0,0);
        \coordinate (B) at (1/2,0);
        \coordinate (C) at (1,0);
        \coordinate (D) at (3/2,0);
        \draw[ultra thick,Red] (A) -- node {\albb} (B);
        \draw[thick,double] (B) -- (C);
        \fill[Red] (A) circle (2.4pt);
        \fill[black] (B) circle (2pt);
        \fill[black] (C) circle (2pt);
    \end{tikzpicture}}- \raisebox{-0.1cm}{\begin{tikzpicture}[scale=0.75]
        \coordinate (A) at (0,0);
        \coordinate (B) at (1/2,0);
        \coordinate (C) at (1,0);
        \coordinate (D) at (3/2,0);
        \draw[thick] (A) -- node {\al} (B);
        \draw[very thick,double, Red] (B) -- (C);
        \fill[black] (A) circle (2pt);
        \fill[Red] (B) circle (2.4pt);
        \fill[Red] (C) circle (2.4pt);
    \end{tikzpicture}} \right)\wedge \phi_{\begin{tikzpicture}[scale=0.75]
        \coordinate (A) at (0,0);
        \coordinate (B) at (1/2,0);
        \coordinate (C) at (1,0);
        \coordinate (D) at (3/2,0);
        \draw[thick,double] (B) -- (C);
        \draw[thick,double] (A) -- (B);
        \fill[black] (A) circle (2pt);
        \fill[black] (B) circle (2pt);
        \fill[black] (C) circle (2pt);
    \end{tikzpicture}}. \notag 
\end{align}
This time a single new function is generated, which corresponds to the bulk of the polytope, with differential given by
\begin{align}
-\d_\kin \phi_{\begin{tikzpicture}[scale=0.8]
        \coordinate (A) at (0,0);
        \coordinate (B) at (1/2,0);
        \coordinate (C) at (1,0);
        \coordinate (D) at (3/2,0);
        \draw[thick,double] (B) -- (C);
        \draw[thick,double] (A) -- (B);
        \fill[black] (A) circle (2pt);
        \fill[black] (B) circle (2pt);
        \fill[black] (C) circle (2pt);
    \end{tikzpicture}} &\simeq (\alpha_1+\alpha_2+\alpha_3) \ \raisebox{0.0cm}{\begin{tikzpicture}[scale=0.8]
        \coordinate (A) at (0,0);
        \coordinate (B) at (1/2,0);
        \coordinate (C) at (1,0);
        \coordinate (D) at (3/2,0);
        \draw[very thick,Red,double] (A) -- (B) ;
        \draw[very thick,Red,double] (B) -- (C);
        \fill[Red] (A) circle (2.4pt);
        \fill[Red] (B) circle (2.4pt);
        \fill[Red] (C) circle (2.4pt);
    \end{tikzpicture}}\ \wedge \phi_{\begin{tikzpicture}[scale=0.8]
        \coordinate (A) at (0,0);
        \coordinate (B) at (1/2,0);
        \coordinate (C) at (1,0);
        \coordinate (D) at (3/2,0);
        \draw[thick,double] (B) -- (C);
        \draw[thick,double] (A) -- (B);
        \fill[black] (A) circle (2pt);
        \fill[black] (B) circle (2pt);
        \fill[black] (C) circle (2pt);
    \end{tikzpicture}}.
\end{align}
In this last step no new elements of the cut basis are generated and the kinematic flow for this sector terminates.

\subsubsection{Three-cycle}
\begin{figure}[]
\centering
\includegraphics[width=\textwidth]{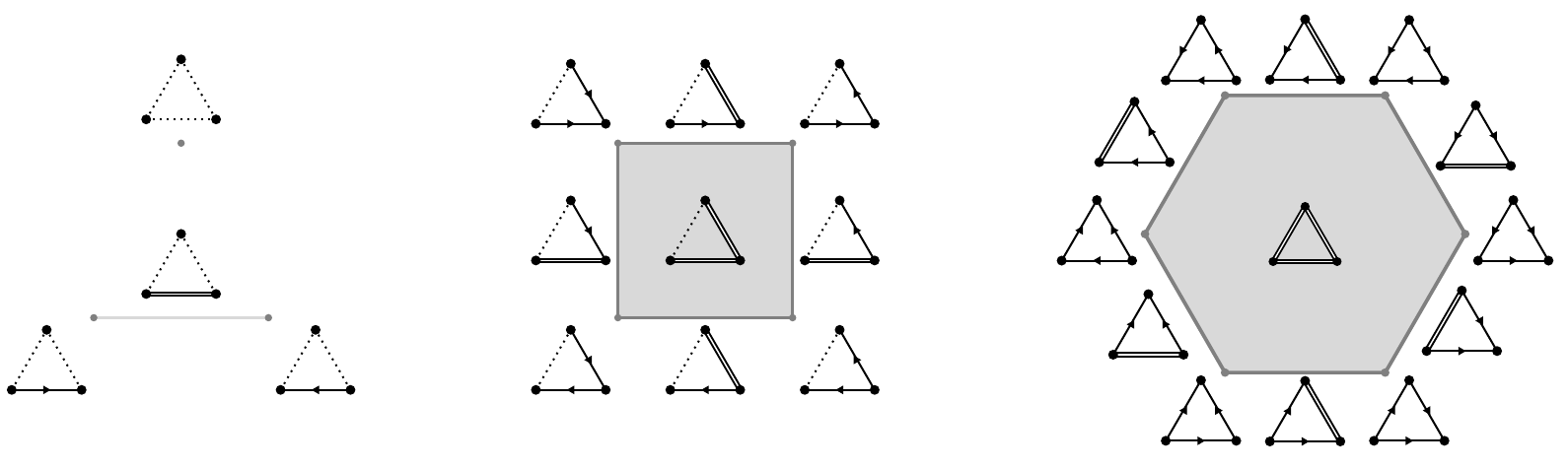}
\caption{A subset of the graphical zonotopes which describe the kinematic flow for the three-cycle.}
\label{fig:zon_cycle3}
\end{figure}

The first seven blocks of the kinematic flow for the three-cycle exhibit a flow structure analogous to those found for the three-site chain, with graphical zonotopes corresponding to points, line segments and squares, as demonstrated in figure~\ref{fig:zon_cycle3}. The final block, however, introduces a novel $(13 \times 13)$ system of differential equations, whose kinematic flow is characterized by the hexagon shown in Figure~\ref{fig:zon_cycle3}. For example, consider the right most vertex of the hexagon, its differential is given by
\begin{align}
-\d_\kin \phi_{\includegraphics[scale=0.9]{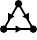}} &\simeq \left(\alpha_1 \raisebox{-0.3cm}{ \includegraphics[scale=1]{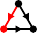}}+\alpha_2\raisebox{-0.3cm}{ \includegraphics[scale=1]{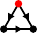}}+\alpha_3\raisebox{-0.3cm}{ \includegraphics[scale=1]{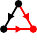}} \right) \wedge \phi_{\includegraphics[scale=0.9]{figures/c3_vert.pdf}} \notag \\
&+\frac{\alpha_1 \alpha_3}{\alpha_1 + \alpha_3}\left( \raisebox{-0.3cm}{ \includegraphics[scale=1]{figures/c3_vert_let_3.pdf}}-\raisebox{-0.3cm}{ \includegraphics[scale=1]{figures/c3_vert_let_1.pdf}} \right)\wedge \phi_{\includegraphics[scale=0.9]{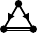}}+\frac{\alpha_1 \alpha_2}{\alpha_1 + \alpha_2}\left( \raisebox{-0.3cm}{ \includegraphics[scale=1]{figures/c3_vert_let_1.pdf}}-\raisebox{-0.3cm}{ \includegraphics[scale=1]{figures/c3_vert_let_2.pdf}} \right) \wedge \phi_{\includegraphics[scale=0.9]{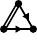}}.
\end{align}
As for the three-site chain, two new terms have been generated corresponding to the two line segments which intersect at the chosen vertex. Applying the derivative to the first element of the cut basis on the second line we find 
\begin{align}
    {-}\d_\kin \phi_{
        \includegraphics[scale=0.9]{figures/c3_vert_line_1.pdf}
    } 
    \simeq \left( 
        (\alpha_1 {+} \alpha_3) 
        \raisebox{-0.23cm}{ \includegraphics[scale=1]{figures/c3_line1_let1}} 
        + \alpha_2 
        \raisebox{-0.23cm}{ \includegraphics[scale=1]{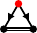}} 
    \right) {\wedge} \phi_{
        \includegraphics[scale=0.9]{figures/c3_vert_line_1.pdf}
    } 
    {+} \frac{(\alpha_1+\alpha_3)\alpha_2}{\alpha_1+\alpha_2+\alpha_3} 
    \left( 
        \raisebox{-0.23cm}{ 
            \includegraphics[scale=1]{figures/c3_line1_let1}
        } 
        {-} \raisebox{-0.23cm}{
            \includegraphics[scale=1]{figures/c3_line1_let2}
        }  
    \right) {\wedge} \phi_{
        \includegraphics[scale=0.9]{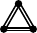}
    }.
\end{align}
Again, a single new function is generated, associated to the bulk of the hexagon. The kinematic flow for this block terminates with 
\begin{align}
-\d_\kin \phi_{\includegraphics[scale=0.9]{figures/c3_bulk.pdf}} \simeq \left(\alpha_1 + \alpha_2 + \alpha_3 \right) \raisebox{-0.23cm}{\includegraphics[scale=1]{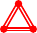}} \wedge \phi_{\includegraphics[scale=0.9]{figures/c3_bulk.pdf}}.
\end{align}

\subsubsection{Four-cycle}
\begin{figure}[]
\centering
\includegraphics[scale=0.85]{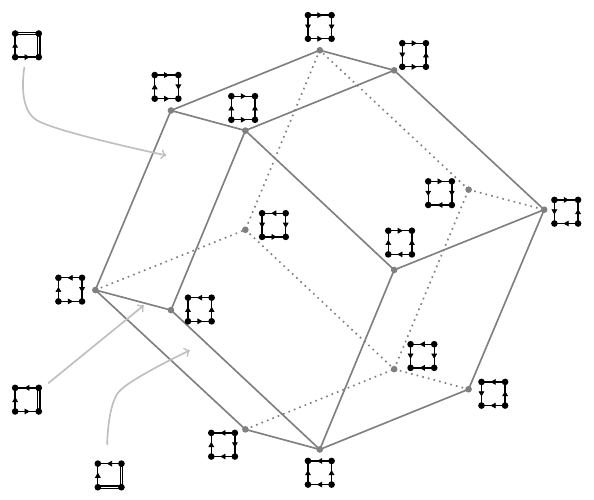}
\caption{The graphical zonotope (rhombic dodecahedron) describing the kinematic flow for a block in the differential equations for the four-cycle.}
\label{fig:rhombic}
\end{figure}

All but one of the blocks in the kinematic flow for the four-cycle are described by zonotopes given by points, line segments, squares and cubes. The structure of the final block is captured by the graphical zonotope displayed in figure~\ref{fig:rhombic}. For illustration, consider the function associated with the top front vertex of the zonotope, we have
\begin{align}
    \d_\kin \phi_{\includegraphics[scale=0.8]{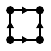}} &\simeq  \left(\alpha_1 \raisebox{-0.25cm}{\includegraphics[scale=0.8]{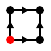}}+\alpha_2\raisebox{-0.25cm}{\includegraphics[scale=0.8]{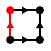}}+\alpha_3\raisebox{-0.25cm}{\includegraphics[scale=0.8]{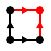}}+\alpha_4\raisebox{-0.25cm}{\includegraphics[scale=0.8]{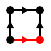}}  \right) \wedge  
    \phi_{\includegraphics[scale=0.8]{figures/cycle4_14}} 
    \notag \\
    &+ \frac{\alpha_1 \alpha_4}{\alpha_1+\alpha_4} \left( \raisebox{-0.25cm}{\includegraphics[scale=0.8]{figures/cycle4_14_let_4.pdf}}-\raisebox{-0.25cm}{\includegraphics[scale=0.8]{figures/cycle4_14_let_1.pdf}}  \right) \wedge  
    \phi_{\includegraphics[scale=0.8]{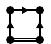}} 
    + \frac{\alpha_1 \alpha_2}{\alpha_1+\alpha_2}\left( \raisebox{-0.25cm}{\includegraphics[scale=0.8]{figures/cycle4_14_let_2.pdf}}-\raisebox{-0.25cm}{\includegraphics[scale=0.8]{figures/cycle4_14_let_1.pdf}} \right) \wedge  
    \phi_{\includegraphics[scale=0.8]{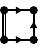}}  
    \notag \\
    &+ \frac{\alpha_2 \alpha_3}{\alpha_2+\alpha_3}\left( \raisebox{-0.25cm}{\includegraphics[scale=0.8]{figures/cycle4_14_let_3.pdf}}-\raisebox{-0.25cm}{\includegraphics[scale=0.8]{figures/cycle4_14_let_2.pdf}}  \right)  \wedge
    \phi_{\includegraphics[scale=0.8]{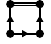}}  + \frac{\alpha_3 \alpha_4}{\alpha_3+\alpha_4} \left( \raisebox{-0.25cm}{\includegraphics[scale=0.8]{figures/cycle4_14_let_3.pdf}}-\raisebox{-0.25cm}{\includegraphics[scale=0.8]{figures/cycle4_14_let_4.pdf}} \right) \wedge 
    \phi_{\includegraphics[scale=0.8]{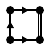}}.
\end{align}
The appearance of four new functions on the right-hand side reflects the fact that the corresponding vertex of the graphical zonotope is non-simple, lying at the intersection of four line segments. The derivative of the first function on the second line yields
\begin{align}
\d_\kin  \phi_{\includegraphics[scale=0.8]{figures/cycle4_14_d}} &\simeq \left( \left(\alpha_1 +\alpha_4 \right)\raisebox{-0.17cm}{\includegraphics[scale=0.8]{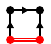}}+\alpha_2\raisebox{-0.17cm}{\includegraphics[scale=0.8]{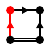}}+\alpha_3\raisebox{-0.17cm}{\includegraphics[scale=0.8]{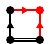}} \right) \wedge \phi_{\includegraphics[scale=0.8]{figures/cycle4_14_d}} \notag \\[0.5em]
&+\frac{(\alpha_1+\alpha_4)\alpha_2}{\alpha_1+\alpha_2+\alpha_4} \left( \raisebox{-0.17cm}{\includegraphics[scale=0.8]{figures/cycle4_14_d_let_2}}-\raisebox{-0.17cm}{\includegraphics[scale=0.8]{figures/cycle4_14_d_let_1}} \right) \wedge \phi_{\includegraphics[scale=0.8]{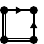}}+\frac{\alpha_2\alpha_3}{\alpha_2+\alpha_3}  \left( \raisebox{-0.17cm}{\includegraphics[scale=0.8]{figures/cycle4_14_d_let_3}}-\raisebox{-0.17cm}{\includegraphics[scale=0.8]{figures/cycle4_14_d_let_2}}\right) \wedge \phi_{\includegraphics[scale=0.8]{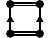}}.
\end{align}
In this case we generate two new functions corresponding to the two codimension-one faces intersecting on this edge. Applying the derivative to the first results in the following 
\begin{align}
    \d_\kin \phi_{\includegraphics[scale=0.8]{figures/cycle4_14_d_l}} 
    &\simeq  
    \left( 
        (\alpha_1{+}\alpha_2{+}\alpha_4)   
        \raisebox{-0.18cm}{\includegraphics[scale=0.8]{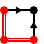}}
        {+} \alpha_3 \raisebox{-0.18cm}{\includegraphics[scale=0.8]{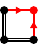}} 
    \right) 
    {\wedge} \phi_{\includegraphics[scale=0.8]{figures/cycle4_14_d_l}} 
    {+} \frac{
        (\alpha_1{+}\alpha_2{+}\alpha_4 )
        \alpha_3
    }{
        \alpha_1{+}\alpha_2{+}\alpha_3{+}\alpha_4
    } 
    \left( 
        \raisebox{-0.18cm}{\includegraphics[scale=0.8]{figures/cycle4_14_d_l_let_2}}
        {-} \raisebox{-0.18cm}{\includegraphics[scale=0.8]{figures/cycle4_14_d_l_let_1}} 
    \right) 
    {\wedge} \phi_{\includegraphics[scale=0.8]{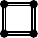}}. 
\end{align}
This function corresponds to a codimension-one face of the graphical zonotope, therefore, its derivative generates only one new function, corresponding to the bulk of the polytope. Finally, applying the derivative to this last function produces no further terms, and the kinematic flow terminates with
\begin{align}
\d_\kin \phi_{\includegraphics[scale=0.8]{figures/cycle4_double}} &\simeq   \left( \alpha_1+\alpha_2+\alpha_3+\alpha_4 \right) \raisebox{-0.15cm}{ \includegraphics[scale=0.8]{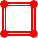}} \wedge 
\phi_{\includegraphics[scale=0.8]{figures/cycle4_double}}.
\end{align}

\section{Conclusion}

It was recently proposed that the independent cuts of the physical FRW-form naturally define a set of forms known as the \emph{cut basis} \cite{De:2024zic}. Building on this insight, we have shown that the cut basis and its associated {\it kinematic flow} are governed by rich combinatorial structures. Each element of the cut basis corresponds one-to-one with a {\it positive geometry} associated with a cut of the physical FRW-from. To systematically label elements of this basis, we introduced the notion of an {\it acyclic minor} and provided a simple prescription for assigning to each a {\it logarithmic} differential form. With the cut basis at hand, we employed the framework of relative twisted cohomology and intersection theory to derive the corresponding system of differential equations. Crucially, at no point in our analysis have we relied on input from bulk physics to derive these equations. 

We provide a number of explicit examples of the master formula \eqref{eq:kin_flow} in the text. 
Additionally, \texttt{Mathematica} code for these and other examples (such as the five-site chain and the two-loop/cycle kite) can be found at the github repository \github\;  or in the ancillary files. 

A remarkable feature of our differential equations is that only acyclic minors related by a \emph{merger} couple to one another. As a consequence, the kinematic flow is naturally organized into exponentially many sectors, each with the structure of a graphical zonotope. While this zonotopal organization had been observed previously \cite{Baumann:2025qjx}, it was interpreted merely as a combinatorial artifact of the differential equations. In contrast, we uncover its true geometric origin: the zonotopes reside within the cosmological hyperplane arrangement, and their canonical forms compute residues of the physical FRW-form. Therefore, these zonotopes not only reflect the geometry underlying the differential equations but also govern the cut combinatorics---{\it flow of cuts}---of both the physical FRW-form and the cut basis itself. Thus, the combinatorics of the kinematic flow emerge as a natural consequence of the geometric and combinatorial structure of the flow of cuts.

At the level of the symbol, the kinematic flow controls which letters can be added to the end of the $\mathcal{O}(\alpha_\bullet^k)$ component of the symbol to generate the $\mathcal{O}(\alpha_\bullet^{k+1})$ component of the symbol for a physical FRW integral in the $\alpha_\bullet \to 0$ expansion. 
On the other hand, the flow of cuts controls the sequential discontinuities of the physical FRW integrals.
At the level of the symbol, a discontinuity annihilates all terms that do not have the compatible letter as their first entry. 
For terms with the compatible letter, the discontinuity simply removes the first entry. 
With these two handles on the analytic structure of FRW integrals, it would be interesting to look for symmetries (such as antipoldal duality) in the symbols of FRW integrals. 
Perhaps it is even possible to assign a cluster(-like) structure to the symbol. 

The differential equations derived here are relevant for a single Feynman graph contribution to the wavefunction coefficient. It would be interesting to consider differential equations for the wavefunction as a whole and to make connection to recently discovered geometries \cite{Arkani-Hamed:2024jbp,Glew:2025otn,Forcey:2025voc}.

It would also be interesting to understand how to the story changes when the scalar field has a generic mass (i.e., not conformally coupled). Do zonotopes play an important role there too? 
Could the cut-first approach help make an invariant definition of the physical subspace in this case? 
Which combinatorial parts of the differential equations survive after loop integration, if any?
We leave these any many other interesting questions for future research.

\section*{Acknowledgments}
RG would like to thank Arnau Padrol, Stefan Forcey and Tomasz Lukowski. 
AP would like to thank D. Baumann,H. Goodhew, H. Lee,  A. McLeod, and L. Ren for useful discussions.
This work was supported in part by the US Department of Energy under contract DE-SC0010010 Task F.

\appendix

\section{The two-cycle}
\label{app:twoCycle}
In this appendix, we explicitly provide the cut basis and their differential equations. 
The cut basis for the two-cycle is 
\begin{align}
\phi_{\begin{tikzpicture}[scale=0.6]
\draw[thick,dotted] (0,-1) to[out=90,in=90] (1,-1);
\draw[thick,dotted] (0,-1) to[out=-90,in=180+90] (1,-1);
\fill[black] (0,-1) circle (2.5pt);
\fill[black] (1,-1) circle (2.5pt);
\end{tikzpicture}}& = \dlog B_{\raisebox{-0.18cm}{\begin{tikzpicture}[scale=0.6]
\draw[thick] (0,-1) to[out=90,in=90] (1,-1);
\draw[thick] (0,-1) to[out=-90,in=180+90] (1,-1);
\fill[black] (0,-1) circle (2.5pt);
\fill[black] (1,-1) circle (2.5pt);
\draw[black,thick] (0,-1) circle (5pt);
\end{tikzpicture}}} \wedge \dlog B_{\raisebox{-0.18cm}{\begin{tikzpicture}[scale=0.6]
\draw[thick] (0,-1) to[out=90,in=90] (1,-1);
\draw[thick] (0,-1) to[out=-90,in=180+90] (1,-1);
\fill[black] (0,-1) circle (2.5pt);
\fill[black] (1,-1) circle (2.5pt);
\draw[black,thick] (1,-1) circle (5pt);
\end{tikzpicture}}}, && \phi_{\begin{tikzpicture}[scale=0.6]
\draw[thick,double] (0,-1) to[out=90,in=90] (1,-1);
\draw[thick,dotted] (0,-1) to[out=-90,in=180+90] (1,-1);
\fill[black] (0,-1) circle (2.5pt);
\fill[black] (1,-1) circle (2.5pt);
\end{tikzpicture}} = \dlog B_{\raisebox{-0.18cm}{\begin{tikzpicture}[scale=0.6]
\fill[black] (0,-1) circle (2.5pt);
\fill[black] (1,-1) circle (2.5pt);
\draw[thick] (0,-1) to[out=90,in=90] (1,-1);
\draw[thick] (0,-1) to[out=-90,in=180+90] (1,-1);
\draw[thick] (0-0.2,-1) to[out=90,in=90] (1+0.2,-1);
\draw[thick] (0+0.2,-1) to[out=90,in=90] (1-0.2,-1);
\draw[thick] (1-0.2,-1) to[out=90+180,in=180] (1,-1.2) to[out=0,in=90+180] (1+0.2,-1);
\draw[thick] (0-0.2,-1) to[out=90+180,in=180] (0,-1.2) to[out=0,in=90+180] (0+0.2,-1);
\end{tikzpicture}}} \wedge \dlog \frac{x_2}{x_1}, \notag \\
 \phi_{\begin{tikzpicture}[scale=0.6]
\draw[thick] (0,-1) to[out=90,in=90] node {\ar} (1,-1);
\draw[thick,dotted] (0,-1) to[out=-90,in=180+90] (1,-1);
\fill[black] (0,-1) circle (2.5pt);
\fill[black] (1,-1) circle (2.5pt);
\end{tikzpicture}} &=    \dlog B_{\raisebox{-0.18cm}{\begin{tikzpicture}[scale=0.6]
\draw[thick] (0,-1) to[out=90,in=90] (1,-1);
\draw[thick] (0,-1) to[out=-90,in=180+90] (1,-1);
\fill[black] (0,-1) circle (2.5pt);
\fill[black] (1,-1) circle (2.5pt);
\draw[black,thick] (0,-1) circle (5pt);
\end{tikzpicture}}} \wedge \dlog B_{\raisebox{-0.18cm}{\begin{tikzpicture}[scale=0.6]
\fill[black] (0,-1) circle (2.5pt);
\fill[black] (1,-1) circle (2.5pt);
\draw[thick] (0,-1) to[out=90,in=90] (1,-1);
\draw[thick] (0,-1) to[out=-90,in=180+90] (1,-1);
\draw[thick] (0-0.2,-1) to[out=90,in=90] (1+0.2,-1);
\draw[thick] (0+0.2,-1) to[out=90,in=90] (1-0.2,-1);
\draw[thick] (1-0.2,-1) to[out=90+180,in=180] (1,-1.2) to[out=0,in=90+180] (1+0.2,-1);
\draw[thick] (0-0.2,-1) to[out=90+180,in=180] (0,-1.2) to[out=0,in=90+180] (0+0.2,-1);
\end{tikzpicture}}}, &&   \phi_{\begin{tikzpicture}[scale=0.6]
\draw[thick] (0,-1) to[out=90,in=90] node {\al} (1,-1);
\draw[thick,dotted] (0,-1) to[out=-90,in=180+90] (1,-1);
\fill[black] (0,-1) circle (2.5pt);
\fill[black] (1,-1) circle (2.5pt);
\end{tikzpicture}} = \dlog B_{\raisebox{-0.18cm}{\begin{tikzpicture}[scale=0.6]
\fill[black] (0,-1) circle (2.5pt);
\fill[black] (1,-1) circle (2.5pt);
\draw[thick] (0,-1) to[out=90,in=90] (1,-1);
\draw[thick] (0,-1) to[out=-90,in=180+90] (1,-1);
\draw[thick] (0-0.2,-1) to[out=90,in=90] (1+0.2,-1);
\draw[thick] (0+0.2,-1) to[out=90,in=90] (1-0.2,-1);
\draw[thick] (1-0.2,-1) to[out=90+180,in=180] (1,-1.2) to[out=0,in=90+180] (1+0.2,-1);
\draw[thick] (0-0.2,-1) to[out=90+180,in=180] (0,-1.2) to[out=0,in=90+180] (0+0.2,-1);
\end{tikzpicture}}} \wedge \dlog B_{\raisebox{-0.18cm}{\begin{tikzpicture}[scale=0.6]
\draw[thick] (0,-1) to[out=90,in=90] (1,-1);
\draw[thick] (0,-1) to[out=-90,in=180+90] (1,-1);
\fill[black] (0,-1) circle (2.5pt);
\fill[black] (1,-1) circle (2.5pt);
\draw[black,thick] (1,-1) circle (5pt);
\end{tikzpicture}}}, \notag \\
\phi_{\begin{tikzpicture}[scale=0.6]
\draw[thick,dotted] (0,-1) to[out=90,in=90] (1,-1);
\draw[thick,double] (0,-1) to[out=-90,in=180+90] (1,-1);
\fill[black] (0,-1) circle (2.5pt);
\fill[black] (1,-1) circle (2.5pt);
\end{tikzpicture}} &=  \dlog B_{\raisebox{-0.28cm}{
\begin{tikzpicture}[scale=0.6]
\fill[black] (0,-1) circle (2.5pt);
\fill[black] (1,-1) circle (2.5pt);
\draw[thick] (0,-1) to[out=90,in=90] (1,-1);
\draw[thick] (0,-1) to[out=-90,in=180+90] (1,-1);
\draw[thick] (0-0.2,-1) to[out=-90,in=180+90] (1+0.2,-1);
\draw[thick] (0+0.2,-1) to[out=-90,in=180+90] (1-0.2,-1);
\draw[thick] (1-0.2,-1) to[out=90,in=180] (1,-0.8) to[out=0,in=90] (1+0.2,-1);
\draw[thick] (0-0.2,-1) to[out=90,in=180] (0,-0.8) to[out=0,in=90] (0+0.2,-1);
\end{tikzpicture}}} \wedge \dlog \frac{x_2}{x_1}, && \phi_{\begin{tikzpicture}[scale=0.6]
\draw[thick,dotted] (0,-1) to[out=90,in=90] (1,-1);
\draw[thick] (0,-1) to[out=-90,in=180+90] node {\ar} (1,-1);
\fill[black] (0,-1) circle (2.5pt);
\fill[black] (1,-1) circle (2.5pt);
\end{tikzpicture}} =   \dlog  B_{\raisebox{-0.18cm}{\begin{tikzpicture}[scale=0.6]
\draw[thick] (0,-1) to[out=90,in=90] (1,-1);
\draw[thick] (0,-1) to[out=-90,in=180+90] (1,-1);
\fill[black] (0,-1) circle (2.5pt);
\fill[black] (1,-1) circle (2.5pt);
\draw[black,thick] (0,-1) circle (5pt);
\end{tikzpicture}}} \wedge \dlog B_{\raisebox{-0.28cm}{
\begin{tikzpicture}[scale=0.6]
\fill[black] (0,-1) circle (2.5pt);
\fill[black] (1,-1) circle (2.5pt);
\draw[thick] (0,-1) to[out=90,in=90] (1,-1);
\draw[thick] (0,-1) to[out=-90,in=180+90] (1,-1);
\draw[thick] (0-0.2,-1) to[out=-90,in=180+90] (1+0.2,-1);
\draw[thick] (0+0.2,-1) to[out=-90,in=180+90] (1-0.2,-1);
\draw[thick] (1-0.2,-1) to[out=90,in=180] (1,-0.8) to[out=0,in=90] (1+0.2,-1);
\draw[thick] (0-0.2,-1) to[out=90,in=180] (0,-0.8) to[out=0,in=90] (0+0.2,-1);
\end{tikzpicture}}}, \notag \\
 \phi_{\begin{tikzpicture}[scale=0.6]
\draw[thick,dotted] (0,-1) to[out=90,in=90] (1,-1);
\draw[thick] (0,-1) to[out=-90,in=180+90] node {\al} (1,-1);
\fill[black] (0,-1) circle (2.5pt);
\fill[black] (1,-1) circle (2.5pt);
\end{tikzpicture}} &=  \dlog B_{\raisebox{-0.28cm}{
\begin{tikzpicture}[scale=0.6]
\fill[black] (0,-1) circle (2.5pt);
\fill[black] (1,-1) circle (2.5pt);
\draw[thick] (0,-1) to[out=90,in=90] (1,-1);
\draw[thick] (0,-1) to[out=-90,in=180+90] (1,-1);
\draw[thick] (0-0.2,-1) to[out=-90,in=180+90] (1+0.2,-1);
\draw[thick] (0+0.2,-1) to[out=-90,in=180+90] (1-0.2,-1);
\draw[thick] (1-0.2,-1) to[out=90,in=180] (1,-0.8) to[out=0,in=90] (1+0.2,-1);
\draw[thick] (0-0.2,-1) to[out=90,in=180] (0,-0.8) to[out=0,in=90] (0+0.2,-1);
\end{tikzpicture}}} \wedge  \dlog B_{\raisebox{-0.18cm}{\begin{tikzpicture}[scale=0.6]
\draw[thick] (0,-1) to[out=90,in=90] (1,-1);
\draw[thick] (0,-1) to[out=-90,in=180+90] (1,-1);
\fill[black] (0,-1) circle (2.5pt);
\fill[black] (1,-1) circle (2.5pt);
\draw[black,thick] (1,-1) circle (5pt);
\end{tikzpicture}}},  && \phi_{\begin{tikzpicture}[scale=0.6]
\draw[thick,double] (0,-1) to[out=90,in=90] (1,-1);
\draw[thick,double] (0,-1) to[out=-90,in=180+90] (1,-1);
\fill[black] (0,-1) circle (2.5pt);
\fill[black] (1,-1) circle (2.5pt);
\end{tikzpicture}} = \dlog B_{\raisebox{-0.26cm}{\begin{tikzpicture}[scale=0.6]
\fill[black] (0,-1) circle (2pt);
\fill[black] (1,-1) circle (2pt);
\draw[thick] (0,-1) to[out=90,in=90] (1,-1);
\draw[thick] (0,-1) to[out=-90,in=180+90] (1,-1);
\draw[thick, black] (0.5,-1) ellipse (0.8cm and 0.5cm);
\end{tikzpicture}}} \wedge \dlog  \frac{x_2}{x_1}, \notag \\
\phi_{\begin{tikzpicture}[scale=0.6]
\draw[thick] (0,-1) to[out=90,in=90] node {\ar} (1,-1);
\draw[thick] (0,-1) to[out=-90,in=180+90] node {\ar} (1,-1);
\fill[black] (0,-1) circle (2.5pt);
\fill[black] (1,-1) circle (2.5pt);
\end{tikzpicture}} &= \dlog B_{\raisebox{-0.18cm}{\begin{tikzpicture}[scale=0.6]
\draw[thick] (0,-1) to[out=90,in=90] (1,-1);
\draw[thick] (0,-1) to[out=-90,in=180+90] (1,-1);
\fill[black] (0,-1) circle (2.5pt);
\fill[black] (1,-1) circle (2.5pt);
\draw[black,thick] (0,-1) circle (5pt);
\end{tikzpicture}}} \wedge \dlog B_{\raisebox{-0.26cm}{\begin{tikzpicture}[scale=0.6]
\fill[black] (0,-1) circle (2pt);
\fill[black] (1,-1) circle (2pt);
\draw[thick] (0,-1) to[out=90,in=90] (1,-1);
\draw[thick] (0,-1) to[out=-90,in=180+90] (1,-1);
\draw[thick, black] (0.5,-1) ellipse (0.8cm and 0.5cm);
\end{tikzpicture}}}, && \phi_{\begin{tikzpicture}[scale=0.6]
\draw[thick] (0,-1) to[out=90,in=90] node {\al} (1,-1);
\draw[thick] (0,-1) to[out=-90,in=180+90] node {\al} (1,-1);
\fill[black] (0,-1) circle (2.5pt);
\fill[black] (1,-1) circle (2.5pt);
\end{tikzpicture}} =  \dlog B_{\raisebox{-0.26cm}{\begin{tikzpicture}[scale=0.6]
\fill[black] (0,-1) circle (2pt);
\fill[black] (1,-1) circle (2pt);
\draw[thick] (0,-1) to[out=90,in=90] (1,-1);
\draw[thick] (0,-1) to[out=-90,in=180+90] (1,-1);
\draw[thick, black] (0.5,-1) ellipse (0.8cm and 0.5cm);
\end{tikzpicture}}} \wedge  \dlog B_{\raisebox{-0.18cm}{\begin{tikzpicture}[scale=0.6]
\draw[thick] (0,-1) to[out=90,in=90] (1,-1);
\draw[thick] (0,-1) to[out=-90,in=180+90] (1,-1);
\fill[black] (0,-1) circle (2.5pt);
\fill[black] (1,-1) circle (2.5pt);
\draw[black,thick] (1,-1) circle (5pt);
\end{tikzpicture}}}.
\label{eq:bub_basis}
\end{align}
The differential equations decouple into $2^2$ blocks:
\begin{align}
 \d_\kin \phi_{\begin{tikzpicture}[scale=0.6]
\draw[thick,dotted] (0,-1) to[out=90,in=90] (1,-1);
\draw[thick,dotted] (0,-1) to[out=-90,in=180+90] (1,-1);
\fill[black] (0,-1) circle (2.5pt);
\fill[black] (1,-1) circle (2.5pt);
\end{tikzpicture}} \simeq \left( \alpha_1 \raisebox{-0.18cm}{\begin{tikzpicture}[scale=0.7]
\draw[thick,dotted] (0,-1) to[out=90,in=90] (1,-1);
\draw[thick,dotted] (0,-1) to[out=-90,in=180+90] (1,-1);
\fill[red] (0,-1) circle (2.5pt);
\fill[black] (1,-1) circle (2.5pt);
\end{tikzpicture}}+\alpha_2 \raisebox{-0.18cm}{\begin{tikzpicture}[scale=0.7]
\draw[thick,dotted] (0,-1) to[out=90,in=90] (1,-1);
\draw[thick,dotted] (0,-1) to[out=-90,in=180+90] (1,-1);
\fill[black] (0,-1) circle (2.5pt);
\fill[red] (1,-1) circle (2.5pt);
\end{tikzpicture}}\right) \phi_{\begin{tikzpicture}[scale=0.6]
\draw[thick,dotted] (0,-1) to[out=90,in=90] (1,-1);
\draw[thick,dotted] (0,-1) to[out=-90,in=180+90] (1,-1);
\fill[black] (0,-1) circle (2.5pt);
\fill[black] (1,-1) circle (2.5pt);
\end{tikzpicture}},
\end{align}
together with
\begin{align}
\d_\kin \left( 
\begin{matrix} 
\phi_{\begin{tikzpicture}[scale=0.5]
\draw[thick] (0,-1) to[out=90,in=90] node {\ar} (1,-1);
\draw[thick,dotted] (0,-1) to[out=-90,in=180+90] (1,-1);
\fill[black] (0,-1) circle (2.5pt);
\fill[black] (1,-1) circle (2.5pt);
\end{tikzpicture}} \\ \phi_{\begin{tikzpicture}[scale=0.5]
\draw[thick] (0,-1) to[out=90,in=90] node {\al} (1,-1);
\draw[thick,dotted] (0,-1) to[out=-90,in=180+90] (1,-1);
\fill[black] (0,-1) circle (2.5pt);
\fill[black] (1,-1) circle (2.5pt);
\end{tikzpicture}} \\ \phi_{\begin{tikzpicture}[scale=0.5]
\draw[thick,double] (0,-1) to[out=90,in=90] (1,-1);
\draw[thick,dotted] (0,-1) to[out=-90,in=180+90] (1,-1);
\fill[black] (0,-1) circle (2.5pt);
\fill[black] (1,-1) circle (2.5pt);
\end{tikzpicture}}
\end{matrix}
\right) &\simeq\left(\begin{matrix}
\alpha_1 \raisebox{-0.2cm}{\begin{tikzpicture}[scale=0.7]
\draw[thick] (0,-1) to[out=90,in=90] node{\ar} (1,-1);
\draw[thick,dotted] (0,-1) to[out=-90,in=180+90] (1,-1);
\fill[black,red] (0,-1) circle (2.5pt);
\fill[black] (1,-1) circle (2.5pt);
\end{tikzpicture}}+\alpha_2\raisebox{-0.2cm}{\begin{tikzpicture}[scale=0.7]
\draw[thick,red] (0,-1) to[out=90,in=90] node{\ar} (1,-1);
\draw[thick,dotted] (0,-1) to[out=-90,in=180+90] (1,-1);
\fill[black] (0,-1) circle (2.5pt);
\fill[black,red] (1,-1) circle (2.5pt);
\end{tikzpicture}} & 0 & \frac{\alpha_1 \alpha_2}{\alpha_1+\alpha_2} ( \raisebox{-0.2cm}{\begin{tikzpicture}[scale=0.7]
\draw[thick,red] (0,-1) to[out=90,in=90] node{\ar} (1,-1);
\draw[thick,dotted] (0,-1) to[out=-90,in=180+90] (1,-1);
\fill[black] (0,-1) circle (2.5pt);
\fill[black,red] (1,-1) circle (2.5pt);
\end{tikzpicture}}-\raisebox{-0.2cm}{\begin{tikzpicture}[scale=0.7]
\draw[thick] (0,-1) to[out=90,in=90] node{\ar} (1,-1);
\draw[thick,dotted] (0,-1) to[out=-90,in=180+90] (1,-1);
\fill[black,red] (0,-1) circle (2.5pt);
\fill[black] (1,-1) circle (2.5pt);
\end{tikzpicture}}) \\
0 & \alpha_1\raisebox{-0.2cm}{\begin{tikzpicture}[scale=0.7]
\draw[thick,red] (0,-1) to[out=90,in=90] node{\al} (1,-1);
\draw[thick,dotted] (0,-1) to[out=-90,in=180+90] (1,-1);
\fill[black,red] (0,-1) circle (2.5pt);
\fill[black] (1,-1) circle (2.5pt);
\end{tikzpicture}}+\alpha_2\raisebox{-0.2cm}{\begin{tikzpicture}[scale=0.7]
\draw[thick] (0,-1) to[out=90,in=90] node{\al} (1,-1);
\draw[thick,dotted] (0,-1) to[out=-90,in=180+90] (1,-1);
\fill[black] (0,-1) circle (2.5pt);
\fill[black,red] (1,-1) circle (2.5pt);
\end{tikzpicture}} & \frac{\alpha_1 \alpha_2}{\alpha_1+\alpha_2}(\raisebox{-0.2cm}{\begin{tikzpicture}[scale=0.7]
\draw[thick,red] (0,-1) to[out=90,in=90] node{\al} (1,-1);
\draw[thick,dotted] (0,-1) to[out=-90,in=180+90] (1,-1);
\fill[black,red] (0,-1) circle (2.5pt);
\fill[black] (1,-1) circle (2.5pt);
\end{tikzpicture}}-\raisebox{-0.2cm}{\begin{tikzpicture}[scale=0.7]
\draw[thick] (0,-1) to[out=90,in=90] node{\al} (1,-1);
\draw[thick,dotted] (0,-1) to[out=-90,in=180+90] (1,-1);
\fill[black] (0,-1) circle (2.5pt);
\fill[black,red] (1,-1) circle (2.5pt);
\end{tikzpicture}}) \\
0 & 0 & \left(\alpha_1 +\alpha_2\right) \raisebox{-0.2cm}{\begin{tikzpicture}[scale=0.7]
\draw[thick,double,red] (0,-1) to[out=90,in=90] (1,-1);
\draw[thick,dotted] (0,-1) to[out=-90,in=180+90] (1,-1);
\fill[red] (0,-1) circle (2.5pt);
\fill[red] (1,-1) circle (2.5pt);
\end{tikzpicture}}
\end{matrix} \right) \left( 
\begin{matrix} 
\phi_{\begin{tikzpicture}[scale=0.5]
\draw[thick] (0,-1) to[out=90,in=90] node {\ar} (1,-1);
\draw[thick,dotted] (0,-1) to[out=-90,in=180+90] (1,-1);
\fill[black] (0,-1) circle (2.5pt);
\fill[black] (1,-1) circle (2.5pt);
\end{tikzpicture}} \\ \phi_{\begin{tikzpicture}[scale=0.5]
\draw[thick] (0,-1) to[out=90,in=90] node {\al} (1,-1);
\draw[thick,dotted] (0,-1) to[out=-90,in=180+90] (1,-1);
\fill[black] (0,-1) circle (2.5pt);
\fill[black] (1,-1) circle (2.5pt);
\end{tikzpicture}} \\ \phi_{\begin{tikzpicture}[scale=0.5]
\draw[thick,double] (0,-1) to[out=90,in=90] (1,-1);
\draw[thick,dotted] (0,-1) to[out=-90,in=180+90] (1,-1);
\fill[black] (0,-1) circle (2.5pt);
\fill[black] (1,-1) circle (2.5pt);
\end{tikzpicture}}
\end{matrix}
\right), \notag \\
\d_\kin \left( 
\begin{matrix} 
\phi_{\begin{tikzpicture}[scale=0.5]
\draw[thick,dotted] (0,-1) to[out=90,in=90]  (1,-1);
\draw[thick] (0,-1) to[out=-90,in=180+90] node {\ar} (1,-1);
\fill[black] (0,-1) circle (2.5pt);
\fill[black] (1,-1) circle (2.5pt);
\end{tikzpicture}} \\ \phi_{\begin{tikzpicture}[scale=0.5]
\draw[thick,dotted] (0,-1) to[out=90,in=90] (1,-1);
\draw[thick] (0,-1) to[out=-90,in=180+90] node {\al} (1,-1);
\fill[black] (0,-1) circle (2.5pt);
\fill[black] (1,-1) circle (2.5pt);
\end{tikzpicture}} \\ \phi_{\begin{tikzpicture}[scale=0.5]
\draw[thick,dotted] (0,-1) to[out=90,in=90] (1,-1);
\draw[thick,double] (0,-1) to[out=-90,in=180+90] (1,-1);
\fill[black] (0,-1) circle (2.5pt);
\fill[black] (1,-1) circle (2.5pt);
\end{tikzpicture}}
\end{matrix}
\right) &\simeq\left(\begin{matrix}
\alpha_1\raisebox{-0.3cm}{\begin{tikzpicture}[scale=0.7]
\draw[thick,dotted] (0,-1) to[out=90,in=90] (1,-1);
\draw[thick] (0,-1) to[out=-90,in=180+90] node {\ar} (1,-1);
\fill[black,red] (0,-1) circle (2.5pt);
\fill[black] (1,-1) circle (2.5pt);
\end{tikzpicture}}+\alpha_2\raisebox{-0.3cm}{\begin{tikzpicture}[scale=0.7]
\draw[thick,dotted] (0,-1) to[out=90,in=90] (1,-1);
\draw[thick,red] (0,-1) to[out=-90,in=180+90] node {\ar} (1,-1);
\fill[black] (0,-1) circle (2.5pt);
\fill[black,red] (1,-1) circle (2.5pt);
\end{tikzpicture}} & 0 & \frac{\alpha_1 \alpha_2}{\alpha_1 + \alpha_2} ( \raisebox{-0.3cm}{\begin{tikzpicture}[scale=0.7]
\draw[thick,dotted] (0,-1) to[out=90,in=90] (1,-1);
\draw[thick,red] (0,-1) to[out=-90,in=180+90] node {\ar} (1,-1);
\fill[black] (0,-1) circle (2.5pt);
\fill[black,red] (1,-1) circle (2.5pt);
\end{tikzpicture}}-\raisebox{-0.3cm}{\begin{tikzpicture}[scale=0.7]
\draw[thick,dotted] (0,-1) to[out=90,in=90] (1,-1);
\draw[thick] (0,-1) to[out=-90,in=180+90] node {\ar} (1,-1);
\fill[black,red] (0,-1) circle (2.5pt);
\fill[black] (1,-1) circle (2.5pt);
\end{tikzpicture}}) \\
0 & \alpha_1\raisebox{-0.3cm}{\begin{tikzpicture}[scale=0.7]
\draw[thick,dotted] (0,-1) to[out=90,in=90] (1,-1);
\draw[thick,red] (0,-1) to[out=-90,in=180+90] node {\al} (1,-1);
\fill[black,red] (0,-1) circle (2.5pt);
\fill[black] (1,-1) circle (2.5pt);
\end{tikzpicture}}+ \alpha_2 \raisebox{-0.3cm}{\begin{tikzpicture}[scale=0.7]
\draw[thick,dotted] (0,-1) to[out=90,in=90] (1,-1);
\draw[thick] (0,-1) to[out=-90,in=180+90] node {\al} (1,-1);
\fill[black] (0,-1) circle (2.5pt);
\fill[black,red] (1,-1) circle (2.5pt);
\end{tikzpicture}} & \frac{\alpha_1 \alpha_2}{\alpha_1 + \alpha_2}(\raisebox{-0.3cm}{\begin{tikzpicture}[scale=0.7]
\draw[thick,dotted] (0,-1) to[out=90,in=90] (1,-1);
\draw[thick,red] (0,-1) to[out=-90,in=180+90] node {\al} (1,-1);
\fill[black,red] (0,-1) circle (2.5pt);
\fill[black] (1,-1) circle (2.5pt);
\end{tikzpicture}}-\raisebox{-0.3cm}{\begin{tikzpicture}[scale=0.7]
\draw[thick,dotted] (0,-1) to[out=90,in=90] (1,-1);
\draw[thick] (0,-1) to[out=-90,in=180+90] node {\al} (1,-1);
\fill[black] (0,-1) circle (2.5pt);
\fill[black,red] (1,-1) circle (2.5pt);
\end{tikzpicture}}) \\
0 & 0 & \left( \alpha_1 + \alpha_2 \right) \raisebox{-0.2cm}{\begin{tikzpicture}[scale=0.7]
\draw[thick,dotted] (0,-1) to[out=90,in=90] (1,-1);
\draw[thick,double,red] (0,-1) to[out=-90,in=180+90] (1,-1);
\fill[red] (0,-1) circle (2.5pt);
\fill[red] (1,-1) circle (2.5pt);
\end{tikzpicture}}
\end{matrix} \right)
\left( 
\begin{matrix} 
\phi_{\begin{tikzpicture}[scale=0.5]
\draw[thick,dotted] (0,-1) to[out=90,in=90]  (1,-1);
\draw[thick] (0,-1) to[out=-90,in=180+90] node {\ar} (1,-1);
\fill[black] (0,-1) circle (2.5pt);
\fill[black] (1,-1) circle (2.5pt);
\end{tikzpicture}} \\ \phi_{\begin{tikzpicture}[scale=0.5]
\draw[thick,dotted] (0,-1) to[out=90,in=90] (1,-1);
\draw[thick] (0,-1) to[out=-90,in=180+90] node {\al} (1,-1);
\fill[black] (0,-1) circle (2.5pt);
\fill[black] (1,-1) circle (2.5pt);
\end{tikzpicture}} \\ \phi_{\begin{tikzpicture}[scale=0.5]
\draw[thick,dotted] (0,-1) to[out=90,in=90] (1,-1);
\draw[thick,double] (0,-1) to[out=-90,in=180+90] (1,-1);
\fill[black] (0,-1) circle (2.5pt);
\fill[black] (1,-1) circle (2.5pt);
\end{tikzpicture}}
\end{matrix}
\right),
\end{align}
and
\begin{align}
\d\kin \left( 
\begin{matrix} 
\phi_{\begin{tikzpicture}[scale=0.5]
\draw[thick] (0,-1) to[out=90,in=90] node {\ar} (1,-1);
\draw[thick] (0,-1) to[out=-90,in=180+90] node {\ar} (1,-1);
\fill[black] (0,-1) circle (2.5pt);
\fill[black] (1,-1) circle (2.5pt);
\end{tikzpicture}} \\ \phi_{\begin{tikzpicture}[scale=0.5]
\draw[thick] (0,-1) to[out=90,in=90] node {\al} (1,-1);
\draw[thick] (0,-1) to[out=-90,in=180+90] node {\al} (1,-1);
\fill[black] (0,-1) circle (2.5pt);
\fill[black] (1,-1) circle (2.5pt);
\end{tikzpicture}} \\ 
\phi_{\begin{tikzpicture}[scale=0.5]
\draw[thick,double] (0,-1) to[out=90,in=90] (1,-1);
\draw[thick,double] (0,-1) to[out=-90,in=180+90] (1,-1);
\fill[black] (0,-1) circle (2.5pt);
\fill[black] (1,-1) circle (2.5pt);
\end{tikzpicture}}
\end{matrix}
\right) &\simeq\left(\begin{matrix}
\alpha_1\raisebox{-0.3cm}{\begin{tikzpicture}[scale=0.7]
\draw[thick] (0,-1) to[out=90,in=90] node{\ar} (1,-1);
\draw[thick] (0,-1) to[out=-90,in=180+90] node {\ar} (1,-1);
\fill[black,red] (0,-1) circle (2.5pt);
\fill[black] (1,-1) circle (2.5pt);
\end{tikzpicture}}+\alpha_2\raisebox{-0.3cm}{\begin{tikzpicture}[scale=0.7]
\draw[thick,red] (0,-1) to[out=90,in=90] node{\ar} (1,-1);
\draw[thick,red] (0,-1) to[out=-90,in=180+90] node {\ar} (1,-1);
\fill[black] (0,-1) circle (2.5pt);
\fill[black,red] (1,-1) circle (2.5pt);
\end{tikzpicture}} & 0 & \frac{\alpha_1 \alpha_2}{\alpha_1+\alpha_2} ( \raisebox{-0.3cm}{\begin{tikzpicture}[scale=0.7]
\draw[thick,red] (0,-1) to[out=90,in=90] node{\ar} (1,-1);
\draw[thick,red] (0,-1) to[out=-90,in=180+90] node {\ar} (1,-1);
\fill[black] (0,-1) circle (2.5pt);
\fill[black,red] (1,-1) circle (2.5pt);
\end{tikzpicture}}-\raisebox{-0.3cm}{\begin{tikzpicture}[scale=0.7]
\draw[thick] (0,-1) to[out=90,in=90] node{\ar} (1,-1);
\draw[thick] (0,-1) to[out=-90,in=180+90] node {\ar} (1,-1);
\fill[black,red] (0,-1) circle (2.5pt);
\fill[black] (1,-1) circle (2.5pt);
\end{tikzpicture}}) \\
0 & \alpha_1\raisebox{-0.3cm}{\begin{tikzpicture}[scale=0.7]
\draw[thick,red] (0,-1) to[out=90,in=90] node{\al} (1,-1);
\draw[thick,red] (0,-1) to[out=-90,in=180+90] node {\al} (1,-1);
\fill[black,red] (0,-1) circle (2.5pt);
\fill[black] (1,-1) circle (2.5pt);
\end{tikzpicture}}+\alpha_2 \raisebox{-0.3cm}{\begin{tikzpicture}[scale=0.7]
\draw[thick] (0,-1) to[out=90,in=90] node{\al} (1,-1);
\draw[thick] (0,-1) to[out=-90,in=180+90] node {\al} (1,-1);
\fill[black] (0,-1) circle (2.5pt);
\fill[black,red] (1,-1) circle (2.5pt);
\end{tikzpicture}} & \frac{\alpha_1 \alpha_2}{\alpha_1+\alpha_2}(\raisebox{-0.3cm}{\begin{tikzpicture}[scale=0.7]
\draw[thick,red] (0,-1) to[out=90,in=90] node{\al} (1,-1);
\draw[thick,red] (0,-1) to[out=-90,in=180+90] node {\al} (1,-1);
\fill[black,red] (0,-1) circle (2.5pt);
\fill[black] (1,-1) circle (2.5pt);
\end{tikzpicture}}-\raisebox{-0.3cm}{\begin{tikzpicture}[scale=0.7]
\draw[thick] (0,-1) to[out=90,in=90] node{\al} (1,-1);
\draw[thick] (0,-1) to[out=-90,in=180+90] node {\al} (1,-1);
\fill[black] (0,-1) circle (2.5pt);
\fill[black,red] (1,-1) circle (2.5pt);
\end{tikzpicture}}) \\
0 & 0 & \left( \alpha_1 +\alpha_2 \right) \raisebox{-0.2cm}{\begin{tikzpicture}[scale=0.7]
\draw[thick,double,red] (0,-1) to[out=90,in=90] (1,-1);
\draw[thick,double,red] (0,-1) to[out=-90,in=180+90] (1,-1);
\fill[red] (0,-1) circle (2.5pt);
\fill[red] (1,-1) circle (2.5pt);
\end{tikzpicture}}
\end{matrix} \right) 
\left( 
\begin{matrix} 
\phi_{\begin{tikzpicture}[scale=0.5]
\draw[thick] (0,-1) to[out=90,in=90] node {\ar} (1,-1);
\draw[thick] (0,-1) to[out=-90,in=180+90] node {\ar} (1,-1);
\fill[black] (0,-1) circle (2.5pt);
\fill[black] (1,-1) circle (2.5pt);
\end{tikzpicture}} \\ \phi_{\begin{tikzpicture}[scale=0.5]
\draw[thick] (0,-1) to[out=90,in=90] node {\al} (1,-1);
\draw[thick] (0,-1) to[out=-90,in=180+90] node {\al} (1,-1);
\fill[black] (0,-1) circle (2.5pt);
\fill[black] (1,-1) circle (2.5pt);
\end{tikzpicture}} \\ \phi_{\begin{tikzpicture}[scale=0.5]
\draw[thick,double] (0,-1) to[out=90,in=90] (1,-1);
\draw[thick,double] (0,-1) to[out=-90,in=180+90] (1,-1);
\fill[black] (0,-1) circle (2.5pt);
\fill[black] (1,-1) circle (2.5pt);
\end{tikzpicture}}
\end{matrix}
\right).
\end{align}

\section{Derivation of kinematic flow from intersection theory}
\label{app:AmatDets}

In this appendix, we derive closed form formulas for the intersection matrix (appendix \ref{app:intMat}) and the matrix elements of differential equation (appendix \ref{app:Adaig} and \ref{app:AmatOffDiag}).

\subsection{Intersection matrix}
\label{app:intMat}

In this appendix, we illustrate how to compute the intersection matrix \eqref{eq:intMat} from the localization formula \eqref{eq:intNum}. 
Recalling the definition of our basis forms and the fact that our basis of FRW- and dual-forms contain one cut (i.e., $\res_{\C_\g}[\phi_{\g^\prime}] \propto |\C_\g| \delta_{\g\g^\prime}$), 
\begin{align}
    \la \check{\phi}_{\mathfrak{g}} \vert \phi_{\mathfrak{g}^\prime} \ra 
    &= \res_{\lfrak_{|\mathbb{L}_\g|}} \circ \cdots \res_{\lfrak_1} 
    \circ \res_{\C_\mathfrak{g}}[
        \dlog\C_{\g^\prime} \wedge 
        \tilde{\Omega}_{\g^\prime}
    ]
    \,,
    \\
    &= |\C_\g| \delta_{\g\g^\prime}
    \res_{\lfrak_{|\mathbb{L}_\g|}} \circ \cdots \res_{\lfrak_1} [\Omega_\g]
    \,,
\end{align}
where $|\C_\g|$ is the number of tubings that belong to this cut; it will always cancel in the matrix elements of the DEQ.
Then, recalling that $\Omega_\g = \bigwedge_{\lfrak \in \mathbb{L}_\g} \Omega_\lfrak$,  
\begin{align}\label{eq:expandedOmega}
    \Omega_\lfrak &= 
    \bigwedge_{v\in V_\lfrak \setminus \overline{V}_\lfrak}
    \dlog \frac{x_v}{x_{\overline{V}_\lfrak}}
    \bigg\vert_{f_\lfrak(\mbf{x}+\mbf{X},\mbf{Y})=0}
    = 
    \sum_{v\in V_\lfrak}
    \mathrm{sig}(v,V_\lfrak\setminus v)
    \bigwedge_{u \in V_\lfrak\setminus v} 
    \dlog x_u
    \bigg\vert_{f_\lfrak(\mbf{x}+\mbf{X},\mbf{Y})=0}
    \,,
\end{align}
and the definition of $\res_\lfrak$ (equation \eqref{eq:resl}) yields
\vspace{-1em}
\be \label{eq:ReslOmegal}
    \res_{\lfrak}[\Omega_\lfrak] 
    &=
    \sum_{v,w \in V_\lfrak}
    \frac{
        \alpha_{v} 
        \mathrm{sig}(v,V_\lfrak {\setminus} v) 
        \mathrm{sig}(w,V_\lfrak {\setminus} w)
    }{
        \prod_{l\in V_\lfrak} \alpha_{l}
    }
    \overset{\delta_{vw}}{\overbrace{
        \underset{u \in V_\lfrak \setminus v}{\bigcirc}
        \res_{x_v=0}\left.
            {\bigwedge_{y \in V_\lfrak \setminus y}}
            \dlog x_y
        \right\vert_{f_\lfrak(\mbf{x}+\mbf{X},\mbf{Y})=0}
    }}
    \\
    &= \frac{
        \sum_{l\in V_\lfrak} \alpha_l 
    }{
        \prod_{l\in V_\lfrak} \alpha_l
    }
    \,.
\ee
Therefore, 
\begin{align}
    C_{\g\g^\prime} := 
    \la \check{\vphi}_{\mathfrak{g}} \vert \vphi_{\mathfrak{g}^\prime} \ra 
    = |\C_\g| \delta_{\mathfrak{g}\mathfrak{g}^\prime} \prod_{
        \lfrak \in \mathbb{L}_{\mathfrak{g}} 
    } 
    \res_{\lfrak}[\Omega_\lfrak] 
    = |\C_\g| \delta_{\mathfrak{g}\mathfrak{g}^\prime} \prod_{
        \lfrak \in \mathbb{L}_{\mathfrak{g}} 
    } 
    \frac{
        \sum_{l\in V_\lfrak} \alpha_l 
    }{
        \prod_{l\in V_\lfrak} \alpha_l
    }
    \,.
\end{align}

\subsection{The DEQ: diagonal elements}
\label{app:Adaig}

To derive equation \eqref{eq:AmatDiag}, note that
\begin{align}
    \label{eq:AdiagResC}
    &\res_{\C_\mathfrak{g}}
    \left[
        \omega \wedge \dlog\C_{\mathfrak{g}} \wedge \tilde\Omega_{\mathfrak{g}}
    \right]
    = (-1)^{|V_\g|} |\C_\g|\;   \Omega_\g \wedge \omega\vert_{\C_g}
    \,,
    \\
    &\omega\vert_{\C_g} = 
    \sum_{\lfrak \in \mathbb{L}_g}
    \left[\sum_{l\in V_\lfrak} \alpha_l\, \dlog x_l\right]_{f_\lfrak(\mbf{x}+\mbf{X},\mbf{Y})=0}
    := \sum_{\lfrak \in \mathbb{L}_g} \omega_\lfrak
    \,,
\end{align}
where we recall that $f_\lfrak$ only depends (linearly) on the $X_{l\in V_\lfrak}$: $f_\lfrak(\mbf{x}+\mbf{X},\mbf{Y}) = \sum_{l \in V_\lfrak} x_l + f_\lfrak(\mbf{X},\mbf{Y})$.
Collecting the $x_{l\in V_{\lfrak}}$-dependence and separating the residues while being careful of the sign generated by the wedge product yields
\begin{align}\begin{aligned}
    &\res_{\lfrak_{|\mathbb{L}_\g|}} \circ \cdots \circ \res_{\lfrak_1}[\Omega_\g \wedge \omega ] \Big\vert_{\C_g}
    \\
    &= 
    \sum_{i=1}^{|\mathbb{L}_\g|}\left[ 
    \left(
        \prod_{j\neq i} \res_{\lfrak_j}[\Omega_{\lfrak_j}]
        \bigg\vert_{f_{\lfrak_j}(\mbf{x}+\mbf{X},\mbf{Y})=0}
    \right)
    \left(
        \res_{\lfrak_i}[\Omega_{\lfrak_i} \wedge \omega_{\lfrak_i}]
        \bigg\vert_{f_{\lfrak_i}(\mbf{x}+\mbf{X},\mbf{Y})=0}
    \right)
    \right] \,.
    \label{eq:AdiagResT}
\end{aligned}\end{align}
The factors in the first brackets above were computed already in \eqref{eq:ReslOmegal}. 
To evaluate the remaining factors, note that 
\be
    \Omega_{\lfrak} \wedge \omega_{\lfrak}
    = \underset{(-1)^{|V_\lfrak|-1}}{\underbrace{
        \mathrm{sig}(\overline{V}_\lfrak, V_\lfrak\setminus\overline{V}_\lfrak)
        \mathrm{sig}(V_\lfrak\setminus\overline{V}_\lfrak, \overline{V}_\lfrak)
    }}
    \left(\sum_{l\in V_\lfrak} \alpha_l\right)
    \left(
        \bigwedge_{v\in V_\lfrak}
        \dlog x_{v} 
        \bigg\vert_{\mbf{x}+f_\lfrak(\mbf{X},\mbf{Y})=0}
    \right)
    \,.
\ee
Then, since each term of $\res_\lfrak$ sets all but one $x_{l_i \in V_\lfrak}$ to zero
\begin{align}
    \res_\lfrak[\Omega_\lfrak\wedge\omega_\lfrak]
    &= 
    \frac{
        (-1)^{|V_\lfrak|-1}
        \sum_{l\in V_\lfrak} \alpha_l
    }{
        \prod_{l\in V_\lfrak} \alpha_l
    }   
    \sum_{v\in V_\lfrak} \alpha_v 
    \mathrm{sig}(v,V_\lfrak\setminus v)
    \underset{u \in V_\lfrak \setminus v}{\bigcirc}
    \res_{x_v=0}\left.
        {\bigwedge_{w \in V_\lfrak}}
        \dlog x_w
    \right\vert_{f_\lfrak(\mbf{x}+\mbf{X},\mbf{Y})=0}
    \nn\\ 
    &= 
    \frac{
        (-1)^{|V_\lfrak|-1}
        \sum_{l\in V_\lfrak} \alpha_l
    }{
        \prod_{l\in V_\lfrak} \alpha_l
    }   
    \left.
    \sum_{v\in V_\lfrak} \alpha_v 
    \mathrm{sig}(v,V_\lfrak\setminus v)
    \mathrm{sig}(V_\lfrak\setminus v,v)
    \dlog x_v
    \right\vert_{f_\lfrak(\mbf{x}+\mbf{X},\mbf{Y})=0, x_{w\in V_\lfrak\setminus v}=0}
    \nn\\
    &= \frac{
        \sum_{l\in V_\lfrak} \alpha_l
    }{
        \prod_{l\in V_\lfrak} \alpha_l
    } 
    \left(\sum_{v\in V_\lfrak} \alpha_v \right)
    \dlog f_\lfrak(\mbf{X},\mbf{Y})
    \,.
\end{align}
Substituting into \eqref{eq:AdiagResT}, yields 
\begin{align}
    \eqref{eq:AdiagResT}
    = 
    \left[
        \prod_{\lfrak \in \mathbb{L}_\g|}
       \left(\frac{ 
        \sum_{l\in V_{\lfrak}} \alpha_l 
        }{
            \prod_{l\in V_{\lfrak}} \alpha_l 
        }\right)
    \right] 
    \left[
    \sum_{\lfrak \in \mathbb{L}_\g|}
    \left( {\textstyle \sum_{l\in V_\lfrak} } \alpha_l\right) \dlog f_\lfrak(\mbf{X},\mbf{Y})
    \right] 
    \,.
\end{align}
Combining \eqref{eq:Amat} with $\g = \g^\prime$, \eqref{eq:AdiagResT} and \eqref{eq:AdiagResC} yields \eqref{eq:AmatDiag}. 

\subsection{The DEQ: off-diagonal elements}
\label{app:AmatOffDiag}

Restoring the sign in \eqref{eq:AmatOffDiagWithSgn} yields
\be
    A_{\g\g^\prime}
    &{=} \frac{
        (-1)^{|V_G|}
    }{
        \la \check{\phi}_{\gp} 
        \vert \phi_{\gp} \ra
    }
    \sum_{(\tau,\tau^\prime) \in \C_{\g\g^\prime}}
    \hspace{-1em}
    \sgn_\tau \sgn_{\tau^\prime} \sgn_{\tau \tau^\prime}
    \res_{\lfrakp_{|\mathbb{L}_\gp|}} {\circ} 
    {\cdots} {\circ} \res_{\lfrakp_1} 
    \left[
       \dlog B_{\tau \setminus \tau^\prime} 
       {\wedge} \tilde\Omega_{\mathfrak{g}}
       {\wedge} \omega
    \right]
    \bigg\vert_{\mathcal{C}_{\g'}}
    \,,
    \\
    &{=} \frac{
        |\mathcal{C}_{\gp}|
    }{
        \la \check{\phi}_{\gp} 
        \vert \phi_{\gp} \ra
    }
    \sgn_{\g\gp} 
    \res_{\lfrakp_{|\mathbb{L}_\gp|}} {\circ} 
    {\cdots} {\circ} \res_{\lfrakp_1} 
    \left[
       \tilde\Omega_{\mathfrak{g}}
       {\wedge} 
       \omega
       {\wedge}
       \dlog B_{\g \setminus \g^\prime} 
    \right]
    \bigg\vert_{\mathcal{C}_{\g'}}
    \,,
\ee
where $\sgn_{\tau\tau'}$ compares the relative ordering of the tubes for a pair $(\tau,\tau^\prime)\in \mathcal{C}_{\g\gp}$, 
and $\sgn_{\g\g'}$ is the same for all $(\tau,\tau^\prime)\in\C_{\g\g^\prime}$.
Explicitly, 
\be
    \sgn_\tau \sgn_{\tau'} \sgn_{\tau\tau^\prime} 
    &:= \mathrm{sig}(\la \overline{V}_\g \ra_\tau)
    \mathrm{sig}(\la \overline{V}_{\g'} \ra_\tau,\la \overline{V}_{\g} \setminus \overline{V}_{\g'} \ra_\tau)
    \,,
    \\
    \sgn_{\g\gp} & := (-1)^{|\tau'|} \sgn_{\g} \sgn_{\g^\prime} \sgn_\tau \sgn_{\tau^\prime} \sgn_{\tau \tau^\prime}
    \,.
\ee   
Similarly, 
\be 
    \dlog B_{\g\setminus\gp}
    &:= 
    \dlog B_{\tau\setminus\tau^\prime}
    \bigg\vert_{B_{t\in\tau^\prime}=0}
    \,,
\ee
is also the same for all pairs $(\tau,\tau^\prime)\in\C_{\g\g^\prime}$. 
While we do not provide an explicit expression for this intermediate object, it has following properties that we be used later 
\be
    \dlog B_{\g\setminus\g'}\vert_{x_{i\in V_{\underline{\lfrak}_\uparrow}} = 0}
    &=  f_{\underline{\lfrak}_\uparrow}(\mbf{X},\mbf{Y})
    \,,
    &
    \dlog B_{\g\setminus\g'}\vert_{x_{i\in V_{\underline{\lfrak}_\downarrow}} = 0}
    &=  f_{\underline{\lfrak}_\downarrow}(\mbf{X},\mbf{Y})
    \,.
\ee
We have also anti-commuted $\dlog B_{\g\setminus\gp}$ to the end of the wedge product which produces a sign $(-1)^{|V_G|-|\tau'|}$ and exposes the canonical forms $\Omega_\lfrak$ on which $\res_{\lfrakp}$ acts. 

Then, performing the $\res_{\lfrak^\prime \in \mathbb{L}_{\g'}}$ yields%
\footnote{It is possible that $\lfrak_\uparrow$ and $\lfrak_\downarrow$ may not be adjacent in $\mathbb{L}_\g$ when ordered by their minimal vertex (as dictated in the definition of $\tilde\Omega_\g$), for a given labeling. 
In such cases, another sign could be generated
\be
    \sgn_{\uparrow\downarrow} 
    &= \mathrm{sig}(
    \overline{V}_{\mathbb{L}_\g}, 
    {V}_{\mathbb{L}_\g} \setminus \overline{V}_{\mathbb{L}_\g} 
    ) 
    \mathrm{sig}(
        \overline{V}_{\mathbb{L}_\g}, {V}_{\mathbb{L}_{\g'}^\uparrow} \setminus \overline{V}_{\mathbb{L}_{\g'}^\uparrow},
        V_{\lfrak_\uparrow} \setminus \overline{V}_{\lfrak_\uparrow},
        V_{\lfrak_\downarrow} \setminus \overline{V}_{\lfrak_\downarrow},
        {V}_{\mathbb{L}_{\g'}^\downarrow} \setminus \overline{V}_{\mathbb{L}_{\g'}^\downarrow}
    )
    \,,
    \\
    V_{\mathbb{L}_{\g'}^{\uparrow}} 
    &= (V_{\lfrak_1'^\uparrow}, \dots, V_{\lfrak_{|\mathbb{L}_{\g'|}}'^\uparrow})
    \,,
    \qquad
    \overline{V}_{\mathbb{L}_{\g'}^{\uparrow}} 
    = (\overline{V}_{\lfrak_1'^\uparrow}, \dots, \overline{V}_{\lfrak_{|\mathbb{L}_{\g'|}}'^\uparrow})
    \,.
\ee
Here, one exchanges $\uparrow$ with $\downarrow$ in the obvious way to get $V_{\mathbb{L}_{\g'}^{\downarrow}}$ and $\overline{V}_{\mathbb{L}_{\g'}^{\downarrow}}$. 
In all checked cases, $\sgn_{\uparrow\downarrow}=1$. 
Yet, we have not been able to show this in full generality from the above formula. 
}
\be
    A_{\g\gp}
    &{=} \frac{
        \sgn_{\g\g'}
    }{
        \prod_{
            \lfrak' \in (
                \underline{\lfrak} 
                \cup \mathbb{L}_{\gp}^\downarrow
            )
        }
        \res_{\lfrak'}[\Omega_{\lfrak'}]
    }
    \res_{\mathbb{L}_{\gp}^\downarrow} {\circ}
    \res_{\underline{\lfrak}}
    \bigg[
        \tilde\Omega_{\lfrak_\uparrow}
        {\wedge} \tilde\Omega_{\lfrak_\downarrow}
        {\wedge} \tilde\Omega_{\mathbb{L}_{\g'}^\downarrow}
        {\wedge} \omega 
        {\wedge} \dlog B_{\g\setminus\g'}
    \bigg]
    \!
    \bigg\vert_{\C_{\g'}}
    \!\!\!,
\ee
where 
\be
    \res_{\mathbb{L}_{\g'}^\downarrow} &:= 
    \res_{ \lfrak_{|\mathbb{L}_{\g'}^\downarrow|}'^\downarrow}
    \circ \cdots \circ
    \res_{ \lfrak_{1}'^\downarrow }
    \,,
    &
    \tilde\Omega_{\mathbb{L}_{\g'}^\downarrow} &:= 
    \Omega_{ \lfrak_{1}'^\downarrow }
    \wedge \cdots \wedge
    \Omega_{\lfrak'^\downarrow_{|\mathbb{L}_{\g'}^\downarrow|}}
    \,.
\ee
For this to be non-zero, $\omega$ must donate some $\dlog x_\bullet$'s to $\tilde\Omega_{\lfrak_\uparrow}\wedge \tilde\Omega_{\lfrak_\downarrow}$. 
Anti-commuting the parts of $\omega$ that survive the restriction of the wedge product to the cut, $\omega_{{\lfrak}_\uparrow} = \sum_{v\in V_{\lfrak_\uparrow}} \alpha_v \dlog x_v$ and $\omega_{{\lfrak}_\downarrow} = \sum_{v\in V_{\lfrak_\downarrow}} \alpha_v \dlog x_v$, 
behind $\tilde\Omega_{{\lfrak}_\uparrow}$ and $\tilde\Omega_{{\lfrak}_\downarrow}$ at the cost of another sign
\begin{align}
    \label{eq:AmatOffDiagTemp}
    A_{\g\g^\prime}
    &{=} \frac{
        \sgn_{\g \g^\prime}
        (-1)^{\sum_{\lfrak'\in \mathbb{L}_{\g'}^\downarrow} (|V_{\lfrak'}|-1) }
        (-1)^{|V_{{\lfrak}_\downarrow}|-1}
    }{
        \prod_{
            \lfrak' \in (
                \underline{\lfrak}
                \cup \mathbb{L}_{\gp}^\downarrow
            )
        }
        \res_{\lfrak'}[\Omega_{\lfrak'}]
    }
    \\&\times
    \res_{\mathbb{L}_{\g'}^\downarrow} \circ
    \res_{\underline{\lfrak}}
    \bigg[
        \left(
            \tilde\Omega_{{\lfrak}_\uparrow}
            \wedge \omega_{{\lfrak}_\uparrow}
            \wedge \tilde\Omega_{{\lfrak}_\downarrow}
            + 
            (-1)^{|V_{{\lfrak}_\downarrow}|-1}
            \tilde\Omega_{{\lfrak}_\uparrow}
            \wedge \tilde\Omega_{{\lfrak}_\downarrow}
            \wedge \omega_{{\lfrak}_\downarrow}
        \right)
        \wedge \tilde\Omega_{\mathbb{L}_\g^\downarrow}
        \wedge \dlog B_{\g\setminus\g'}
    \bigg]
    \bigg\vert_{\C_\gp}
    \,.
    \nn
\end{align}
Next, note that 
\be
    &\res_{\underline{\lfrak}}[
        \tilde\Omega_{{\lfrak}_\uparrow}
        \wedge \omega_{{\lfrak}_L}
        \wedge \tilde\Omega_{{\lfrak}_\downarrow}
        \wedge \tilde\Omega_{\mathbb{L}_\g^\downarrow}
        \wedge \dlog B_{\g\setminus\g'}
    ]\vert_{\C_{\g'}}
    \\&\qquad\qquad
    = - \mathrm{sig}(V_{\lfrak_\uparrow}, V_{\lfrak_\downarrow})
    \res_{\lfrak_\uparrow}[\Omega_{\lfrak_\uparrow}]
    \,
    \res_{\lfrak_\downarrow}[\Omega_{\lfrak_\downarrow}]
    \,
    \tilde\Omega_{\mathbb{L}_\g^\downarrow}
    \wedge \underset{ 
        \dlog f_{{\lfrak}_\uparrow}(\mbf{X},\mbf{Y})
    }{\underbrace{
        \dlog B_{\g\setminus\g'}
        \vert_{\C_{\g'}, x_{l\in V_{{\lfrak}_\uparrow}=0}}
    }}
    \,,
    \\
    &\res_{\underline{\lfrak}}[
        \tilde\Omega_{{\lfrak}_\uparrow}
        \wedge \tilde\Omega_{\underline{\lfrak}_\downarrow}
        \wedge \omega_{\underline{\lfrak}_\downarrow}
        \wedge \tilde\Omega_{\mathbb{L}_\g^\downarrow}
        \wedge \dlog B_{\g^\prime\setminus\g}
    ]\vert_{\C_\g}
    \\&\qquad
    = \mathrm{sig}(V_{\lfrak_\uparrow}, V_{\lfrak_\downarrow})
    (-1)^{ |V_{\underline{\lfrak}_\downarrow}| -1}
    \res_{\lfrak_\uparrow}[\Omega_{\lfrak_\uparrow}]
    \,
    \res_{\lfrak_\downarrow}[\Omega_{\lfrak_\downarrow}]
    \,
    \tilde\Omega_{\mathbb{L}_\g^\downarrow}
    \wedge \underset{ 
        \dlog f_{\underline{\lfrak}_\downarrow}(\mbf{X},\mbf{Y})
    }{\underbrace{
        \dlog B_{\g^\prime\setminus\g}
        \vert_{\C_\g, x_{l\in V_{\underline{\lfrak}_\downarrow}=0}}
    }}
    \,.
\ee
Then, equation \eqref{eq:AmatOffDiagTemp} becomes
\vspace{-1em}
\be
    A_{\g\g^\prime}
    &{=} \mathfrak{s}(\g,\gp)
    \frac{
        \res_{\lfrak_\uparrow}[\Omega_{\lfrak_\uparrow}]
        \res_{\lfrak_\downarrow}[\Omega_{\lfrak_\downarrow}]
    }{
        \res_{\underline{\lfrak}}[\Omega_{\underline{\lfrak}}]
    }\;
    \overset{1}{\overbrace{
    \frac{
        \res_{\mathbb{L}_{\g'}^\downarrow}
        \left[\Omega_{\mathbb{L}_\g^\uparrow}\right]
    }{
        \prod_{
            \lfrak' \in \mathbb{L}_{\gp}^\downarrow
        }
        \res_{\lfrak'}[\Omega_{\lfrak'}]
    }
    }}
    \;
    \dlog \frac{
        f_{{\lfrak}_\uparrow}(\mbf{X},\mbf{Y})
    }{
        f_{{\lfrak}_\downarrow}(\mbf{X},\mbf{Y})
    }
    \,,
    \\
    \mathfrak{s}(\g,\gp) &:=
    - \mathrm{sig}(V_{\lfrak_\uparrow}, V_{\lfrak_\downarrow})
    \sgn_{\g \g^\prime}
    (-1)^{
        \sum_{\lfrak'\in \mathbb{L}_{\g'}^\downarrow} (|V_{\lfrak'}|-1) 
        + (|V_{{\lfrak}_\downarrow}|-1)
    } 
    \,.
\ee
One can show that the sign above simplifies to 
\begin{align}
    \mathfrak{s}(\g,\g') 
    &:= \begin{cases}
        \phantom{-} 1 
        & \text{if oriented edges connecting $\lfrak_\uparrow$ to $\lfrak_\downarrow$ point away from $\lfrak_\uparrow$}
        \\
        -1 
        & \text{if oriented edges connecting $\lfrak_\uparrow$ to $\lfrak_\downarrow$ point towards $\lfrak_\uparrow$}
    \end{cases}
    \,.
\end{align}
Finally, taking the remaining residues yields
\be
    A_{\g\g^\prime}
    &= \mathfrak{s}(\g,\gp)
    \frac{
        \Big(\sum_{v \in V_{\lfrak_\uparrow}} \alpha_v\Big)
        \Big(\sum_{v \in V_{\lfrak_\downarrow}} \alpha_v\Big)
    }{
        \prod_{v \in V_{\lfrak_\uparrow} \cup V_{\lfrak_\downarrow}} \alpha_v
    }
    \dlog \frac{
        f_{{\lfrak}_\uparrow}(\mbf{X},\mbf{Y})
    }{
        f_{{\lfrak}_\downarrow}(\mbf{X},\mbf{Y})
    }
    \,.
\ee

\bibliographystyle{JHEP}
\bibliography{main.bbl}

\end{document}